\PassOptionsToPackage{dvipsnames}{xcolor}
\documentclass[11pt]{article}
\usepackage{jheppub}

\usepackage[space]{grffile}
\usepackage{soul}
\usepackage{orcidlink}
\usepackage{pdfpages}
\usepackage{adjustbox}
\usepackage{slashed,physics}
\usepackage{graphicx}
\usepackage{amsmath,amssymb,graphicx}
\usepackage{textcomp} 
\usepackage{gensymb} 
\usepackage{epsf,color}
\usepackage{cancel}
\usepackage{framed}
\usepackage{hyperref}
\usepackage{float}
\usepackage{multirow}
\usepackage{subfigure}
\usepackage[compat=1.1.0]{tikz-feynman}
\usepackage{contour}
\usepackage[normalem]{ulem}
\usetikzlibrary{shapes.geometric}
\tikzfeynmanset{warn luatex=false}
\usepackage{dsfont}

\usepackage{diagbox}
\usepackage{makecell}
\usepackage{colortbl}
\definecolor{colora1}{rgb}{0.996,0.890,0.569}
\definecolor{colora2}{rgb}{0.996,0.769,0.310}
\definecolor{colora3}{rgb}{0.984,0.604,0.161}
\definecolor{colora4}{rgb}{0.925,0.439,0.0784}
\definecolor{colora5}{rgb}{0.800,0.298,0.00784}

\definecolor{nicered}{rgb}{0.5,0.,0.}
\definecolor{nicegreen}{rgb}{0.,0.5,0.}
\definecolor{niceblue}{rgb}{0.,0.,0.5}
\hypersetup{colorlinks,citecolor=nicegreen,linkcolor=nicered,urlcolor=niceblue}
\numberwithin{equation}{section}
\newcommand{\beq}{\begin{equation}}
\newcommand{\eeq}{\end{equation}}
\newcommand{\bea}{\begin{eqnarray}}
\newcommand{\eea}{\end{eqnarray}}
\newcommand{\bear}{\begin{eqnarray}}
\newcommand{\eear}{\end{eqnarray}}

\newcommand{\GeV}{\text{GeV}}

\newcommand{\TeV}{\text{TeV}}
\newcommand{\fb}{\textrm{fb}}
\newcommand{\iab}{\textrm{ab}^{-1}}

\newcommand{\ii}{\mathrm{i}}

\newcommand{\calS}{\mathcal{S}}
\newcommand{\calL}{\mathcal{L}}

\newcommand{\calM}{\mathcal{M}}
\newcommand{\mmu}{m_{\mu}}
\newcommand{\LL}{\mathcal{L}}
\newcommand{\pd}{\partial}

\newcommand{\parameters}{$\alpha\beta$-parameters }

\newcommand{\ba}{\begin{array}}
\newcommand{\ea}{\end{array}}

\newcommand{\hc}{\textrm{h.c.}}

\newcommand\aNLO{{\sc\small MadGraph5\_aMC@NLO}}
\newcommand\WZ{\sc\small Whizard}

\newcommand{\MadLoop}{{\sc\small MadLoop}}

\newcommand{\HEFTm}{{\rm HEFT_{-}}}
\newcommand{\HEFTp}{{\rm HEFT_{+}}}
\newcommand{\SMEFTs}{{\rm SMEFT_{6}}}
\newcommand{\SMEFTe}{{\rm SMEFT_{8}}}

\title {Probing Higgs-muon interactions at a multi-TeV muon collider}

\date{\today}

\author[a,b,\orcidlink{0000-0001-9236-0844}]{Eugenia Celada,}
\affiliation[a]{Dipartimento di Fisica e Astronomia, Universit\`a di Bologna, via Irnerio 46, 40126 Bologna, Italy}
\affiliation[b]{Department of Physics and Astronomy, University of Manchester, Oxford Road, Manchester M13 9PL, United Kingdom}

\author[c,\orcidlink{0000-0002-5543-0716}]{Tao Han,}
\affiliation[c]{Pittsburgh Particle Physics, Astrophysics, and Cosmology Center, 
Department of Physics and Astronomy,
University of Pittsburgh, 3941 O'Hara St
Pittsburgh, PA 15260, USA}

\author[d,\orcidlink{0000-0001-5521-5277}]{Wolfgang Kilian,}
\author[d,\orcidlink{0000-0001-9042-1076}]{Nils Kreher,}
\affiliation[d]{Center for Particle Physics Siegen, University of Siegen, Walter-Flex-Str. 3,
    57072 Siegen, Germany}

\author[e,\orcidlink{0000-0002-9419-6598}]{Yang Ma,}
\affiliation[e]{INFN,
Sezione di Bologna, via Irnerio 46, 40126 Bologna, Italy}

\author[a,e,f,\orcidlink{0000-0003-4890-0676}]{Fabio Maltoni,}
\author[e,\orcidlink{0000-0002-0553-1105}]{Davide Pagani,}
\affiliation[f]{Center for Cosmology, Particle Physics and Phenomenology, Universit\'e catholique de Louvain, B-1348 Louvain-la-Neuve, Belgium}

\author[g,\orcidlink{0000-0003-1866-0157}]{J\"urgen Reuter,}
\affiliation[g]{Deutsches Elektronen-Synchrotron DESY, Notkestr. 85, 22607 Hamburg,
Germany}

\author[d,\orcidlink{0009-0009-2463-5899}]{Tobias Striegl,}
\author[c,h,\orcidlink{0000-0003-4261-3393}]{and Keping Xie}
\affiliation[h]{Department of Physics and Astronomy, Michigan State University, 567 Wilson Road, East Lansing, MI 48824, USA}

\emailAdd{eugenia.celada@postgrad.manchester.ac.uk}
\emailAdd{than@pitt.edu}
\emailAdd{kilian@physik.uni-siegen.de}
\emailAdd{nils.kreher@uni-siegen.de}
\emailAdd{yang.ma@bo.infn.it}
\emailAdd{fabio.maltoni@cern.ch}
\emailAdd{davide.pagani@bo.infn.it}
\emailAdd{juergen.reuter@desy.de}
\emailAdd{tobias.striegl@physik.uni-siegen.de}
\emailAdd{xiekepi1@msu.edu}

\preprint{\\DESY 23-222, PITT-PACC-2325, SI-HEP-2023-33, P3H-23-103,  IRMP-CP3-23-74, MSUHEP-23-034, COMETA-2023-04.}

\abstract{We study the capabilities of a muon collider, at 3 and 10 TeV center-of-mass energy, of probing the interactions of the Higgs boson with the muon.
We consider all the possible processes involving the direct production of EW bosons ($W,Z$ and $H$) with up to five particles in the final state. We study these processes both in the HEFT and SMEFT frameworks, assuming that the dominant BSM effects originate from the muon Yukawa sector. 
Our study shows that a Muon Collider has sensitivity beyond the high-luminosity LHC, especially as it does not rely on the Higgs-decay branching fraction to muons.
A 10 TeV muon collider provides a unique sensitivity on muon and (multi-) Higgs interactions, significantly better than the 3 TeV option. Particularly, we find searches based purely on multi-Higgs production to be particularly effective in probing these couplings.}

\begin{document}
\maketitle

\section{Introduction}
\label{sec:intro}

With the milestone discovery of the Higgs boson at the CERN Large Hadron Collider (LHC) \cite{ATLAS:2012yve,CMS:2012qbp}, the last predicted building block of the Standard Model (SM) of particle physics has been revealed.
Up to the highest energy scale accessible by the current experiments, the SM provides a consistent and accurate description of Nature \cite{ParticleDataGroup:2022pth}, governed by the gauge principle and parameterized by a small set of independent numerical parameters. While the SM is a triumphant achievement as it is, studying the patterns of the model parameters and exploring the underlying principles would not only help to understand nature at short distances as well as  the early universe cosmology, 
but also shed light on the possible new physics beyond the SM (BSM) at higher energy scales.

One of the most intriguing puzzles in the SM is the ``flavor problem", namely the fermion masses and the flavor mixings.
While the dimensionless gauge-coupling and Higgs self-coupling parameters have rather natural values 
of the order of unity, the Yukawa couplings of the quarks and leptons to the Higgs field exhibit a pronounced hierarchy over many orders of magnitude, which is not explained by any known simple symmetry principle~\cite{Froggatt:1978nt}.
Within the SM paradigm, the Yukawa couplings are directly related to the fermion masses. The current experimental observations are consistent with this structure to good precision,
as augmented by a theoretical structure $-$ the ``minimal flavor violation'' \cite{DAmbrosio:2002vsn}. However, in any extension of the SM, the relation between Yukawa couplings and fermion masses is typically modified, reflecting the underlying mass generation mechanism for the fermions. 
In order to test the validity of the SM predictions and to seek new physics beyond the SM, Yukawa couplings must be determined independently in processes that directly involve the Higgs either as a final-state particle or in the internal propagators.

The LHC data on Higgs properties lead us to conclude that the charged fermions of the third generation couple to the Higgs boson with about the strength as expected in the SM.  Current measurements \cite{ATLAS:2020fcp,ATLAS:2020jwz, ATLAS:2021qou,ATLAS:2022yrq, ATLAS:2022vkf,ATLAS:2023cbt,CMS:2020djy,CMS:2021sdq,CMS:2022dwd,CMS:2023tfj} have reached an accuracy of about $(10,20,10)\%$ for the $(t,b,\tau)$, respectively. While our experimental knowledge about the Yukawa couplings of the second and first generations is still limited to quite loose upper bounds~\cite{ATLAS:2022ers,CMS:2022psv,CMS:2022urr}, {\it cf.}~also \cite{Bishara:2016jga,Balzani:2023jas}, the muon Yukawa coupling arises as the next opportunity to scrutinize. The observed signal at the LHC for the $H\to \mu^+\mu^-$ decay \cite{ATLAS:2020fzp,CMS:2020xwi} appears to be in line with the Higgs-muon Yukawa coupling 
of the SM prediction. However, this is a measurement of the decay branching fraction, depending on the assumption of the total width to be the SM value.
Especially, it does not provide information on the sign of the coupling from the decay (or the phase).
Significant deviation with respect to the SM prediction 
is still permissible. It is thus necessary to go beyond the LHC as well as the future hadron colliders, where only the branching fractions are measured.

In the SM, the light-fermion Yukawa couplings are very weak due to the chiral suppression, and thus difficult to directly probe. 
This, on the other hand, opens a window of opportunity for observing large deviation from the SM owing to new physics in the Yukawa sector. For instance, it is conceivable that new physics at the scale of a few TeV exists which couples left- and right-handed fermion components, in addition to the normal Higgs-fermion couplings. Such effects would induce only tiny corrections to the interactions of the third generation since these are dominated by the SM Higgs.  By contrast, for the muon where the SM Yukawa coupling is highly suppressed, the relative correction due to BSM physics could be sizable, within the bounds set by the current measurement. Simultaneously,
the muon magnetic moment, which is also a 
chirality-flip interaction, would be affected.  Such a correction could contribute to the current tension in the muon $g-2$ measurement  \cite{Capdevilla:2020qel,Yin:2020afe,Capdevilla:2021rwo,Dermisek:2021mhi}.
It is thus imperative to scrutinize the muon Yukawa coupling and check the consistency with the SM mass generation mechanism.
In this paper, instead of working on a particular theoretical model, we adopt the framework of the effective field theory (EFT) to investigate the accessible sensitivity to test the BSM couplings and to probe the corresponding new physics scale. To make our study cover as broad a scope as possible, we work within both the Standard Model Effective Field Theory (SMEFT) \cite{Buchmuller:1985jz,Hagiwara:1992eh,Hagiwara:1993ck, Grzadkowski:2010es,Henning:2014wua} and the Higgs Effective Field Theory (HEFT) \cite{Appelquist:1980vg,Longhitano:1980iz,Longhitano:1980tm,Appelquist:1993ka}.  In general, SMEFT is more suitable for a weakly coupled theory to the SM sector below a new physics scale $\Lambda$, while HEFT features the EW Goldstone boson interactions with $\Lambda\sim 4\pi v$, where $v\approx 246$ GeV  is the vacuum expectation value of the SM Higgs field. Given an underlying BSM theory characterized by the new physics scale $\Lambda$, these two formulations should be equivalent if one sums up sufficient terms in the power expansion, but they may differ if we only consider the effects from the leading dimensional operators and thus lead to different predictions.

There has been much interest in recent years to explore the physics potential \cite{MuonCollider:2022xlm,Aime:2022flm,Black:2022cth,Maltoni:2022bqs,Belloni:2022due,Accettura:2023ked} with access to the multi-TeV energy regime and the possibility for a high-energy muon collider  \cite{Delahaye:2019omf,Bartosik:2020xwr,Schulte:2021hgo,Long:2020wfp,MuonCollider:2022nsa,MuonCollider:2022ded,MuonCollider:2022glg}. 
Along with the great opportunities at the energy frontier, a muon collider will be in a unique position of investigating new physics effects in the muon sector, in particular the helicity-flip muon interactions associated with the fermion mass generation mechanism. 
In a previous paper \cite{Han:2021lnp}, we found that $\mathcal{O}(1)$ deviations in the low-energy Yukawa coupling lead to enhancements in multi-boson production at multi-TeV energies which is potentially observable at a high-energy  muon collider. Intuitively, the effects would be more dramatic if a Higgs boson is directly involved in the process. 
We thus extend the study and focus on vector-boson production in association with a Higgs particle, where helicity-flip interactions consistently overtake the SM helicity-conserving production amplitude at high energy if the effective Yukawa coupling deviates from the SM. 
We systematically consider the direct multi-Higgs ($H$) production, associated with multi EW gauge bosons ($V$)
\begin{equation}
\mu^+\mu^-\to mV + nH\, . \label{eq:firsteq}
\end{equation} 
The phenomenology about the $\bar \mu{\mu}H^n\ (n\geq2)$ couplings 
 introduced by such SMEFT or HEFT operators remains largely unexplored.
This class of processes is particularly interesting because there is no helicity-conserving contribution, so the SM prediction is severely suppressed by the muon Yukawa coupling itself. Furthermore, the precision measurement at the high energies will enable us to disentangle the operator structure of new interactions and could point to a particular version of linear or non-linear Higgs representation as the most natural parameterization, below a scale where actual new degrees of freedom would be excited. In combination, a measurement of all accessible multi-boson processes allows us to not just establish new physics but also to study its structure in more detail. In our work we consider processes with multiplicity up to $m+n=5$ in Eq.~\eqref{eq:firsteq}. We notice that the case  $m+n\le 3$ has also been recently considered in Ref.~\cite{Dermisek:2023rvv}, highlighting in particular the $HH$, $HHH$ and $ZHH$ processes. 

Many of the processes of Eq.~\eqref{eq:firsteq} are in principle sensitive also to Higgs-self couplings of the form  $H^n\ (n\geq3)$, which, although they would be probed with high precision at a high-energy muon collider (see {\it e.g.}~\cite{Costantini:2020stv}), are still largely unconstrained \cite{ATLAS:2024fkg}. We  consistently take into account also their possible effects in the study of the $\bar \mu{\mu}H^n$ couplings.

\medskip

First of all we find that the impact of the anomalous Higgs self-couplings is in general negligible in our analysis, which primarily targets the $\bar \mu{\mu}H^n$ couplings. The only exception is the $ZHH$ production process, which however is only one of the several different processes considered in this work.  

We find that a 10 TeV muon collider can be very sensitive to probe the $\bar{\mu}\mu H^n$ anomalous interactions with $n=3,4,5$: at the level of few percents. Assuming a SMEFT scenario where effects from operators of dimension-8 or higher are not unnaturally large, the process $\mu^+\mu^-\to 3H$ leads to an indirect measurement of the strength of the $\bar{\mu}\mu H$ interaction, the only free BSM parameter in this scenario, at the 5\% level. In the general HEFT framework, where much more freedom for the different $\bar{\mu}\mu H^n$ is possible, the precision is only slightly worse, at the level of 10\%. This bound originates from the combination of many different processes, which also lead to a constraint of 20\% on the $\bar{\mu}\mu H^2$ interaction.

We find instead that a 3 TeV muon collider is much less powerful in testing the Higgs-muon coupling than a 10 TeV collider. Bounds on  $\bar{\mu}\mu H^n$ with $n=2, 3,4, 5$ degrade to respectively 40, 30, 50 and 90\%. Also in the aforementioned SMEFT scenario, the  strength of the $\bar{\mu}\mu H$ interaction can be probed only at the 25\% level.

The difference of the bounds between the general HEFT framework and SMEFT scenario of dimension-6 only is due to the fact that the latter is a single particular direction in the multidimensional parameter space of the former. Along this direction the dependence of the  $\mu^+\mu^-\to mV + nH $ cross sections on the different $\bar{\mu}\mu H^n$ interactions is very different than in the rest of the parameter space. In turn, this affects the sensitivity of a muon collider on such interactions. In our study we discuss also other  directions in the parameter space that can impact bounds, such as the one imposed on  $\bar{\mu}\mu H$ and $\bar{\mu}\mu H^2$
by the unitarization condition for the $WWH$, $ZZH$ and $ZHH$. 

The difference between the 3 and 10 TeV results are instead due to the large growth with the energy that is induced by anomalous $\bar{\mu}\mu H^n$ interactions, especially if effects of higher dimensional operators in SMEFT are taken into account, as effectively done in the general HEFT framework. Given the presence of this large growth with the energy we also inspect possible problems with perturbative unitarity violation. We conclude that at 3 and 10 TeV such problems are not present for the processes studied in this work, but would start to appear already at energies as, {\it e.g.}, 30 TeV.  

In order to study perturbative unitarity we calculate high-energy limits in the HEFT framework. We provide a general formalism and explicit formulas for any process considered in this work. These formulas are then also used to understand the behavior of the Monte Carlo simulations based on exact matrix elements, which we use in our phenomenological study in order to extract bounds.

\medskip

This paper is organized as follows. In Sec.~\ref{sec:theory}, we first lay out the theoretical framework and establish the convention in terms of the SMEFT and HEFT. We also demonstrate the motivation and the relevance of those operators of our current interest in connection with predictive UV-complete models. In Sec.~\ref{sec:Unitary} we discuss the high-energy approximation in the HEFT and  introduce four benchmark scenarios (two for HEFT and two for SMEFT) that we exploit to investigate  perturbative unitarity for the processes discussed in our study. In Sec.~\ref{sec:pheno} we  present our predictions for multiple gauge boson and Higgs boson production. We discuss the phenomenology in the HEFT framework and in a dimension-six SMEFT scenario, deriving bounds on anomalous $\bar{\mu}\mu H^n$ interactions. At the same time we verify the negligible contributions from anomalous self-couplings ($H^n$ interactions), in this context.  We summarize our results and draw the conclusions in Sec.~\ref{sec:conclusions}. Technical details and supplementary information are attached in the  appendices at the end, where we also provide two UV models supporting our EFT parameterizations.


\section{Theoretical Framework}
\label{sec:theory}

As mentioned above, the smallness of the first and second generation Yukawa couplings has until now prevented us from revealing that the mass generation mechanism via the Higgs boson is the one realized in the SM. There exist different kinds of BSM scenarios that provide alternative mechanisms for the first two generations, {\it e.g.}, flavored Higgs sectors or multi-Higgs doublet models~\cite{Dermisek:2021mhi,Dermisek:2023rvv}. We will be discussing our analyses of searching for anomalous Higgs-muon couplings in terms of EFT frameworks. For illustration, in Appendix \ref{sec:uvcomplete_models} 
we present two specific model setups in an appendix that generate the relevant EFT operators for these anomalous Higgs-muon couplings: the first class of models mixes the SM muon with a heavy weak singlet (and the Higgs with a scalar singlet), the second class features muon recurrences as they occur in models of partial muon compositeness or Kaluza-Klein extra dimensions. We now turn to the discussion of the EFT framework used in the rest of the analysis. 

\subsection{Linear and non-linear realizations of electroweak symmetry}

In the absence of concrete knowledge for a theory beyond the Standard Model, it is convenient to use an EFT approach to parameterize the effects from BSM physics at a higher scale $\Lambda$.  Following the Wilsonian approach, the EFT provides a local operator expansion in powers of $1/\Lambda^n$, which respects an assumed pattern of symmetry in the Higgs sector.  The expansion is valid below the appearance of new degrees of freedom in the spectrum characterized by $\Lambda$. 

Depending on the treatment of the Higgs sector, there are two formalisms in realizing the electroweak gauge symmetry: 
\begin{itemize}
\item
The Standard Model Effective Field Theory (SMEFT) treats the Higgs field as an SU(2)$_L$ doublet to realize the SM gauge symmetry linearly  in the higher dimensional  operator expansion in $1/\Lambda^n$. This formalism is more suitable for a weakly coupled theory to the SM sector.
It implements the decoupling assumption that the BSM interactions can be formally decoupled by sending a dimensionful parameter $\Lambda$ to infinity \cite{Appelquist:1974tg}.  
\item
The Higgs Effective Field Theory (HEFT) \cite{Appelquist:1980vg,Longhitano:1980iz,Longhitano:1980tm,Appelquist:1993ka}, on the other hand, treats 
 the physical Higgs field as an electroweak singlet and realizes the electroweak gauge symmetry non-linearly on the Goldstone boson triplet. 
The HEFT ansatz retains the electroweak scale $v$, the Higgs vacuum expectation value, and the relevant scale in the parameterization is $\Lambda = 4\pi v \approx 3\,\TeV$, in the hope to describe residual non-decoupling new physics that has not been incorporated in the SM.
In this sense, the HEFT is more suitable to parameterize a strongly-interacting Higgs sector not too far above the EW scale. Example models for these are models of compositeness~\cite{Farhi:1980xs,Dimopoulos:1981xc,Kaplan:1983fs,Kaplan:1983sm} or partial compositeness~\cite{Kaplan:1991dc}, or its dual picture of warped or deconstructed extra dimensions~\cite{Randall:1999ee,Hill:2002ap}.
\end{itemize}

If one resums all orders of the power expansion in the effective field theories, the SMEFT and HEFT must be equivalent to reproducing the underlying UV-complete theory \cite{Weinberg:1978kz}.
In particular, SMEFT and HEFT can be continuously transformed into each other \cite{Kilian:1998bh,Kilian:2003pc}
with an appropriate power-counting.
In practice, however, any EFT power series is truncated at finite order, both to capture the leading effects, and for practical purposes.
As such, the truncated SMEFT and HEFT parameterizations depend on a finite number of parameters and are distinguishable. This is the essential point to present our analyses in both EFT realizations:
any observed pattern of new effects would be better represented by one truncated series or the other, and thus hint at the nature of the underlying theory which, in the context of the current interest, would modify the SM Yukawa sector at multi-TeV energies.
Currently, the absence of new strong dynamics from the Higgs property measurements at the LHC seems to be in favor of a parameterization for a weakly coupled theory in the EW sector. However, as already discussed earlier, the mechanism for fermion mass generation is largely untested and any new underlying dynamics associated with the fermion masses and mixings could be probed by determining the Higgs Yukawa interactions at the higher energy regime.
In this work, we will present results obtained in both these parameterizations in describing the helicity-flip muon interactions
to the Higgs boson and longitudinal gauge bosons (via Goldstone bosons).
We first briefly present our conventions for the SMEFT and HEFT formalisms, respectively, and introduce the frameworks for non-standard Higgs-muon interaction setups.
We use both frameworks to introduce parameterized vertices for analytic results, as well as for Monte Carlo implementations. We then discuss  parameter choices for scenarios in our study before we turn to checks for preservation of tree-level perturbative unitarity.


\subsection{The SMEFT framework}
\label{sec:smeft_framework}
The SMEFT Lagrangian consists of an infinite series of local operators built from the minimal field content with the same multiplet structure as in the SM.  We are particularly interested in those which are relevant for our study of muon  interactions.  The SM contribution reads
  \begin{align}
    \begin{split}
    \mathcal{L}_{\text{EW}}=&-\frac{1}{2} \operatorname{tr}{W_{\mu \nu}  W^{\mu \nu}}-\frac{1}{4}B_{\mu \nu} B^{\mu \nu} + (D_\mu \varphi)^{\dagger}(D^\mu \varphi)+\mu^2 \varphi^{\dagger}\varphi-\lambda (\varphi^{\dagger}\varphi)^2\\ &+ \sum_{f\in\{\ell_L,\mu_R\}} i \bar f^i \slashed D f^i -\left(\bar \ell_L  y_{\mu} \varphi \mu_R + \text{h.c.} \right) + \mathcal{L}_{\text{gauge-fix}}, \label{eq:LEW}
  \end{split}
\end{align}
where $\varphi$ and $\ell_L$ stand for $SU(2)$ doublets
\begin{eqnarray}
    \varphi=\frac{1}{\sqrt 2}
      \begin{pmatrix}
    \sqrt 2 \phi^+ \\
    v+H +i \phi^0
    \end{pmatrix}, ~~~~
    \ell_L=
    \begin{pmatrix}
     \nu_\mu \\ \mu
    \end{pmatrix}_L.
\end{eqnarray}
We add the following series of SMEFT operators,
\begin{equation}
    \mathcal{L}_{\text{SMEFT},\mu \varphi} = - \sum_{n=1}^\infty
    \frac{c_\varphi^{(2n+4)}}{\Lambda^{2n}}
    (\varphi^\dagger\varphi)^{n+2}
    - \sum_{n=1}^\infty
    \frac{c_{\ell\varphi}^{(2n+4)}}{\Lambda^{2n}}
    (\varphi^\dagger\varphi)^n
    ({\bar \ell}_L\varphi \mu_R+\hc),
    \label{eq:SMEFT}
\end{equation}
where $c_{\ell \varphi}^{(D)}$ and $c_\varphi^{(D)}$  denote the Wilson coefficients of Yukawa and Higgs self-coupling operators with mass dimension $D$, respectively. The  $c_{\ell \varphi}^{(D)}$ coefficients parameterize the anomalous muon-Higgs interactions, which are the main subject of our work. Instead, the  $c_\varphi^{(D)}$ coefficients parameterize the Higgs self couplings, which also contribute to the same processes considered in this work, as already anticipated in the Introduction, and therefore cannot be in principle neglected. The parameter $\Lambda$ is a constant with the dimension of mass, which is left unspecified at this point.  The Wilson coefficients themselves are dimensionless numbers. Generically speaking, if the EFT approach effectively parameterizes the new physics at a high scale $\Lambda$, then the Wilson coefficients $c_\varphi$ should be the order of unity for a weakly coupled theory, or $4\pi$ for a strongly interacting theory. 
Note that this naive expectation might be wrong for a parameter like the muon Yukawa coupling which is unnaturally small in the SM. Here, a ``correction'' of a factor of several tens could be possible even in a weakly coupled scenario if the underlying mass generation mechanism differs from the SM. 

After inserting the Higgs vacuum expectation value (VEV) $v$, the Wilson coefficients in Eq.~\eqref{eq:SMEFT} contribute to observable masses and couplings and thus modify the formal relations between their observed values and the corresponding SM parameters. We define a modified VEV $\bar v$ as the minimum of the complete Higgs potential of $\mathcal{L}_{\text{SM}}+\mathcal{L}_{\text{SMEFT},\mu\varphi}$, which is given by the solution of the implicit equation
\begin{equation}
\label{eq:vevbeforect}
    0=\left(\mu^2-\lambda \bar v^2 -\sum_{n=1}^\infty
    \frac{c_{\varphi}^{(2n+4)}}{\Lambda^{2n}}(n+2)\left(\frac{\bar v^2}{2}\right)^{n+1}\right).
\end{equation}
Once the Higgs field $\phi$ is expanded around the VEV, a Higgs mass square is generated of the form
\begin{equation}
    \bar m_H^2=\left(2\mu^2+2\sum_{n=1}^\infty
    \frac{c_{\varphi}^{(2n+4)}}{\Lambda^{2n}}n(n+2)\left(\frac{\bar v^2}{2}\right)^{n+1}\right),
\end{equation}
where we have used the condition Eq.~\eqref{eq:vevbeforect} to obtain the result.
The muon mass introduced from $ \mathcal{L}_{\text{SM}}+\mathcal{L}_{\text{SMEFT},\mu\phi}$ reads
\begin{align}
    \bar m_{\mu}&=\frac{\bar v}{\sqrt 2}\left(y_{\mu }+\sum_{n=1}^\infty
    \frac{c_{\ell\varphi}^{(2n+4)}}{\Lambda^{2n}}
    \left(\frac{\bar v^2}{2}\right)^n\right) .
\end{align}
However, we regard the experimentally measured masses $m_\mu$ and $m_H$ and the value of $v$ determined, {\it e.g.}, from the Fermi constant as input and want to retain their relations within the SM-only Lagrangian. This can be achieved by a finite renormalization of fields and parameters in the broken phase.  Alternatively, we may add the following EW-symmetric tree-level counterterm Lagrangian
\begin{align}
    \mathcal{L}_{ct}=&
    \sum_{n=1}^\infty
    \frac{c_{\varphi}^{(2n+4)}}{\Lambda^{2n}}\left(\frac{v^2}{2}\right)^{n}
    \left[\left(\frac{v^2}{2}\right)^{2}+
    \begin{pmatrix}n+2 \\ 1 
    \end{pmatrix}
    \frac{v^2}{2}
    \left(\varphi^{\dagger}\varphi-\frac{v^2}{2}\right)+
    \begin{pmatrix} n+2 \\ 2
    \end{pmatrix}
    \left(\varphi^{\dagger}\varphi-\frac{v^2}{2}\right)^2\right]\notag\\
    & + \sum_{n=1}^\infty
    \frac{c_{\ell\varphi}^{(2n+4)}}{\Lambda^{2n}}
    \left(\frac{v^2}{2}\right)^n
    ({\bar \ell}_L\varphi \mu_R+\hc) 
    \quad ,
    \label{eq:SMEFT-ct}
\end{align}
which depends on the $D>4$ Wilson coefficients, but contains only terms of dimension $\leq 4$.  The counterterms do not add any new effects but rather restore the familiar SM relations
\begin{align}
    v&=\sqrt{\frac{\mu^2}{\lambda}},
    &m^2_{H}&=2\mu^2,
    &m_{\mu}&=\frac{y_{\mu } v}{\sqrt{2}}.
    \label{eq:SM-input}
\end{align}
The numerical values of $v,m_H,m_\mu$ thus remain fixed in the presence of higher-dimensional Lagrangian terms while the $D\geq 6$ operators in Eq.~\eqref{eq:SMEFT} describe corrections to the observable triple-Higgs and muon-Yukawa couplings, as well as new vertices which are not present in the SM.

Now we can expand the sum of the effective Lagrangian (Eq.~\eqref{eq:SMEFT}) and the tree-level counter terms (Eq.~\eqref{eq:SMEFT-ct}) up to dimension eight:
\begin{align}
\label{eq:SMEFTandct}
\begin{split}
  &\mathcal{L}^{(6)+(8)}_{\text{SMEFT},\mu\varphi} + \mathcal{L}^{(6)+(8)}_{ct}\\
  &=-
    \frac{c_\varphi^{(6)}}{\Lambda^{2}}
    \left[(\varphi^\dagger\varphi)^{3}-\left(\frac{v^2}{2}\right)^{3}-3\frac{v^2}{2}
    \left(\varphi^{\dagger}\varphi-\frac{v^2}{2}\right)
    \varphi^{\dagger}\varphi\right]\\
    &\quad- 
    \frac{c_\varphi^{(8)}}{\Lambda^{4}}
    \left[(\varphi^\dagger\varphi)^{4}-\left(\frac{v^2}{2}\right)^{4}-4\left(\frac{v^2}{2}\right)^{2}
    \left(\varphi^{\dagger}\varphi-\frac{v^2}{2}\right)\left[\frac{v^2}{2}+\frac{3}{2}
    \left(\varphi^{\dagger}\varphi-\frac{v^2}{2}\right)\right]\right]\\
    &\quad- 
    \frac{c_{\ell\varphi}^{(6)}}{\Lambda^{2}}
    \left[(\varphi^\dagger\varphi)-\frac{v^2}{2}\right]
    ({\bar \ell}_L\varphi \mu_R+\hc) -
    \frac{c_{\ell\varphi}^{(8)}}{\Lambda^{4}}
    \left[(\varphi^\dagger\varphi)^2-\left(\frac{v^2}{2}\right)^{2}\right]
    ({\bar \ell}_L\varphi \mu_R+\hc)  
\end{split}
\end{align}
These terms set the stage for our phenomenological studies interpreted in terms of the SMEFT framework. 

\subsection{The HEFT framework}
\label{sec:heft_framework}

As an alternative to the above SMEFT parameterization, the HEFT Lagrangian works with a non-linear parameterization of the Higgs sector, separating out the Higgs boson $H$ as a singlet and the Goldstone bosons $\phi^a$ as a triplet under $SU(2)$ transformations. The Higgs-dependent part of the HEFT Lagrangian is written as
\begin{equation}
    \mathcal L_{Uh} =\frac{v^2}{4}\operatorname{tr}[D_{\mu}U^{\dagger}D^{\mu}U]
    F_U(H)+\frac{1}{2}\partial_{\mu}H\partial^{\mu}H-V(H) 
    -\frac{v}{\sqrt{2}}\left[\bar \ell_L Y_\ell(H) UP_-\ell_R+\operatorname{h.c.}\right],
  \label{eq:HEFT}
\end{equation} 
where the matrix-valued field $U$ is defined by
\begin{equation}
  U=e^{2i\phi^aT_a/v} \quad \text{with} \quad \phi^aT_a=\frac{1}{\sqrt 2}\begin{pmatrix}
    \frac{\phi^0}{\sqrt 2} &  \phi^+\\
    \phi^-& -\frac{\phi^0}{\sqrt 2} 
  \end{pmatrix} ,
\end{equation}
or alternatively
\begin{equation}\label{eq:Utrigonemetric}
 U
 = \cos\left(\frac{\rho}{v}\right)\mathds 1+\frac{2 i}{\rho} \sin\left(\frac{\rho}{v}\right)\phi^aT_a
 \quad \text{with} \quad \rho=\sqrt{\phi_0^2+2\phi_+\phi_-}\,.
\end{equation}
The covariant derivative of $U$ is defined as
\begin{equation}
  D_{\mu}U=\partial_{\mu}U+igW_{\mu}U-ig'B_{\mu}UT_3,
\end{equation}
$T_a=\frac{\tau_a}{2}$ represents the generators of  $\operatorname{SU}(2)$ in the fundamental representation. $\{\phi^+,\phi^-,\phi^0\}$ are the Goldstone bosons which correspond to the physical vector bosons $\{W^+_L,W^-_L,Z_L\}$.
Furthermore, we define $\ell_R=(0,\mu_R)^T$ and $P_{\pm}=\tfrac{1}{2}(1  \pm \tau_3)$.

The functions $F_U(H)$, $V(H)$ and $Y_\ell(H)$ can be expressed as series expansions,
\begin{align}  F_U(H)&=1+\sum_{k\geq1}f_{U,k}\left(\frac{H}{v}\right)^k ,\\  V(H)&=v^4\sum_{k\geq2}f_{V,k}\left(\frac{H}{v}\right)^k,
\\
  Y_{\ell}(H)&=\frac{\sqrt 2  m_\ell}{v}+\sum_{k\geq 1}y_{\ell,k}\left(\frac{H}{v}\right)^k
  \quad .
\end{align}
The HEFT Lagrangian reduces to the SM for the specific choice
\begin{align}\label{eq:HiggsParameters}
  f_{V,2}&= \frac{m_H^2}{2v^2}\,,
  &f_{V,3}&=\lambda\,,
  &f_{V,4}&=\frac{\lambda}{4}\,,\notag\\
  f_{U,1}&=2\,,
  &f_{U,2}&=1\,,\notag\\
  y_{\ell,1}&=\frac{\sqrt 2 m_{\ell}}{v}\,.
\end{align}
For generic parameter values, the HEFT Lagrangian again describes deviations from the SM predictions for the Higgs self-couplings and the muon Yukawa coupling, as well as new vertices not present in the SM. The expansion of the $U$ matrix field generates infinite towers 
of multiple Goldstone-emission amplitudes. For specific parameter choices, in particular for the SM point~\eqref{eq:HiggsParameters}, the multi-particle emission amplitudes mostly cancel, leading to a weakly interacting high-energy limit. Without such cancellations, a generic HEFT scenario typically leads to strongly interacting physics at high energy.

\subsection{Parameterized couplings}
\label{eq:paramvertices}

In both the SMEFT and the HEFT frameworks, there is the possibility of choosing the unitary gauge for all electroweak calculations, which is actually exploited by standard calculation tools.  The formalism allows us to eliminate unphysical modes altogether (ghost and Goldstone bosons) and to reduce the gauge symmetry to its unbroken part, QED and QCD.  The basic principles of $S$-matrix theory guarantee that observables and their mutual relation are not affected by the choice of calculation method, so simplifying either Lagrangians in this way does not alter physical predictions.

The most general parameterization of interactions in terms of observable states involves the so-called form factors, {\it i.e.}, momentum-dependent coupling matrices of the Lorentz-covariant fields which are associated with observable particles.   In the absence of singularities, {\it i.e.}, extra undiscovered light degrees of freedom, form factors admit a low-energy expansion.  For simplicity, we may keep the leading momentum-independent term in this expansion for each distinct coupling.  We parameterize couplings of type $\bar{\mu}_L\mu_RH^n$ by $\alpha_n$ and couplings of type $H^n$ by $\beta_n$.  In terms of a Lagrangian built from physical fields only, we have
\begin{eqnarray}
  \mathcal{L}\supset  -\frac{m_H^2}{2}H^2 - m_\mu \bar{\mu}\mu - \sum_{n=3}^\infty \beta_n \frac{\lambda}{v^{n-4}} H^n - \sum_{n=1}^\infty \alpha_n \frac{m_\mu}{v^n}\bar{\mu}\mu H^n. \label{eq:LextendedK}
\end{eqnarray}
The SM parameter set is given by $\alpha_1=1$ and $\alpha_n=0$ for $n>1$ in the Yukawa sector, and $\beta_3=1$, $\beta_4=1/4$ but $\beta_n=0$ for $n>4$ in the Higgs sector.

As expected, this Lagrangian coincides with the unitary-gauge version of the HEFT Lagrangian~\eqref{eq:HEFT} if the operator expansion is kept up to a sufficiently high order.  We obtain the relations
\begin{align}
   y_{\mu,n} &=\frac{{\sqrt 2} m_\mu}{v} \alpha_n\,,
   &f_{V,n} &= \beta_n \lambda\,,\label{eq:alphaandbeta}
\end{align}
which allows us to express the HEFT parameters in terms of the $\alpha_i$ and $\beta_j$.  Analogously, the Lagrangian~\eqref{eq:LextendedK} also coincides with the SMEFT Lagrangian evaluated in unitary gauge if we again expand sufficiently high in operator dimension. 

We can thus relate all three parameter sets, 

\begin{equation}\label{eq:map}
\begin{aligned}
  &\alpha_1= \frac{v}{\sqrt 2 m_\mu} y_{\mu,1} = 1+ \frac{v^3}{\sqrt 2 m_\mu} \frac{c_{\ell \varphi}^{(6)}}{\Lambda^2}+ \frac{v^5}{\sqrt 2 m_\mu} \frac{c_{\ell \varphi}^{(8)}}{\Lambda^4}+ \frac{3v^7}{4\sqrt 2 m_\mu} \frac{c_{\ell \varphi}^{(10)}}{\Lambda^6}\,, \\
  &\alpha_2= \frac{v}{\sqrt 2 m_\mu} y_{\mu,2} =  \frac{3v^3}{2\sqrt 2 m_\mu} \frac{c_{\ell \varphi}^{(6)}}{\Lambda^2} +\frac{5 v^5}{2\sqrt 2 m_\mu} \frac{c_{\ell \varphi}^{(8)}}{\Lambda^4}+\frac{21 v^7}{8\sqrt 2 m_\mu} \frac{c_{\ell \varphi}^{(10)}}{\Lambda^6}\,,  \\
&\alpha_3 = \frac{v}{\sqrt 2 m_\mu} y_{\mu,3} = \frac{v^3}{2\sqrt 2 m_\mu} \frac{c_{\ell \varphi}^{(6)}}{\Lambda^2} +\frac{ 5 v^5}{ 2\sqrt 2m_\mu} \frac{c_{\ell \varphi}^{(8)}}{\Lambda^4}+\frac{35v^7}{8\sqrt 2 m_\mu}\frac{c_{\ell\varphi}^{(10)}}{\Lambda^6}\,,  \\
&\alpha_4 = \frac{v}{\sqrt 2 m_\mu} y_{\mu,4} = \frac{ 5 v^5}{ 4\sqrt 2m_\mu} \frac{c_{\ell \varphi}^{(8)}}{\Lambda^4}+\frac{35v^7}{8\sqrt 2 m_\mu}\frac{c_{\ell\varphi}^{(10)}}{\Lambda^6}\,,  \\
\end{aligned}
\end{equation}
\begin{equation*}
\begin{aligned}
 &\alpha_5 = \frac{v}{\sqrt 2 m_\mu} y_{\mu,5} = \frac{v^5}{ 4\sqrt 2m_\mu} \frac{c_{\ell \varphi}^{(8)}}{\Lambda^4}+\frac{21v^7}{8\sqrt 2 m_\mu}\frac{c_{\ell\varphi}^{(10)}}{\Lambda^6}\,,
 \phantom{+\frac{35v^7}{\sqrt 2 m_\mu}\frac{c_{\ell\varphi}^{(10)}}{\Lambda^6}}
 \end{aligned}
\end{equation*}
where in the SMEFT expansion, we have stopped at operator dimension ten. 

Stated differently, any HEFT parameter point is technically equivalent to a corresponding SMEFT parameter point, and vice versa.   Nevertheless, distinct predictions of either formalism emerge if the operator series is truncated at finite order, {\it i.e.}, all subsequent coefficients are assumed to be negligible.
For the Yukawa sector, we list in  Appendix \ref{sec:relations} all the  relations for matching HEFT in terms of SMEFT, and vice versa, considering three independent parameters: $\{c^{(6)}_{\ell\varphi},c^{(8)}_{\ell\varphi},c^{(10)}_{\ell\varphi}\}$ in SMEFT and $\{y_{\mu,1},y_{\mu,2},y_{\mu,3}\}$ or equivalently $\{\alpha_1,\alpha_2,\alpha_3\}$ in HEFT.  
We also provide the corresponding Feynman rules for the vertices with up to six particles in Appendix \ref{sec:feynmanrules}.

The physical implications of truncating either the SMEFT or the HEFT power expansion, are reflected by the free (heavy) mass parameter $\Lambda$ in the former series as opposed to the fixed electroweak scale $v$ in the latter series. A SMEFT parameter point in the decoupling limit $\Lambda\to\infty$ corresponds to exceptional fine-tuning and large UV cancellations in the HEFT expansion.  Conversely, generic HEFT parameter points imply high-multiplicity particle emission and non-converging behavior of high-dimension Wilson coefficients in the SMEFT picture.

In SMEFT, Wilson coefficients are naively expected to be less than $4\pi$ (we already commented on the different possibilities for an unnaturally small SM parameter like the muon Yukawa coupling). Under this assumption,  {\it e.g.}, varying $c_{\ell \varphi}^{(6)}$ in $[-4\pi,4\pi]$ for a given BSM scale $\Lambda = 20\,{\rm TeV}$, which is a reasonable assumption given the potentially accessible energies of a muon collider,  yields the naive parameter range for the $\bar{\mu} \mu H$ couplings $\alpha_1 \in [-2.14,4.14]$. By varying $c_{\ell \varphi}^{(8)}$ in the same range $[-4\pi,4\pi]$, and assuming $c_{\ell \varphi}^{(6)}\simeq c_{\ell \varphi}^{(8)}$,  the range of $\alpha_{1}$ is affected only at the sub-permille level.\footnote{We are here implicitly assuming a ``well'' behaving SMEFT parameterization, not only with higher-order effects in powers of $1/\Lambda^2$ that are suppressed, but also loop effects induced by the SMEFT itself are negligible. In other words, having perturbativity under control. Outside of this regime, a HEFT description is more suitable. Alternatively, one could count the powers of the generic coupling $g_*$ associated with the generic underlying UV model, which for the operators considered is $c_{\ell \varphi}^{(2n+4)}\sim g_*^{2n+1}$, with now $g_*$ bounded by $4\pi$. We notice that following this argument  $|c_{\ell \varphi}^{(8)}/c_{\ell \varphi}^{(6)}|<(4\pi)^2$ and therefore ``sub-permille'' should be replaced by ``per-cent level'' in the main text.} Similar considerations apply to the case of $\alpha_{2}$ and $\alpha_{3}$ and the analogous effects from $c_{\ell \varphi}^{(10)}$ are even smaller.  On the other hand, with the same assumptions, $|\alpha_{4}|,|\alpha_{5}|\ll |\Delta \alpha_{1}|, |\alpha_{2}|, |\alpha_{3}|$, since $\alpha_{4}$ and $\alpha_{5}$ emerge only at dimension eight in SMEFT, with, {\it e.g.}, $|\alpha_{4}| \simeq 0.01 \%  \,\cdot\,|\Delta \alpha_{1}|$. 

The quantitative statements in the previous argument depend on the specific value chosen for $\Lambda$ and especially on the correlation between the size of $c_{\ell \varphi}^{(6)}$ and $c_{\ell \varphi}^{(8)}$ (see Appendix \ref{sec:model-d8} for a UV model precisely invalidating this assumption). Nevertheless, the pattern that we have shown is a clear sign of the different underlying physics that can be captured via either the SMEFT or HEFT framework. While $|\alpha_{1}|\simeq |\alpha_{4}|$ would be unnatural in the SMEFT language, this is not the case in HEFT. Similarly, while having   $\alpha_{1}$, $\alpha_{2}$ and $\alpha_{3}$ correlated is unnatural in HEFT, it is not the case in SMEFT. This last point will be further discussed in the next section and it is manifest in Eqs.~\eqref{eq:dim6lock} and \eqref{eq:dim8lock}.  In our study, we will choose the $\alpha_k$ set for a common parameterization, but we will always stress  the different interpretation within the HEFT or SMEFT framework.  

\subsection{EFT scenarios}
\label{sec:EFT-scenarios}

If we want to reduce the multi-dimensional parameter space of the generic EFT, we can identify specific scenarios that serve as benchmarks.
The choice of the benchmarks is clearly not unique and also in our work, given different motivations, we will make different choices when discussing the unitarity bounds in Sec.~\ref{sec:Unitary} or the phenomenological analysis in  Sec.~\ref{sec:pheno}. Overall, the benchmarks that will be considered are the following:

\begin{enumerate}
\item {\bf SM}: The unmodified SM where $D>4$ operators do not appear.  The SM exhibits the unique feature that all amplitudes respect unitarity bounds, and furthermore all pure multi-Higgs final states in muon collisions are severely suppressed compared to Higgs production associated with vector bosons. 

\item {\bf $\boldsymbol{\SMEFTs}$ and $\boldsymbol{\SMEFTe}$}: In these two scenarios we target new physics of decoupling nature, which is well described by the SMEFT framework.  In general, in the SMEFT the leading operator that competes with the SM Yukawa coupling could be of dimension six, eight, or higher~\cite{Maltoni:2001dc}.  For our study we will specifically consider the cases where the leading BSM contribution is in the Yukawa sector and it is given by the $D=6$ operator  ($\SMEFTs$)  or  alternatively the $D=8$ operator ($\SMEFTe$).  Higher-dimensional corrections are dropped by truncating the SMEFT series.  
For our phenomenological collider studies, we will only consider ($\SMEFTs$), while we will keep the case ($\SMEFTe$) in the discussion of unitarity below.
The latter is precisely invalidating the condition $c_{\ell \varphi}^{(6)}\simeq c_{\ell \varphi}^{(8)}$ and it is of particular interest when studying unitarity constraints.
Indeed, these simplified SMEFT scenarios provide enhancements in multi-boson final states that extend only up to finite multiplicity.   
In Appendix \ref{sec:uvcomplete_models}, we present specific simplified models which realize both the $\SMEFTs$ and $\SMEFTe$ scenarios.

\item {\bf $\boldsymbol{\HEFTm}$ and $\boldsymbol{\HEFTp}$}: In these two scenarios we target non-decoupling and presumably strongly interacting new physics in the TeV energy range, which could also involve more complex patterns of decoupling new physics at various scales (with the scale usually limited by $4\pi v$ in the sector coupled to the EW gauge sector). 
This is represented by generic values of $\alpha_i$ and  $\beta_i$, as  in Eq.~\eqref{eq:alphaandbeta}.  For illustration, we introduce two arbitrarily selected parameter sets:  $\alpha_i=\pm \alpha_1$ with alternating sign; this benchmark point ($\HEFTm$) does not involve cancellations in observables and may thus be labeled as a conservative guess.  All $\alpha_k\equiv\alpha_1$; this benchmark point ($\HEFTp$) results in considerable cancellations in various interactions. For instance, it leads to small enhancements of $ZHH$, $ZZH$, and $WWH$ cross sections compared to the SM, as we will discuss in detail in Sec.~\ref{sec:multiVH}.\end{enumerate}

In the previous scenarios, we did not specify any condition on $\beta_i$ or $c_{\varphi}^{(2n+4)}$ since we have found that their contributions are in general negligible. This will be clear both in Sec.~\ref{sec:Unitary}, where we derive and discuss  the high-energy limit of the cross sections for all the multi-boson production processes, and in  the phenomenological analysis presented in Sec.~\ref{sec:pheno}.  
We recall that any parameter set can be translated to the corresponding HEFT or SMEFT parameterizations if we accept that within the SMEFT power-counting framework, generic $\alpha$ values imply considerable fine-tuning, and low-energy symmetries would have to be qualified as emergent or accidental.  Nevertheless, such a scenario is not excluded by any fundamental principles except unitarity.  In the following chapter, we will therefore review the unitarity bounds which set the ultimate restrictions on BSM contributions.

\medskip

\begin{table}[t!]
\renewcommand{\arraystretch}{1.2} 
\begin{displaymath}
    \begin{array}{l|c|cc|cc}
      \hline
      & \text{SM} & \SMEFTs & \SMEFTe & \HEFTm & \HEFTp \\
      \hline
      \Delta\alpha_1= & 0   
      &\frac{v}{\sqrt2 m_\mu}\frac{v^2}{\Lambda^2}c^{(6)}_{\ell\varphi}
      &\frac{v}{\sqrt2 m_\mu}\frac{v^4}{\Lambda^4}c^{(8)}_{\ell\varphi}
      &\frac{v}{\sqrt2 m_\mu}y_{\mu,1}-1
      &\frac{v}{\sqrt2 m_\mu}y_{\mu,1}-1
      \\
      \hline
      \alpha_1 & 1 & 1 + \Delta\alpha_1 & 1 + \Delta\alpha_1 & 1+\Delta\alpha_1 & 1+\Delta\alpha_1\\
      \alpha_2 & 0 & \frac32\Delta\alpha_1 & \frac52\Delta\alpha_1 & -(1+\Delta\alpha_1) & 1+\Delta\alpha_1 \\
      \alpha_3 & 0 & \frac12\Delta\alpha_1 & \frac52\Delta\alpha_1 & 1+\Delta\alpha_1 &1+ \Delta\alpha_1\\
      \alpha_4 & 0 & 0 & \frac54\Delta\alpha_1 & -(1+\Delta\alpha_1) & 1+\Delta\alpha_1 \\
      \alpha_5 & 0 & 0 & \frac14\Delta\alpha_1 & 1+\Delta\alpha_1 &1+ \Delta\alpha_1 \\
      \alpha_6 & 0 & 0 & 0 & -(1+\Delta\alpha_1) & 1+\Delta\alpha_1 \\
     \hline
    \end{array}
\end{displaymath}
\caption{ Muon couplings to $k$ Higgs bosons ($\alpha_k$)  in the benchmark scenarios defined in the text.  For each scenario except for the SM, there is a single free parameter $\Delta\alpha_1$.  The definition of the parameter $\Delta\alpha_1$ depends on the chosen scenario, {\it cf.}~{\it e.g.}, Eqs.~\eqref{eq:dim6lock} and \eqref{eq:dim8lock}.
\label{tab:benchmark}}
\end{table}

As can be easily deduced from Eq.~\eqref{eq:map}, the presence of 
only the dimension-6 operator in the SMEFT framework is equivalent to having non-vanishing $\alpha_i$ values only for $1\le i \le 3$. The following relation links $\alpha_i$ values for $1\le i \le 3$ leading to
\begin{flalign}
\SMEFTs: \quad
\Delta\alpha_1\equiv \alpha_1-1=\frac{2}{3}\alpha_2=2\alpha_3\,, 
\label{eq:dim6lock}
\quad {\rm and} \quad  
\alpha_4=\alpha_5=0\,. && 
\end{flalign} 
Analogously, the presence of only the dimension-8 operator in the SMEFT framework is equivalent to having non-vanishing $\alpha_i$ values for $1\le i \le 5$ and with the following relations 
\begin{flalign}
\SMEFTe: \qquad
\Delta\alpha_1\equiv \alpha_1-1=\frac{2}{5}\alpha_2=\frac{2}{5}\alpha_3=\frac{4}{5}\alpha_4=4 \alpha_5\,.  &&
\label{eq:dim8lock}
\end{flalign} 
We list in Tab.~\ref{tab:benchmark} the first five $\alpha_i$ parameters, as given by Eq.~\eqref{eq:map}, for the benchmark scenarios defined above.  Each scenario depends on one free parameter, which we choose to be $\Delta\alpha_1$.  We emphasize that while the SMEFT scenarios ($\SMEFTs$ and $\SMEFTe$) are motivated by power-counting arguments (and supported by the simplified models in  Appendix \ref{sec:uvcomplete_models}), the HEFT scenarios ($\HEFTm$ and $\HEFTp$)  are deliberately chosen as representatives of new physics scenarios that for any reason do not follow power-counting expectations at all.  
Ultimately, detailed muon-collider data should allow us to determine all Yukawa couplings independently and thus decide which kind of scenario is actually valid.


\section{Unitarity bounds for the EFT setups}
\label{sec:Unitary}

Within the EFT frameworks, any parameter set that deviates from the SM describes amplitudes that rapidly grow with energy. This high-energy behavior of individual terms becomes more pronounced with each additional order of higher-power or higher-multiplicity operators included in the Lagrangian.  Eventually, the EFT prediction will saturate the bound given by the optical theorem at some energy value, and the approximation will break down at or below this scale. In weakly-coupled theories, for instance, new states normally show up considerably below that scale. The effective description has to give way to a new and more complete BSM theory. Unfortunately, for some SM production and scattering processes at the LHC ({\it e.g.}, vector-boson scattering), this fact leads to a dilemma. If the BSM effect is large enough to be measured within an EFT, the associated scale ($\Lambda$ in SMEFT or $4\pi v$ in HEFT) typically falls within the kinematically accessible energy range \cite{Kilian:2014zja,Chaudhary:2019aim,Lang:2021hnd}, invalidating the EFT frameworks as a model-independent approach. Different approaches how to test and deal with this intrinsic limitation of EFTs have been proposed, see {\it e.g.}, Refs.~\cite{Trott:2021vqa,Brivio:2022pyi}. 

 By contrast, the muon-Yukawa interaction is severely suppressed in the SM.  This fact leaves ample room for a power-law rise of muon-scattering amplitudes from BSM effects which stays away from unitarity bounds.\footnote{This is similar to the case of gauge couplings, for which more room for new physics is left in the scattering of transversal EW gauge bosons w.r.t.~longitudinal ones, due to precisely unitarity bounds \cite{Brass:2018hfw}.} An EFT parameterization could remain valid well into the multi-TeV range that is probed at a muon collider. This observation strongly motivates us to explore the wide parameter space for BSM new physics associated with the Higgs-muon interaction. In the following, we evaluate the upper limit of the allowed energy range where the EFT description is consistent with unitarity bounds.  We account for all relevant multi-particle scattering channels for the $\mu^+\mu^-$ initial state, so the limit applies to the inclusive cross section summed over all multi-boson final states.

\subsection{High-energy limit in HEFT}
\label{sec:HEHEFT}

The optical theorem relates the squared scattering matrix elements $\calM_{iX}$ of initial-final transitions $i\to X$ to the forward-scattering amplitude. As usual, we expand the
elastic scattering matrix element in terms of Legendre polynomials,
\begin{equation}
 \calM_{ii}=16\pi\sum_{l=0}^{\infty}(2l+1)a_l P_{l}(\cos \theta)\,.
\end{equation}
If we denote the inelastic cross section by $\sigma_{iX}$ and indicate the average (sum) over initial (final) helicities by a bar, respectively, the optical theorem can be stated as the Argand-circle condition ({\it cf.}, {\it e.g.}, \cite{ParticleDataGroup:2020ssz})
\begin{equation}
  \overline{\operatorname{Re}(a_l^2)}
  +\left(\overline{\operatorname{Im}(a_l)}-\frac{1}{2}\right)^2
  + \frac{s}{16\pi}\sum_{X\neq i} \bar \sigma_{iX}\bigg|_{l-\text{wave}}
  = \frac14\,.
\end{equation}
This condition leads to the simple inequality for the total inelastic cross section of each partial wave
\begin{equation}
\label{eq:unitarityconditionlwave}
\sum_{X\neq i} \bar \sigma_{iX}\bigg|_{l-\text{wave}}\leq\frac{4\pi}{s}.
\end{equation}

To use the bound above, we need to evaluate this sum for the parameterized EFT amplitudes with all multi-boson final states taken into account, in the high-energy limit.  To this end, we make use of the Goldstone-boson equivalence theorem (GBET)~\cite{Chanowitz:1985hj,Cornwall:1973tb,Cornwall:1974km}.  The GBET can be stated in any formulation of the EFT which implements gauge symmetry.  This includes both the HEFT and SMEFT frameworks.  In the high-energy limit, amplitudes involving longitudinally polarized vector bosons are approximated by the corresponding amplitudes involving final-state Goldstone bosons, up to corrections $\mathcal{O}(m_W^2/s)$.  Since we do not consider EFT operators which specifically involve gauge fields, {\it i.e.}, transversally polarized vector bosons, this description is exhaustive and allows us to reliably approximate each inelastic cross section that contributes to the optical theorem in muon scattering.

For concreteness, we adopt the HEFT framework where Higgs and Goldstone bosons are treated separately.  The results can be easily translated to the SMEFT parameterization by means of Eq.~\eqref{eq:map}.  (For an analogous SMEFT calculation, {\it cf.}\ \cite{Maltoni:2001dc}).  We thus start with the gauged HEFT Lagrangian in  Eq.~\eqref{eq:HEFT}.
Graphically, the amplitudes which we consider have the form
\begin{center}
\raisebox{-.45\height}{\begin{tikzpicture}
  \begin{feynman}
    \vertex (i1) at (-1,1) {\(\mu^+\)};
    \vertex (i2) at (-1,-1) {\(\mu^-\)};
    \vertex[blob] (a)  at (0,0)  {\contour{white}{$\Gamma_{nij}$}};
    \vertex (f1) at (0.25,1) {\(\phi_{+}\phi_{-}\)};
    \vertex (f2) at (1,0.25) {\(\phi_{0}\)};
    \vertex (f3) at (1,-0.25) {\(H_1\)};
    \vertex (f4) at (0.25,-1) {\(H_{n}\)};
    \vertex (f5) at (1.25,1.25);
    \vertex (f6) at (1.25,-1.25);
    \diagram* {
      {[edges=fermion]
        (i2) -- (a) -- (i1),
      },
      {[edges=plain]
        (f1) -- (a),
        (f2) -- (a),
        (f3) -- (a),
        (f4) -- (a),
      },
      {[edges=ghost]
        (f1) -- (f2),
        (f3) -- (f4),
      },
    };
  \end{feynman}
  \draw [decoration={brace}, decorate] (f5.north east) -- (f6.south east)
          node [pos=0.5, right] {\(X_{nij}\)};
\end{tikzpicture}}
  $ \qquad=i ( \Gamma_{L,nij}P_L +\Gamma_{R,nij}P_R)\, ,$ 
\end{center}
where the blob depicts a generic multi-particle interaction which consists of local and non-local contributions. $\Gamma_{L/R,nij}$ denotes all different Lorentz structures that can appear inside in contributing diagrams. The parameter $n$ counts the number of external Higgs bosons, $i$ the number of external neutral Goldstone bosons and $j$ the number of charged Goldstone boson pairs, respectively.

To narrow down the topologies and multiplicities of Feynman diagrams that remain relevant in the high-energy limit, we apply simple power-counting arguments.  In the high-energy limit, all invariants are of the same order $s$, masses are neglected and therefore the partial waves depend only on a single scale $\sqrt{s}$.
We count each momentum appearing in propagators or vertices as a power of $\sqrt s$. Each of the scalar and vector ($S+V$) bosonic propagators provides a factor ${1}/{s}$.  Analogously, each of the $F$ fermionic propagators provides a factor ${1}/{\sqrt{s}}$. Furthermore, the kinetic term of the matrix exponential $U$ in the HEFT Lagrangian introduces momentum-dependent vertices. There are two classes, containing either one gauge boson and one momentum (factor $\sqrt s$), or no gauge bosons but two momenta (factor $s$).  We denote the numbers of such vertices as $P_1$ and $P_2$, respectively.  Momentum-dependent vertices originating from the gauge-kinetic term, also present in the SM, are counted by $P_V$.   Finally, each of the two external fermions provides a factor $s^{{1}/{4}}$.  As argued above, external (transversal) gauge bosons can be neglected in the GBET limit.

Let $\gamma$ be a graph which contributes to the
$\mu^+\mu^-\rightarrow X_{nij}$ amplitude. The associated power of $s$ is given by
\begin{equation}
  d_{\gamma}=s^{P_2+\tfrac{P_1+P_V}{2}-\tfrac{F}{2}-S-V+\tfrac{1}{2}}\, .
\end{equation}
Drawing possible tree-level graphs (and using Euler's identity) shows quickly that 
\begin{equation}
\tfrac{P_1+P_V}{2}-\tfrac{F}{2}-V\leq 0\, .
\end{equation}
In the strict high-energy limit, \emph{i.e.}, if all masses are set to zero, only those graphs survive which satisfy
\begin{equation}
\tfrac{P_1+P_V}{2}-\tfrac{F}{2}-V=0.
\end{equation}
Hence we only need to consider graphs with
\begin{equation}
  d_{\gamma}=s^{P_2-S+\tfrac{1}{2}}\,,
\end{equation}
and internal gauge-boson propagators have dropped out.  The Lagrangian can be reduced to
\begin{align}
  \begin{split}
    \mathcal L_{Uh}&\supset\frac{v^2}{4}\operatorname{tr}[\pd_{\mu}U^{\dagger}\pd^{\mu}U]
    F_U(H)+\frac{1}{2}\partial_{\mu}H\partial^{\mu}H
    -\frac{v}{\sqrt{2}}\left[\bar{\ell}_L Y_\ell(H) UP_-\ell_R+\operatorname{h.c.}\right]\,
  \end{split}\label{eq:HEFTLag}
\end{align}
where pure Higgs self-couplings have disappeared (together with the couplings to transversal gauge bosons), as the power-counting argument above classifies those as subleading as well.  The dominant terms are then given by graphs of the type
\begin{equation}
\label{eq:heft_diags}
\centering
\raisebox{-.45\height}{\begin{tikzpicture}
  \begin{feynman}
    \vertex (i1) at (-1,1) {\(\mu^+\)};
    \vertex (i2) at (-1,-1) {\(\mu^-\)};
    \vertex[blob] (a)  at (0,0)  {\contour{white}{$\Gamma_{nij}$}};
    \vertex (f1) at (0.25,1) {\(\phi_{+}\phi_{-}\)};
    \vertex (f2) at (1,0.25) {\(\phi_{0}\)};
    \vertex (f3) at (1,-0.25) {\(H_1\)};
    \vertex (f4) at (0.25,-1) {\(H_{n}\)};
    \vertex (f5) at (1.25,1.25);
    \vertex (f6) at (1.25,-1.25);
    \diagram* {
      {[edges=fermion]
        (i2) -- (a) -- (i1),
      },
      {[edges=plain]
        (f1) -- (a),
        (f2) -- (a),
        (f3) -- (a),
        (f4) -- (a),
      },
      {[edges=ghost]
        (f1) -- (f2),
        (f3) -- (f4),
      },
    };
  \end{feynman};
\end{tikzpicture}}
  =\raisebox{-.45\height}{\begin{tikzpicture}
    \begin{feynman}
      \vertex (i1) at (-1,1) {\(\mu^+\)};
      \vertex (i2) at (-1,-1) {\(\mu^-\)};
      \vertex (a)  at (0,0)  ;
      \vertex (f1) at (0.25,1) {\(\phi_{+}\phi_{-}\)};
      \vertex (f2) at (1,0.25) {\(\phi_{0}\)};
      \vertex (f3) at (1,-0.25) {\(H_1\)};
      \vertex (f4) at (0.25,-1) {\(H_{n}\)};
      \vertex (f5) at (1.25,1.25);
      \vertex (f6) at (1.25,-1.25);
      \diagram* {
        {[edges=fermion]
          (i2) -- (a) -- (i1),
        },
        {[edges=plain]
          (f1) -- (a),
          (f2) -- (a),
          (f3) -- (a),
          (f4) -- (a),
        },
        {[edges=ghost]
          (f1) -- (f2),
          (f3) -- (f4),
        },
      };
    \end{feynman};
  \end{tikzpicture}}
  +\raisebox{-.45\height}{\begin{tikzpicture}
    \begin{feynman}
      \vertex (i1) at (-1,1) {\(\mu^+\)};
      \vertex (i2) at (-1,-1) {\(\mu^-\)};
      \vertex (a)  at (0,0)  ;
      \vertex (f1) at (0.25,1) {\(\phi_{+}\phi_{-}\)};
      \vertex (f3) at (0.5,0) ;
      \vertex[blob] (f7) at (1,0.5) {\contour{white}{}};
      \vertex[blob] (f8) at (1,-0.5) {\contour{white}{}};
      \vertex (f9) at (2.15,-1) {\(\phi_{+}\phi_{-}\)};
      \vertex (f10) at (2,0.25) {\(\phi_0\)};
      \vertex (f12) at (2,-0.25) {\(H_1\)};
      \vertex (f11) at (2,1) {\(\phi_{0}\)};
      \vertex (f4) at (0.25,-1) {\(H_n\)};
      \vertex (f5) at (1.25,1.25);
      \vertex (f6) at (1.25,-1.25);
      \diagram* {
        {[edges=fermion]
          (i2) -- (a) -- (i1),
        },
        {[edges=plain]
          (f1) -- (a),
          (f3) -- (a),
          (f4) -- (a),
          (f3) -- (f7),
          (f3) -- (f8),
          (f8) -- (f9),
          (f8) -- (f12),
          (f7) -- (f11),
          (f7) -- (f10),
        },
        {[edges=ghost]
          (f11) -- (f10),
          (f9) -- (f12),
        },
      };
    \end{feynman};
  \end{tikzpicture}}
  +\;\cdots 
  \, , 
\end{equation}
{\it i.e.}, terms with insertions of the highest-multiplicity contact terms.

The reasoning above would not work if any $t$-channel diagrams would become singular in the limit of massless exchanged particles, leading to divergent partial-wave amplitudes.  However, for the helicity-flip amplitudes which we consider here, this does not happen. The argument is related to the fact that in the parton picture of initial-state evolution, scalar radiation does not provide a collinear singularity and thus no forward peak (for the electroweak case {\it cf.}, {\it e.g.}, \cite{Chen:2016wkt,Han:2020uid,Han:2021kes,Ma:2022vmy,Garosi:2023bvq}).

We conclude that the Lorentz structure of the leading term reduces to a purely scalar one
\begin{equation}
\label{eq:dirstr}
    i ( \Gamma_{R,nij}P_R +\Gamma_{L,nij}P_L)= i ( C_{nij}P_R +C^*_{nij}P_L) \,.
\end{equation}
The spin-averaged matrix elements will therefore take the form
\begin{equation}
\label{eq:sdep}
  \overline{|\calM_{X_{nij}}|^2}=|C_{nij}(\lbrace s_{l\dots m}\rbrace )|^2\frac{s}{2}\, ,
\end{equation}
where the coefficients $|C_{nij}(\lbrace s_{l\dots m}\rbrace )|^2$ could in principle depend on different invariant masses of partial momentum sums,
\begin{equation}
  s_{l\dots m}=(p_l +\dots+ p_m)^2\,.
\end{equation}

To evaluate the individual coefficients $C_{nij}$, we perform an explicit algebraic calculation using {\sc \small FeynArts}~\cite{Hahn:2000kx}.  Summing over all graphs in the GBET limit for each amplitude, the dependence on the individual invariants disappears completely, as one can expect by considering the $S$-matrix equivalence of different approaches.  Only the explicit $s$ dependence remains, 
\begin{equation}
\label{eq:sindep}
  \overline{|\calM_{X_{nij}}|^2}=|C_{nij}(0)|^2\frac{s}{2}\,.
\end{equation}
The integral over phase space then becomes trivial and yields
\begin{equation}
  \label{eq:sigma-klm}
  \bar{\sigma}_{X_{nij}}=\frac{(2\pi)^4}{4}|C_{nij}(0)|^2\Phi_{X_{nij}}(k_1+k_2;p_1,\dots ,p_{2j+i+n})\,,
\end{equation}
where the volume of massless phase space is given by
\begin{equation}
  \Phi_{X_{nij}}(k_1+k_2;p_1,\dots ,p_{n+i+2j})=\frac{1}{2(2\pi)^4(4\pi)^{2N-3}}\frac{s^{N-2}}{\Gamma(N)\ \Gamma(N-1)}S_{X_{nij}}\delta_{N,n+i+2j}\,. \label{eq:smartformula}
\end{equation}
The factor
\begin{equation}
    S_{X_{nij}}=\frac{1}{n!\ i!\ (j!)^2}
\end{equation}
is the symmetry factor introduced in the phase space of indistinguishable particles.

The final result is a function of $s$ and of the independent parameters 
$\alpha_{m}$ with $m\le n+i+2j$
or, alternatively, the equivalent 
$n+i+2j$-parameter sets in the HEFT or SMEFT expansion.  We have verified that the analogous calculation within the SMEFT framework which proceeds differently, involving the operator coefficients of dimension 6, 8, and 10, eventually yields an identical result. The full list of the multi-particle amplitudes, with up to six particles in the final state,  are provided in Appendix~\ref{sec:HEAmps}.  Moreover, in Sec.~\ref{sec:pheno} we will explicitly report the high-energy approximation of the cross section for each of the processes considered in our analysis. We will also find that this approximation is in good agreement with our Monte Carlo simulations based on exact matrix elements.

\subsection{HEFT and SMEFT unitarity tests}
\label{sec:tests-Unitary}

In the following, we consider the SMEFT and HEFT scenarios introduced in Sec.~\ref{sec:EFT-scenarios} ($\SMEFTs$, $\SMEFTe$, $\HEFTm$ and $\HEFTp$) for setting unitary constraints.  We start considering a generic multiboson production process, where the final state has multiplicity up to $N$ (following the notation introduced in this section, $N=n+i+2j$) and afterwards we consider the specific case of final states with only Higgs bosons ($N=n$).

\begin{figure}[!t]
 \centering 
 \includegraphics[width=.48\textwidth]{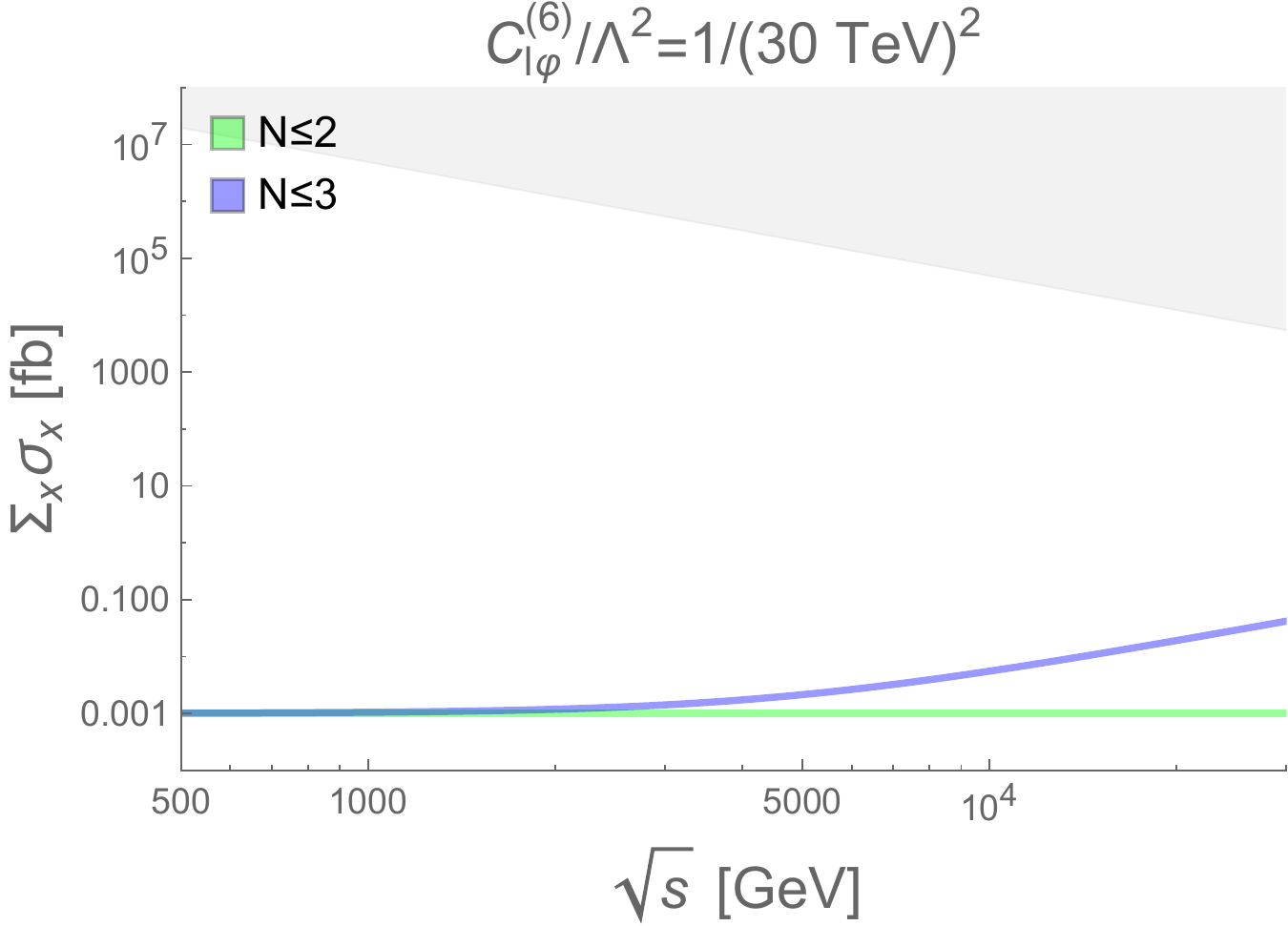}
 \includegraphics[width=.48\textwidth]{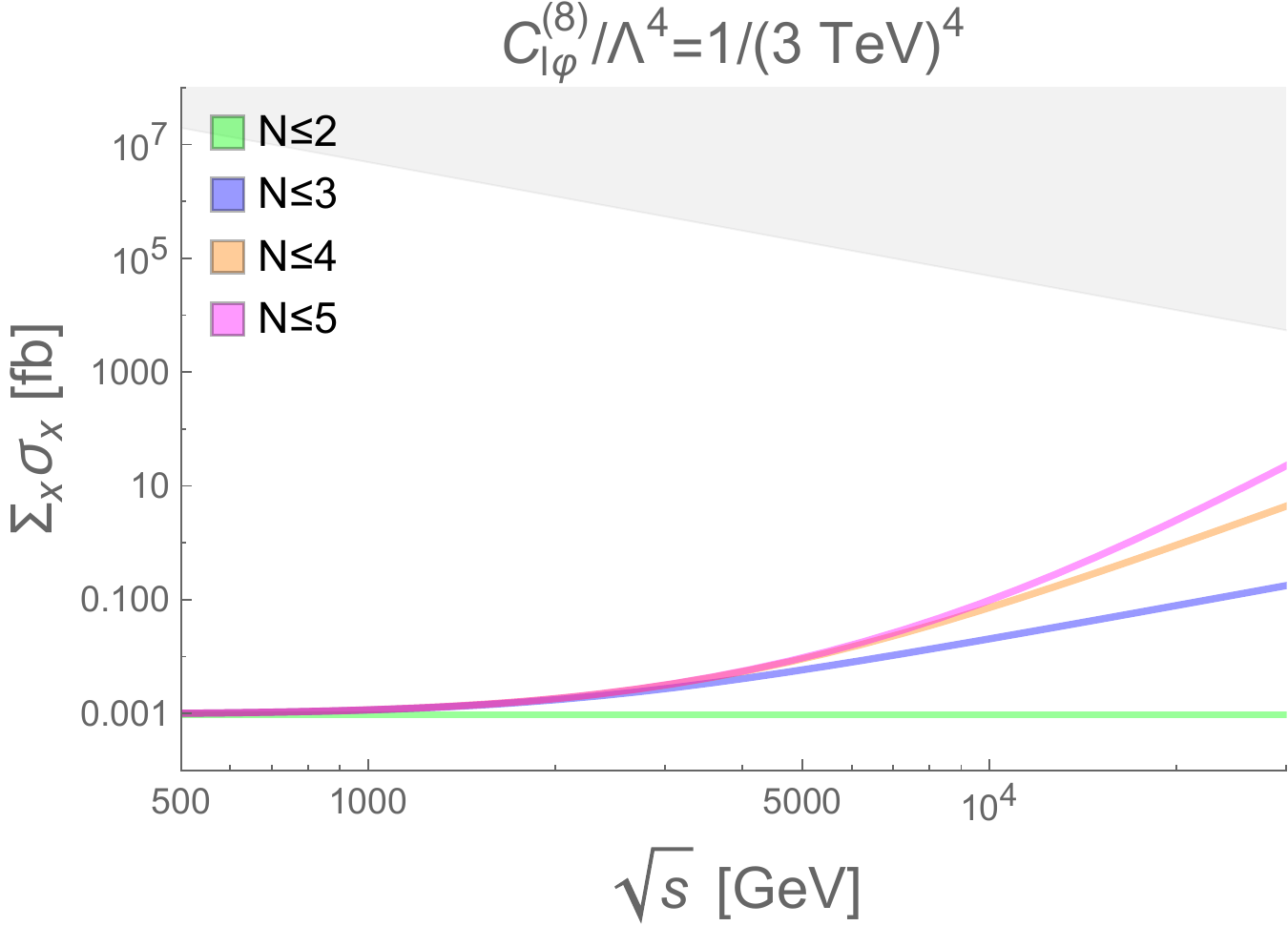}
\caption{Sum of inclusive cross sections for processes $\mu^+\mu^-\rightarrow X_{nij}$ with at most $N=n+i+2j$ final-state particles, in the high-energy approximation. The curves represent two scenarios with a single SMEFT operator: $\SMEFTs$ (left) and $\SMEFTe$  (right). The gray shaded area indicates the region which is excluded by the unitarity requirement.  \label{fig:uni-smeft}}
\end{figure}

\begin{figure}[!t]
\centering
\includegraphics[width=.48\textwidth]
{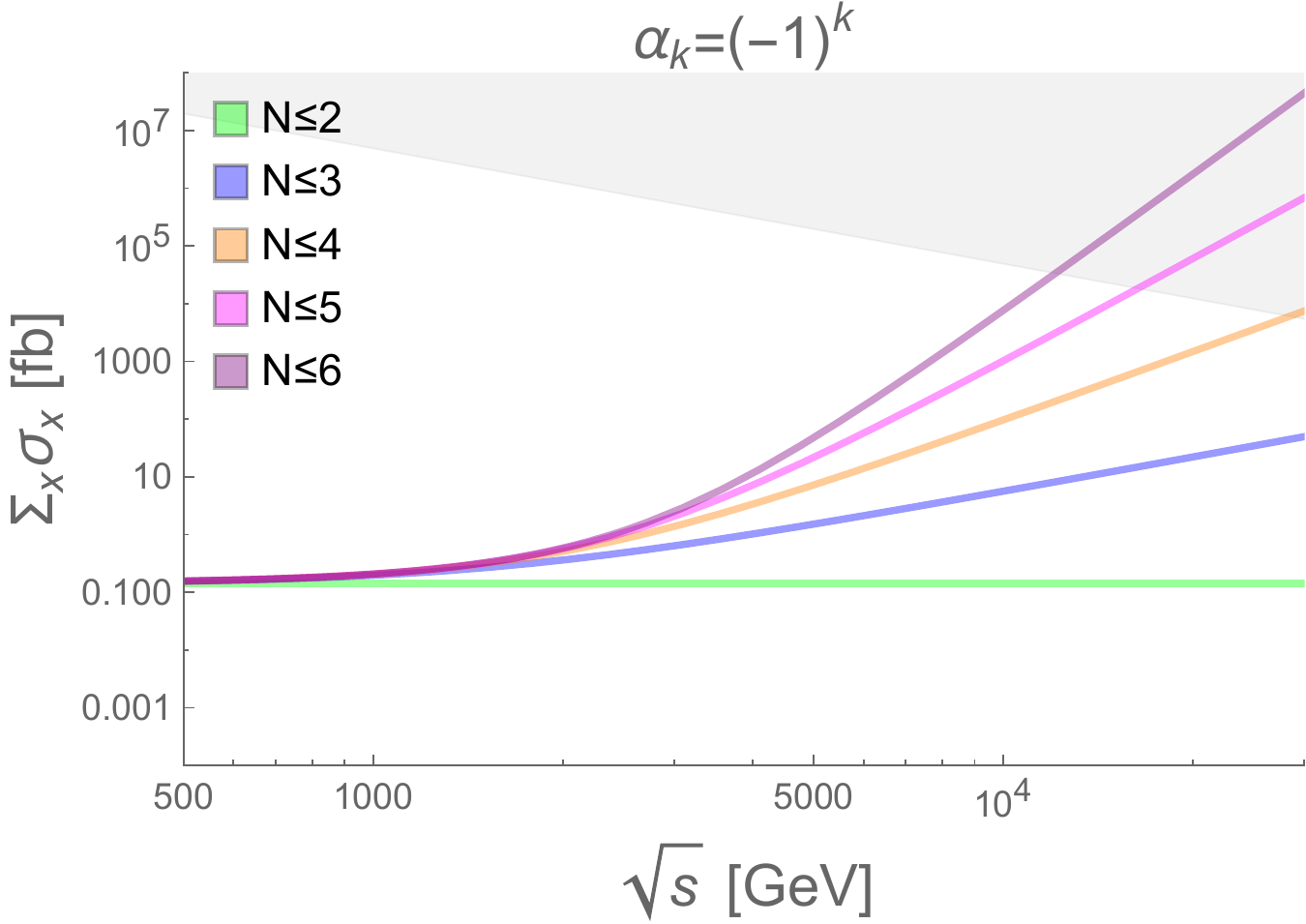}
\includegraphics[width=.48\textwidth]{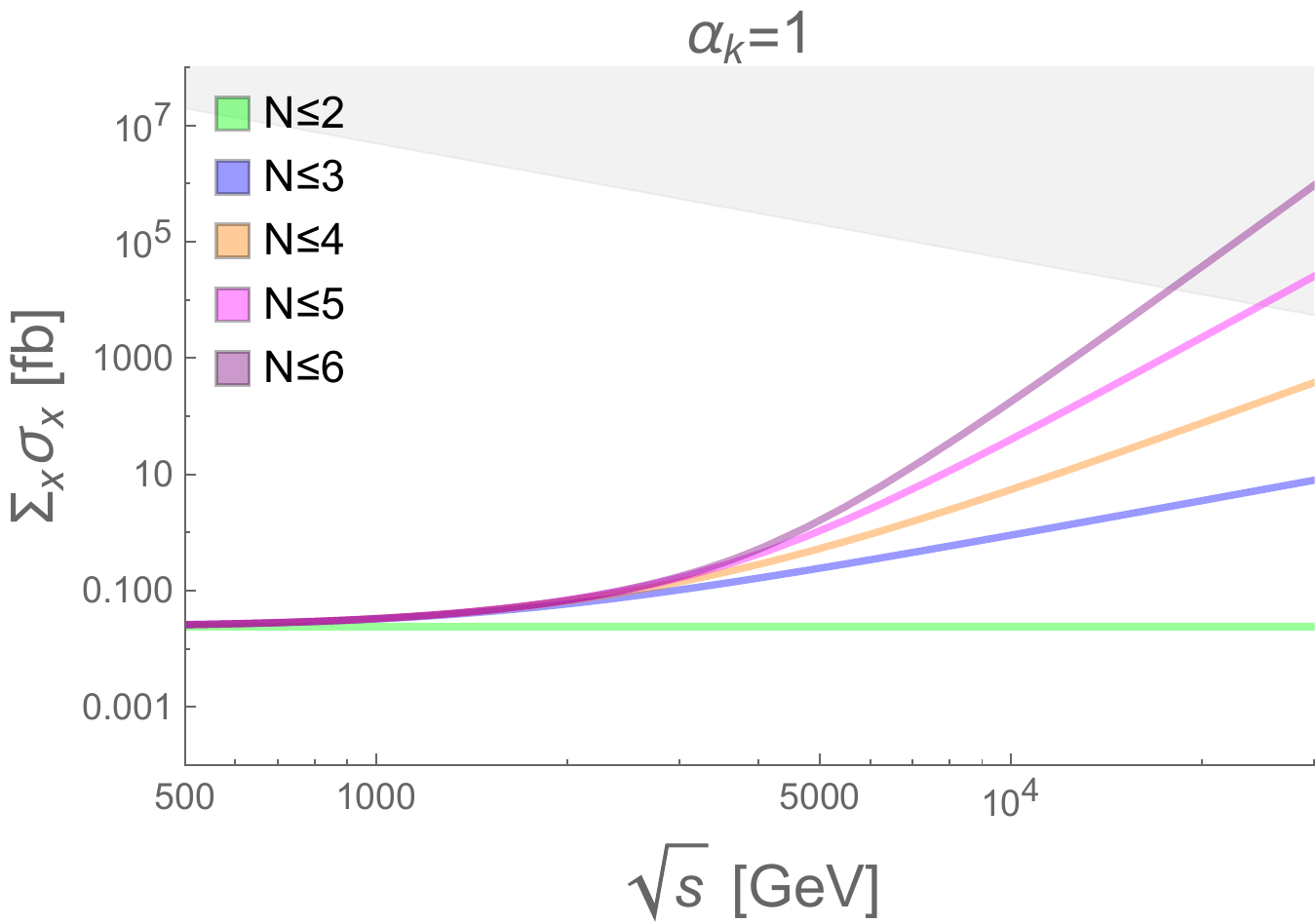}
\caption{Same as Fig.~\ref{fig:uni-smeft}, but for the HEFT.  The curves represent
the two HEFT scenarios with uniform absolute value of the $\alpha_k$ coefficients:  $\HEFTm$ with $\alpha_k=(-1)^k$ (left) and $\HEFTp$  with $\alpha_k=+1$ (right), respectively. The gray shaded area indicates the region which is excluded by the unitarity requirement. \label{fig:uni-heft} }
\end{figure}

 We display the results for the unitarity constraints in Fig.~\ref{fig:uni-smeft} for the SMEFT scenarios and in Fig.~\ref{fig:uni-heft} for the HEFT scenarios.  Using Eq.~\eqref{eq:sigma-klm} we plot, as a function of $\sqrt{s}$, the sum of the total cross sections of any possible final state of the form $nH+i\phi_{0}+j(\phi_{+}\phi_{-})$ up to the multiplicity $N$, where $N=n+i+2j$. We consider all the  cases up to $N\le 6$.

In Fig.~\ref{fig:uni-smeft} the SMEFT scale is set to $\Lambda=30\,\TeV$ ($3\,\TeV$) for the $\SMEFTs$ ($\SMEFTe$) scenario, and $c^{(6)}_{\ell\varphi} \ (c^{(8)}_{\ell\varphi})=1$. Equivalently, we set  $\Lambda=30\,\TeV$ and $c^{(6)}_{\ell\varphi}=1$ in the $\SMEFTs$ scenario and $c^{(8)}_{\ell\varphi}\simeq 10^4$ in the $\SMEFTe$ scenario. 
In this way, we get $\Delta \alpha_1 \simeq 0.1$, which is well within the current experimental bounds. The grey areas in the plots denote the cross-section values for which unitarity is violated, $\sigma>4\pi/s$ ({\it cf.}~Eq.~\eqref{eq:unitarityconditionlwave}). In the left plot of Fig.~\ref{fig:uni-smeft} we can see how, in the $\SMEFTs$ scenario, unitarity can be violated only well beyond the 30 TeV energy scale. The reason why we consider only $N\le 3$ in the $\SMEFTs$ scenario is simple. Using Eq.~\eqref{eq:sigma-klm}, the result for the cross section would be exactly equal to zero for $N>3$. Indeed for $3< N\le 5$ the dominant growth in energy is given in the SMEFT framework from the  dimension-8 operator, which precisely corresponds to the $\SMEFTe$ scenario. We plot the cases $3\le N\le 5$ in the right plot and we see that also in this case unitarity can be violated only beyond the 30 TeV energy scale.\footnote{The manifest difference between the $N=3$ line in the left and right plot is only induced by the small differences in the values of $\alpha_{1}$ and $\alpha_{2}$ between the scenario $\SMEFTs$ with $\Lambda=30\,\TeV$ and the scenario $\SMEFTe$ with $\Lambda=3\,\TeV$.}  Fig.~\ref{fig:uni-smeft} clearly shows the necessity of considering the $\SMEFTe$ in this context. 
Finally, we notice that the case $N=2$ is constant in energy.  This plateau dominates over the helicity-flip inclusive cross section in the SM (not shown), which decreases like $1/s$ and is furthermore suppressed by the muon Yukawa coupling. Also, we can see that with increasing final-state multiplicity $N$, the curves develop a power-law rise with energy.  

In the HEFT scenarios (Fig.~\ref{fig:uni-heft}), the behavior is similar to the case of $\SMEFTe$, with a different normalization given by the chosen parameter values.  In the $\HEFTp$ scenario we choose $\Delta \alpha_1=0$, so the $\bar \mu \mu H$ interaction is as in the SM and compatible to current data. Instead,  in the $\HEFTm$ scenario we choose $\Delta \alpha_1=-2$ and therefore $\alpha_1=-1$, so the $\bar \mu \mu H$ interaction has the opposite sign of the SM but the same strength, which is also compatible to current data. The cancellations in the $\HEFTp$ scenario (right) reduce the size of the cross sections compared to the scenario $\HEFTm$ (left), but do not alter the power-law rise, which as in $\SMEFTe$ depends only on the final-state multiplicity. However, we can clearly see that for $N=6$,\footnote{In the SMEFT case, $N=6$ would receive the dominant contribution from the dimension-10 operator coefficient $c^{(10)}_{\ell \phi}$. In order to reach $\Delta \alpha_1 \simeq 0.1$, satisfying the current bounds on $y_{\mu}$, we would need $\Lambda\simeq v$ and therefore we refrained from considering this scenario, which is better captured by the HEFT formalism.  } slightly above $\sqrt{s}=10~{\rm TeV}$ unitarity can be violated. For this reason in our study, we limit ourselves to the case $N\le 5$.

In conclusion, both for our SMEFT and HEFT scenarios, if we take into account at most variation of $\mathcal{O}(1)$ for a generic $\alpha_{i}$ and we consider processes with $N\le 5$, at 10 TeV, which is the maximum energy considered in the phenomenological study presented in Sec.~\ref{sec:pheno}, we are ascertained that unitarity is not violated.

\medskip 

The plots of Figs.~\ref{fig:uni-smeft} and \ref{fig:uni-heft} have been obtained via Eq.~\eqref{eq:sigma-klm}, which assumes the high-energy limit where all invariants are of the same order $s$ and the masses of the final-state particles can be neglected. For the multiplicities and the maximum energies that we have considered this is clearly a good approximation\footnote{We will comment in Sec.~\ref{sec:pheno} on possible problems in using this formula for performing phenomenological studies. We anticipate already here that while here we use the high-energy approximation, in Sec.~\ref{sec:pheno} we use  Monte Carlo simulations based on exact matrix elements.} and reveal us another interesting feature. In the strict high-energy limit there is no inherent limit on the number of produced Higgs and Goldstone particles. In other words, $N$ is not bounded from above.  Therefore in the HEFT framework, without the SMEFT power counting in place, we may assume that an actual BSM model provides a saturation mechanism for multiplicities
\begin{figure}[!t]
  \centering
  \includegraphics[width=0.8\textwidth]{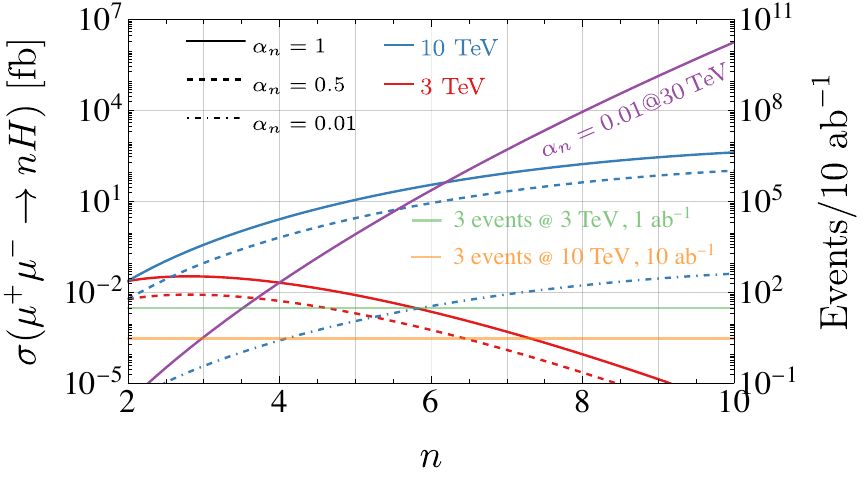}
  \caption{Cross sections $\sigma(\mu^+\mu^-\rightarrow nH)\approx\bar{\sigma}_{X_{n00}}$ in Eq.~\eqref{eq:sigma_n00} as function of the number of final state Higgs bosons $n$. The red (blue) lines refer to at a 3~TeV (10~TeV) collisions. Solid, dashed, and dot-dashed curves are for $\alpha_n=1,\,0.5,\,0.01$, respectively. The green (orange) horizontal lines represent the cross section value necessary for 3 events at the proposed 3~TeV (10~TeV) muon collider.
  The purple solid curve is for $\alpha_n=0.01$ at a 30~TeV muon collider. 
  The axis on the right shows the number of events per 10~ab$^{-1}$ integrated luminosity.
   \label{fig:mmnh}}
\end{figure}
$N\to\infty$, otherwise the EFT description would eventually break down at any fixed energy above the electroweak scale $4\pi v\approx 3\,\TeV$~\cite{Maltoni:2001dc}. 

Nevertheless, in the following we concentrate on the specific case $N=n$, meaning a process with only Higgs bosons in the final state,  $\mu^+\mu^-\to n H$, and we study the dependence on $n$ for different energies. For such processes, Eq.~\eqref{eq:sigma-klm} simplifies to 
\begin{eqnarray}
\label{eq:sigma_n00}
 \sigma(\mu^+\mu^-\rightarrow nH)\approx\bar{\sigma}_{X_{n00}}
=\frac{n \alpha_n^2 m_\mu^2 s^{n-2} }{8\Gamma(n-1)(4\pi)^{2n-3}v^{2n}}\, . \label{eq:mmnhaninsec3}
\end{eqnarray}
As it is manifest from Eq.~\eqref{eq:mmnhaninsec3}, the leading term in $s$ depends only on $\alpha_{n}$. Indeed,  the diagram featuring the $\bar \mu \mu H^{n}$ point interaction is the only one of order  $\mathcal O(s^0)$, while all the others have at least a suppression of $\mathcal O(1/s)$.
In SMEFT, the leading contribution to $\mu^+\mu^- \to nH$ will just be the contact term introduced by the higher dimensional operators. In particular, the leading term in energy in HEFT is the same as $\SMEFTs$ for $n=2,3$ and the same as $\SMEFTe$ for $n=4,5$. We cross-checked the results in the SMEFT calculation and the $\alpha_n$ parameterization gives identical results for the high-energy behavior if Eq.~\eqref{eq:map} is used. In general, the HEFT calculation is equivalent to a SMEFT calculation where the contribution depends on $c_{\ell \varphi}^{(n+4)}$ ($c_{\ell \varphi}^{(n+3)}$)  and the associated operator of dimension $n+4$ ($n+3$) for $n$ even (odd).

In Fig.~\ref{fig:mmnh} we plot for 3 TeV and 10 TeV as red and blue lines, respectively,  $\sigma(\mu^+\mu^-\rightarrow nH)$ from Eq.~\eqref{eq:mmnhaninsec3} as a function of $n$, for $2\le n \le 10$. Solid, dashed, and dot-dashed lines correspond to the choices  $\alpha_n = 1, 0.5, 0.01$, respectively. We also show, only for $\alpha_n = 0.01$, the case of 30 TeV as a violet solid line. Since the unitarity condition (Eq.~\eqref{eq:unitarityconditionlwave}) is energy dependent, we list the following specific condition for each energy considered here:
\begin{eqnarray}
\sigma&\lesssim& {4 \times 10^4~\rm fb}~~{\rm at}~3~{\rm TeV}\,,\\
\sigma&\lesssim& {4 \times10^3~\rm fb}~~{\rm at}~10~{\rm TeV}\,,\\
\sigma&\lesssim &{4 \times 10^2~\rm fb}~~{\rm at}~30~{\rm TeV}\,.
\end{eqnarray}

Assuming $\bar{\sigma}_{X_{n00}}$ as a good approximation of $\sigma(\mu^+\mu^-\rightarrow nH)$, we see that for the case of 3 and 10 TeV unitarity is not violated also for values of $n$ much larger than 5 even with $\alpha_n = 1$. 
For the case of 30 TeV, where we have set $\alpha_n = 0.01$, we see that up to $n=6$ the unitarity is still preserved, while for larger $n$ is not. Especially, we see a different trend in the dependence of $\sigma(\mu^+\mu^-\rightarrow nH)$ on the multiplicity $n$. In fact, for any energy, the cross section grows up to a given multiplicity and then decreases. 
The multiplicity $n$ for which the turning point of this trend takes place is within the interval $[\hat n, \hat n+1]$ that can be determined by requiring
\begin{equation}
\label{eq:turningpoint}
\bar{\sigma}_{X_{\hat n 00}}=\bar{\sigma}_{X_{(\hat n +1) 00}} \Longrightarrow \frac{s}{16 \pi^2 v^2} \frac{\hat n+1}{\hat n(\hat n+1)}=1\, ,
\end{equation}
where we have used the identity $\Gamma(\hat n)=(\hat n-1)!\,$. By solving the r.h.s.~of Eq.~\eqref{eq:turningpoint} one gets that $\hat n \simeq 2.3$ at 3 TeV,  $\hat n \simeq 12.3$ at 10 TeV and  $\hat n \simeq 96.1$ at 30 TeV. These numbers show how changing the energy from 3 to 30 TeV can have a big impact on the information that can be extracted in the HEFT framework from multi-Higgs final states at a muon collider. Notice that for very large energies, and therefore very large $\hat n$, 
\begin{equation}
\label{eq:turningpointhe}
\hat n = \frac{s}{16 \pi^2 v^2} \,.
\end{equation}
All this argument relies on the fact that $\bar{\sigma}_{X_{ n 00}}$ is a good approximation for the $\sigma(\mu^+\mu^-\rightarrow nH)$ cross section.
Performing a full calculation and therefore including all the diagrams and taking into account the mass of the Higgs in the final state, as done in Monte Carlo simulations discussed in the next section, the result can be different. We have verified that for multiplicities up to $5$ and energies down to 3 TeV this approximation can overestimate the exact result by at most a factor 2. This factor is important for extracting bounds, and this is one of the reasons why this approximation is not used in the detailed numerical analyses in the next section, but it is not affecting at all any of the unitarity arguments discussed in this section. On the other hand, Eqs.~\eqref{eq:turningpoint} and especially \eqref{eq:turningpointhe} should be taken as a limiting behavior at the high energy and high multiplicity, since, with very large multiplicity,  threshold effects and especially subleading topologies, which grow factorially with $n$, may be relevant. 

Finally, in the plot of Fig.~\ref{fig:mmnh} we have also shown as a green (orange) line, the cross section that is necessary to obtain 3 events at a 3 (10) TeV muon collider with $\mathcal{L}=1\ (10)~\textrm{ab}^{-1}$ for 3 (10) TeV collisions  \cite{Delahaye:2019omf,Bartosik:2020xwr} integrated luminosity. The axis on the right shows the number of events per 10~ab$^{-1}$ integrated luminosity. This information will be useful when we will discuss in the next section the constraints that can be set on the values of $\alpha_i$.

\section{Multi-boson Phenomenology at multi-TeV muon colliders}
\label{sec:pheno}

\subsection{Overview of the analysis}
\label{sec:phenooverview}

\begin{table}[t!]
  \centering
  \begin{tabular}{|c|c|c|c|c|c|c|}
  \hline
  \diagbox{$H$}{$V$} & \hspace{0.8em} 0   \hspace{0.8em}   &\hspace{0.7em} 1   \hspace{0.7em}   & 2                  & 3                   & 4                        & 5                    \\
  \hline
  0                                 & -     & $Z$    & \cellcolor{colora1}$Z^2$,$W^2$ &\cellcolor{colora1}\thead{$Z^3$\\~~~$W^2Z\hspace{0.26cm}$}& \cellcolor{colora2}\thead{$Z^4$, $W^4$\\~~~~$W^2Z^2\hspace{0.35cm}$}    &\cellcolor{colora2}\thead{$Z^5$,$W^2Z^3$\\$W^4Z$}
  \\ \hline
  1                                 &\cellcolor{colora1} $H$   &\cellcolor{colora1}$ZH$& \cellcolor{colora2}\thead{$W^2H$\\ ~~~$Z^2H$~~~}     & \cellcolor{colora2}\thead{$W^2ZH$\\ ~~~~$Z^3H$~~~}     & \cellcolor{colora3}\thead{$W^4H$, $Z^4H$\\$W^2Z^2H$} & -
  \\ \hline
  2                                 &\cellcolor{colora2} $H^2$ & \cellcolor{colora2}$ZH^2$ & \cellcolor{colora3}\thead{$W^2H^2$\\ ~~\,$Z^2H^2\hspace{0.125cm}$} & \cellcolor{colora3}\thead{$W^2ZH^2$\\ ~~\,$Z^3H^2~~\hspace{0.1ex}$} & -                        & -                     \\ \hline
  3                                 &\cellcolor{colora3}$H^3$ & \cellcolor{colora3}$ZH^3$ & \cellcolor{colora4}\thead{$W^2H^3$\\ ~\,$Z^2H^3\hspace{0.23cm}$} & -                   & -                        & -                     \\ \hline
  4                                 &\cellcolor{colora4}$H^4$ & \cellcolor{colora4}$ZH^4$ & -                  & -                   & -                        & -                     \\ \hline
  5                                 &\cellcolor{colora5}$H^5$ & -      & -                  & -                   & -                        & -                     \\ \hline
  \end{tabular}
  \begin{tabular}{|c|}\hline
\cellcolor{colora1}$\alpha_1$\\ \hline
\cellcolor{colora2}$\alpha_{1,2}$\\ \hline
\cellcolor{colora3}$\alpha_{1,2,3}$\\ \hline
\cellcolor{colora4}$\alpha_{1\cdots4}$\\ \hline
\cellcolor{colora5}$\alpha_{1\cdots5}$\\ \hline
  \end{tabular}
  \caption{The dependence of different multi-boson production processes on the corresponding 
   Higgs-muon effective couplings $\alpha_{n}$, indicated with the corresponding color codes. In the text, besides the $V^m H^n$ notation, we will also interchangeably use the $mVnH$ notation when referring to the various processes. }
  \label{tab:processes}
  \end{table}

In this section, we turn to the phenomenological investigation of multi-boson production processes within the environment of a multi-TeV muon collider as proposed in Refs.~\cite{Bartosik:2020xwr,Schulte:2021hgo,Long:2020wfp}.  The study builds upon and extends our previous work~\cite{Han:2021lnp,Reuter:2022zuv}, where we have shown that the multi-boson production offers a good opportunity to measure the Higgs-muon coupling.
In this work we extend our study in two directions. First, we analyze also the pure multi-Higgs production processes together with multi-gauge boson (in association with Higgs boson) production. Second, we interpret the constraints from expected measurements not only by varying the strength of the $\bar\mu  \mu H$ interaction, but we consider both a general HEFT framework, as described in Sec~\ref{sec:heft_framework}, where all the possible $\bar\mu  \mu H^n$ are present and can vary independently, and the $\SMEFTs$ scenario. The latter has been introduced in Sec.~\ref{sec:EFT-scenarios}, based on the SMEFT framework as described in Sec~\ref{sec:smeft_framework}, and it depends only on one single parameter, $\Delta \alpha_1$. All the others are given by the relation in Eq.~\eqref{eq:dim6lock}.
We stress that considering only the dimension-6 SMEFT operator is not just a simplification. Assuming decoupling for the heavy degrees of freedom, this truncation is for a large class of models equivalent to retaining only the leading effects, as also quantified at the end of Sec.~\ref{eq:paramvertices}. Exceptions are possible, as documented in Appendix \ref{sec:model-d8}, and are not ignored here; they are a specific direction or point in the general HEFT framework, which is also discussed in detail.

As motivated in the previous section based on unitarity arguments, we will consider processes with up to five bosons in the final state and the dependence on all the   $\alpha_i$ with $i\le 5$.
The $\alpha_i$ parameterization  allows us to 
keep the study mostly free of specific theoretical assumptions and express our results  for the HEFT, without further constraints except unitarity. However,  since we are interested also in the $\SMEFTs$ scenario, we also refer to the one-dimensional subspace corresponding to it.  The relation among the parameters of the SMEFT and HEFT frameworks are reported in Eq.~\eqref{eq:map}, and for the specific $\SMEFTs$ scenario lead precisely to Eq.~\eqref{eq:dim6lock}.

Since we will present a plethora of new results concerning 3 and 10 TeV collisions, for several different final states, and we will discuss them both in view of the HEFT and the $\SMEFTs$ scenarios, we have decided to anticipate and summarize in Tab.~\ref{tab:processes} which $\alpha_i$ with $1\le i \le 5$ enters the cross section for each of the processes considered here. The SMEFT case can be easily deduced from Eqs.~\eqref{eq:dim6lock} and \eqref{eq:dim8lock}.  In Tab.~\ref{tab:processes} each row represents a different number of Higgs particles, each column represents a different number of gauge bosons, and  the $\alpha_n$ dependence is indicated by each different color.
In the rest of this section we will discuss in detail how each process depends on the $\alpha_i$ and also what are the bounds that can be achieved in the measurement. However we do anticipate what we will find, in order to help the reader to navigate the results.

At a 10 TeV muon collider, assuming HEFT, we will show that multi-Higgs production is able to set very strong constraints on $\alpha_i$ with $3\le i \le 5$. In particular each of the $n H$ final state gives, independently, the best constraints on the corresponding $\alpha_n$. For $\alpha_1$ and $\alpha_2$, the situation is different. Multi-gauge-boson production (possibly with additional Higgs bosons) give better constraints, but typically each of these processes depends on more than one $\alpha_i$, especially those with the highest sensitivity.  Thus, multi-Higgs production can be used to first constrain the parameters $\alpha_i$ with $3\le i \le 5$, and then the rest of the processes, combining different classes of them, to constrain $\alpha_1$ and $\alpha_2$. 
At 3 TeV, besides a generally lower expected sensitivity on the different $\alpha_i$, the situation is more complex as many of the high-multiplicity processes are not yet far enough above their production thresholds. Hence, subleading terms play a larger role and the dependence of the processes on the $\alpha_i$ parameters is less clean than at 10 TeV.

In the $\SMEFTs$ scenario the situation is obviously much simpler. The previous discussion shows that at 10 TeV it is sufficient to measure $3H$ production only in order to obtain strong constraints on $\Delta \alpha_1$ (see Eqs.~\eqref{eq:dim6lock}) and in turn $c_{\ell \varphi}^{(6)}$ (see Eq.~\eqref{eq:map}). Instead, for 3 TeV,  as we will see in the next sections, the best constraints can be obtained from $2H$ production, and further slightly improved by the measurements of many other multi-boson production processes that do {\it not} feature a dependence on $\alpha_i$ with $i\ge 3$.
We have also investigated the role of anomalous Higgs self couplings,  parametrized as in Eq.~\eqref{eq:LextendedK} via the quantities $\beta_i$. Besides the case of the $ZHH$ production at 3 TeV, the dependence on the $\beta_i$ is largely suppressed and we will focus on the $\alpha_i$ only. On the other hand, this means that modifications in the Higgs potential do not contaminate the measurements of the Higgs-muon couplings.

\medskip

In order to determine the experimental sensitivities, background processes have to be taken into account.  At low energy, helicity-conserving annihilation of $\mu^+\mu^-$ proceeds via a virtual $s$-channel  $Z$ or $\gamma$  and provides the dominant source of multi-boson final states in the SM.  This production mechanism is not affected by the new physics which we consider in this work and thus constitutes a major background.\footnote{In principle, this SM background would be reduced by properly selecting muon and anti-muon polarization in the initial state. }  The SM cross section falls off proportional to $1/s$. On the contrary, what we consider as signal is constant or even rises with $s$.  Thus, at a multi-TeV muon collider, the signal is significantly larger than $s$-channel  $Z$ or $\gamma$ induced backgrounds, as already shown in~\cite{Han:2021lnp,Reuter:2022zuv}. 

Vector-boson fusion (VBF) into multiple bosons in the final state ({\em i.e.} elastic and inelastic vector-boson scattering) can provide in principle a major background~\cite{Han:2020uid,Costantini:2020stv,Ruiz:2021tdt}. The VBF cross section rises with energy and in the multi-TeV range the pattern of electroweak particle splitting and radiation starts to resemble QCD in hadron-collider physics~\cite{Han:2020uid,Han:2021kes,BuarqueFranzosi:2021wrv,Ma:2022vmy}.    However, the energy scale effectively probed by VBF is significantly lower than the energy available in production processes.  For these reasons, cuts can efficiently suppress the VBF part of electroweak processes.

Other SM backgrounds originating from top-quark or pure QCD processes must also be taken into account, but they are of minor relevance.  A more concerning problem is the process-specific identification of final states, which relies on experimental details such as $b$-tagging and the elimination of combinatorial background. For this exploratory study, we do not dive into details but apply generic estimates for cut and detection efficiencies.  Those can be deduced from existing fast- and full-simulation studies which are available in the literature, partly derived from earlier results that apply to high-energy $e^+e^-$ collisions.

The rest of the section is organized as follows. In Sec.~\ref{sec:cutandstat} we describe the common analysis framework for the Monte Carlo simulations and the statistical procedure for determining bounds and their associated confidence levels. Then, in Secs.~\ref{sec:multiH}--\ref{sec:multiVH} we respectively consider the case of multi-Higgs productions ($\mu^+\mu^-\to nH$ with $2 \le n \le 5 $), multi-gauge boson production ($\mu^+\mu^- \to mV$ with $2 \le m \le 5 $), and Higgs associated EW gauge boson production ($\mu^+\mu^-\to mV+nH$ with $2 \le m+n \le 5 $). In each of Secs.~\ref{sec:multiH}--\ref{sec:multiVH} we determine and discuss the different dependencies on the $\alpha_i$ parameters at 3 and 10 TeV collisions, and we calculate the corresponding bounds in the HEFT and SMEFT frameworks.  In Secs.~\ref{sec:combination} we provide bounds from the combination of results from processes of the three different classes.

\subsection{Simulation setup and statistical interpretation}
\label{sec:cutandstat}

We define the signal as the set of direct muon-annihilation processes into multiple Higgs and vector bosons: 
\begin{itemize}
\item $\mu^+\mu^-\to nH$ with $2 \le n \le 5 $, 
\item $\mu^+\mu^- \to mV$ with $2 \le m \le 5 $,
\item $\mu^+\mu^-\to mV+nH$ with $2 \le m+n \le 5 $. 
\end{itemize}
As already anticipated,  we cover moderate multiplicities up to $n+m=5$ because of the unitarity arguments discussed in Sec.~\ref{sec:tests-Unitary}.

Summarizing what has already been anticipated in Sec.~\ref{sec:phenooverview}, two kinds of backgrounds exist for the BSM signal in each of the processes we consider. First, the contribution from the SM ($\alpha_1=1$ and $\alpha_i=0$ for $i\ge2$) for the same process. Second, the vector-boson fusion (VBF) induced production for the same final state. While the former is irreducible, the latter can be suppressed with dedicated cuts.

In order to isolate the annihilation signal, it is important to reduce background from VBF topologies.
Following our previous study~\cite{Han:2021lnp}, we apply the acceptance cuts
\begin{equation}
\label{eq:cuts}
 \theta_{iB} > 10^\circ, \qquad
 \Delta R_{BB}>0.4, \qquad
 M_F > 0.8 \sqrt {s}\,.
\end{equation}
In Eq.~\eqref{eq:cuts},  $\theta_{iB}$ is the smallest angle between any final-state boson $B$ ($B=H,W,Z$) and the initial-state axis. The quantity $\Delta R_{BB}=\sqrt{\Delta\eta^2+\Delta \phi^2}$ is the separation between any two bosons in the rapidity-azimuthal angle
plane, and it is necessary to resolve the final-state bosons within the detector. Instead, as explicitly demonstrated in Ref.~\cite{Han:2021lnp}  and also in the rest of the section, the invariant mass cut, \emph{i.e.}, $M_F>0.8\sqrt{s}$, where $F$ include all the final-state bosons, is sufficient to suppress the VBF backgrounds\footnote{In the VBF channels, an additional cut $M_{\mu^+\mu^-},M_{\nu_\mu\bar{\nu}_\mu}>150~\GeV$ has been imposed to exclude the on-shell $Z$ contribution.}. The annihilation processes are largely unaffected, as the momentum conservation automatically ensures $M_F\simeq\sqrt{s}>0.8\sqrt{s}$~\footnote{Here, the initial-state radiation will take away a small fraction of energy and ends up with $M_{F}\lesssim\sqrt{s}$.}. 

Numerical studies have been performed via {\aNLO} \cite{Alwall:2014hca, Frederix:2018nkq} and {\WZ} \cite{Moretti:2001zz, Kilian:2007gr}. In particular, as we will discuss in Sec.~\ref{sec:multiH}, the dominant SM contribution for multi-Higgs production originates from loop-induced channels, which have been evaluated thanks to the module developed in Ref.~\cite{Hirschi:2015iia}, based on {\MadLoop} \cite{Hirschi:2011pa}.

In the next sections, after having quantified and discussed the dependence of each process on the $\alpha_i$ (and $\beta_i$) parameters, we will also study the sensitivity of each class of processes on them.  In doing so we will assume an integrated luminosity $\mathcal{L}=1~\textrm{ab}^{-1}$ for 3 TeV collisions and $\mathcal{L}=10~\textrm{ab}^{-1}$ for 10 TeV collisions \cite{Delahaye:2019omf,Bartosik:2020xwr}, and we will assume statistical uncertainties as the dominant ones. There are definitely other effects that may be relevant both from the experimental ({\it e.g.}, vector-boson tagging efficiency) and theory ({\it e.g.}, PDF effects or NLO EW corrections~\cite{Bredt:2022dmm}). However, we do not expect them to dramatically impact our conclusions and we leave it for future work.

Assuming uncertainties dominated by statistics, we employ the following approach for determining expected bounds at future muon colliders.
Given an expected number of events $N$, by means of the Poisson distribution we can define a statistical likelihood function as~\cite{ParticleDataGroup:2020ssz,Cowan:2010js}
\begin{eqnarray}
  L(\mu) =\frac{(\mu S+B)^{N}}{N!}e^{-(\mu S+B)},
\end{eqnarray}
where $\mu=(N-B)/S$ is the maximal likelihood estimator, and $S$ and $B$ denote the signal and background event numbers, respectively. The experimental sensitivity can be obtained through the formula
\begin{eqnarray}
\label{eq:sen}
{\mathcal S} = \sqrt{2 \log \frac{L(1)}{L(0)}} = \sqrt{2(S+B)\log(1+\frac{S}{B})-2S}\,,
\end{eqnarray}
where ${\mathcal S} =2~(3)$ corresponds to the $2~(3)~\sigma$ exclusion limit, \emph{i.e.}, the 95\% (99\%) confidence level (CL).
It is interesting to note that in the limit $S\ll B$, the above expression can be simplified with its more commonly known leading-order Taylor expansion
\begin{eqnarray}
  {\mathcal S}\simeq\frac{S}{\sqrt{B}}\,,
\end{eqnarray}
namely the figure of merit which is widely used as a measure of expected discovery significance, \emph{e.g.}, in Ref.~\cite{Han:2021lnp}.

It is important to note that in our sensitivity projection we will stay with the complete definition, Eq.~\eqref{eq:sen}, since in several cases the opposite condition $B\ll S$ occurs. 
 In fact, when the background is too small ($B\ll1$ event),  the previous formulae cannot be used and we will instead use the condition $S = 3$, meaning 3 events,  in order to obtain the $2\sigma$ (95\% CL) exclusion limit \cite{ParticleDataGroup:2020ssz}. 
In general, unless specified differently, we will refer in the text to limits at the 95 \% C.L, derived from the $\calS=2$ contours or the requirement for $S = 3$ in the case of less than one event for the background.

\subsection{Multi-Higgs productions: $\mu^+\mu^-\to nH$}
\label{sec:multiH}
As already discussed in Ref.~\cite{Dermisek:2021mhi}, the multi-Higgs production processes offer opportunities to directly measure the contact vertices $\bar{\mu}\mu H^{n}~(n\geq2)$
Recently, similar investigations have been conducted in Ref.~\cite{Dermisek:2023rvv}, exploring the high-energy muon collider's capacity to study multi-Higgs productions. Our findings align consistently with theirs whenever there is an overlap in the studied scenarios. Similar considerations have also been proposed for the case of light-quark Yukawas at LHC and future hadron colliders \cite{Falkowski:2020znk, Alasfar:2022vqw, Vignaroli:2022fqh}. 

In the SM, due to the smallness of the SM $\bar{\mu}\mu H$ coupling, the leading order cross section $\sigma_{\rm SM}^{\rm LO}$ is highly suppressed. Indeed, the dominant contribution to the SM multi-Higgs annihilation originates from the square of one-loop diagrams ($\sigma_{\rm SM}^{\rm loop}$).\footnote{In our parameterization, the leading order SM contribution $\sigma_{\rm SM}^{\rm LO}$ is  included in $ \sigma_{\rm sub}$. From results presented in Tab.~\ref{tab:mm23h} it is safe to assume that the interferences of the tree-level diagrams entering $\sigma_{\rm SM}^{\rm LO}$ and those at one-loop entering $\sigma_{\rm SM}^{\rm loop}$ can be neglected since it is expected to be of the order $\sqrt{\sigma_{\rm SM}^{\rm loop}\sigma_{\rm SM}^{\rm LO}}$~\cite{Bredt:2022dmm}. In the discussion of Sec.~\ref{sec:Unitary}, $\sigma_{\rm SM}^{\rm loop}$ has been instead ignored.}
Including BSM effects, in the HEFT parameterization, the multi-Higgs production cross sections can be approximated as
\begin{eqnarray}
\sigma_{\rm BSM}(\mu^+\mu^-\to nH)=\sigma_{\rm SM}^{\rm loop} + \sigma_n (\alpha_n^2) +  \sigma_{\rm sub}(\alpha_n,\alpha_{n-1}, \dots
,\alpha_1) \, ,
\end{eqnarray}
where $\sigma_{\rm SM}^{\rm loop}$ is the SM loop-induced contribution, $\sigma_n (\alpha_n^2)$ represents the contribution from the $\bar\mu\mu H^n$ vertex alone and  $\sigma_{\rm sub}(\alpha_n,\alpha_{n-1}, \dots ,\alpha_1)$ contains all the possible $\alpha_i$ vertices with $i\le n$.
In the high-energy limit, as discussed in detail in Sec.~\ref{sec:HEHEFT}, the leading contact-vertex parameterized by $\alpha_n$ dominates the cross section, which can be approximated by $\sigma_n (\alpha_n^2)$ alone. In particular, using Eq.~\eqref{eq:sigma-klm}, one obtains $\sigma_n (\alpha_n^2)\approx \bar{\sigma}_{X_{n00}}$, which is explicitly given  in Eq.~\eqref{eq:mmnhaninsec3}, and leads for $2\le n\le 5$ to  
\begin{eqnarray}
\label{eq:approxmultiH}
&&\sigma_{\rm BSM}(\mu^+\mu^- \to 2H)
\simeq\sigma_{\rm SM}^{\rm loop}(2H) + \frac{\alpha_2^2 m_\mu^2}{16 \pi v^4}\,,\\
&&\sigma_{\rm BSM} (\mu^+\mu^- \to 3H)
\simeq\sigma_{\rm SM}^{\rm loop}(3H) + \frac{3 \alpha_3^2 m_\mu^2 s}{512 \pi^3 v^6 }\,, \nonumber\\
&&\sigma_{\rm BSM}(\mu^+\mu^- \to 4H)
\simeq\sigma_{\rm SM}^{\rm loop}(4H) + \frac{\alpha_4^2 s^2 m_\mu^2}{4096 \pi^5 v^8}\,,  \nonumber\\
&&\sigma_{\rm BSM} (\mu^+\mu^- \to 5H)
\simeq\sigma_{\rm SM}^{\rm loop}(5H) + \frac{5\alpha_5^2 s^3 m_\mu^2}{786432 \pi^7 v^{10}}\,.  \nonumber
\end{eqnarray}

These are, however, approximations that do not take into account mass effects in the final state and subleading diagrams. Moreover, in the $\SMEFTs$ scenario, this approximation would lead to the illusion that there is no dependence on the Wilson coefficient $c_{\ell \varphi}^{(6)}$ ({\it c.f.} Eq.~\eqref{eq:map}) for $4H$ and $5H$ production. As anticipated at the end of Sec.~\ref{sec:tests-Unitary}, in this section we use the exact calculation, performed with the help of {\aNLO}.

\begin{table}[!t]
    \centering
     \scriptsize
\adjustbox{width=1.2\textwidth,center}{
    \begin{tabular}{c|cccc|cccc}
    \hline
$\sqrt{s}$ & \multicolumn{4}{c|}{3 TeV} & \multicolumn{4}{c}{10 TeV}\\
\hline
 & $\alpha_{2(3)}=1^\dagger$ & SM LO & Loop  & VBF &$\alpha_{2(3)}=1^\dagger$ & SM LO & Loop  & VBF \\
\hline
$\sigma~[\fb]$ & \multicolumn{8}{c}{$2H$} \\
\hline
No cut & $2.4\cdot10^{-2}$ &$1.6\cdot10^{-7}$ & $2.6\cdot10^{-3}$ & 0.951 &  $2.4\cdot10^{-2}$ & $1.3\cdot10^{-9}$ & $4.2\cdot10^{-4}$ & 3.80 \\
$M_F>0.8\sqrt{s}$& $2.4\cdot10^{-2}$ & $1.6\cdot10^{-7}$ & $2.6\cdot10^{-3}$ & $6.12\cdot10^{-4}$ & $2.4\cdot10^{-2}$ & $1.3\cdot10^{-9}$ & $4.2\cdot10^{-4}$   &$6.50\cdot10^{-4}$\\
$|\theta_{iB}|>10^\circ$ & $2.3\cdot10^{-2}$ & $1.6\cdot10^{-7}$ & $2.6\cdot10^{-3}$ & $1.18\cdot10^{-4}$ & $2.3\cdot10^{-2}$ & $1.3\cdot10^{-9}$ & $4.1\cdot10^{-4}$ & $3.46\cdot10^{-5}$\\
\hline
event \# & 23 & -- & 2.6 & 0.12 & 230 &-- & 4.1 & 0.3 \\
\hline
$\sigma~[\fb]$ & \multicolumn{8}{c}{$3H$} \\
\hline
No cut & $3.1\cdot10^{-2}$ & $3.0\cdot10^{-8}$ & $1.1\cdot10^{-5}$ & $3.69\cdot10^{-4}$ & $3.7\cdot10^{-1}$ & $2.3\cdot10^{-9}$ & $1.7\cdot10^{-6}$ & $5.52\cdot10^{-3}$\\
$M_F>0.8\sqrt{s}$& $3.1\cdot10^{-2}$  &$3.0\cdot10^{-8}$ & $1.1\cdot10^{-5}$ & $2.84\cdot10^{-6}$ & $3.7\cdot10^{-1}$ & $2.3\cdot10^{-9}$ & $1.7\cdot10^{-6}$ & $7.85\cdot10^{-5}$\\
$|\theta_{iB}|>10^\circ$ & $3.0\cdot10^{-2}$ &$2.8\cdot10^{-8}$ & $1.1\cdot10^{-5}$ & $6.82\cdot10^{-7}$ & $3.5\cdot10^{-1}$ & $2.2\cdot10^{-9}$ & $1.7\cdot10^{-6}$ & $7.37\cdot10^{-5}$\\
$\Delta R_{BB}>0.4$& $2.9\cdot10^{-2}$ & $2.7\cdot10^{-8}$ & $8.1\cdot10^{-6}$ & $6.07\cdot10^{-7}$ & $3.4\cdot10^{-1}$ & $2.1\cdot10^{-9}$ &$6.8\cdot10^{-7}$ & $7.22\cdot10^{-5}$\\
\hline
event \# & 29 & -- & -- & -- & 3400 & -- & -- &   0.7 \\
\hline
\end{tabular}
}
\caption{The cross sections and expected event numbers of $\mu^+\mu^- \to 2H,\,3H$ at 3 and 10 TeV muon colliders.
$^\dagger$In the BSM benchmark points for $nH$ production, the leading Higgs-muon coupling is fixed to its SM value $\alpha_1=1$, while the other couplings except for $\alpha_n$ are set to zero. Cuts are applied consecutively.
}
\label{tab:mm23h}
\end{table}

\begin{table}[]
\centering
    \begin{tabular}{c|ccc|ccc}
    \hline
$\sqrt{s}$ & \multicolumn{3}{c|}{3 TeV} & \multicolumn{3}{c}{10 TeV}\\
\hline
 & $\alpha_{4(5)}=1^\dagger$& SM LO & VBF &$\alpha_{4(5)}=1^\dagger$ & SM LO & VBF \\
\hline
$\sigma~[\fb]$ & \multicolumn{6}{c}{$4H$} \\
\hline
No cut  & $1.7\cdot10^{-2}$ & $3.1\cdot 10^{-10}$ & -- & $2.5$ & $1.6\cdot 10^{-11}$ & --  \\
$M_F>0.8\sqrt{s}$ & $1.7\cdot10^{-2}$ & $3.1\cdot 10^{-10}$ &  $7.2\cdot10^{-8}$ & $2.5$ & $1.6\cdot 10^{-11}$ & $3.8\cdot10^{-4}$  \\
$|\theta_{iB}|>10^\circ$  & $1.6\cdot10^{-2}$ &$2.9\cdot 10^{-10}$&  $6.7\cdot10^{-8}$ & $2.4$ & $1.5\cdot 10^{-11}$ & $3.5\cdot10^{-4}$  \\
$\Delta R_{BB}>0.4$ & $1.4\cdot10^{-2}$ & $2.1\cdot 10^{-10}$ & $6.2\cdot10^{-8}$ & $2.2$ & $5.6\cdot 10^{-12}$ & $3.2\cdot10^{-4}$  \\
\hline
event \# & 14 & -- &-- & 22000 & -- & 3.2 \\
\hline
$\sigma~[\fb]$ & \multicolumn{6}{c}{$5H$} \\
\hline
No cut  & $5.1\cdot10^{-3}$  & $4.8\cdot10^{-12}$ &  -- & 11 & $3.4\cdot10^{-13}$ & --  \\
$M_F>0.8\sqrt{s}$ & $5.1\cdot10^{-3}$ & $4.8\cdot10^{-12}$ &  -- & 11 & $3.4\cdot10^{-13}$ & --  \\
$|\theta_{iB}|>10^\circ$  & $4.7\cdot10^{-3}$ & $4.5\cdot10^{-12}$ &  -- & 9.7 & $3.2\cdot10^{-13}$ & --  \\
$\Delta R_{BB}>0.4$ & $4.0\cdot10^{-3}$ & $3.1\cdot10^{-12}$ &  $7.9\cdot10^{-9}$ & 8.3 & $1.6\cdot10^{-13}$ & $8.8\cdot10^{-4}$  \\
\hline
event \# & 4 & -- & -- & 83000 & -- & 8.8 \\
\hline
\end{tabular}
\caption{The cross sections and expected event numbers of $\mu^+\mu^- \to 4H,5H$ at 3 and 10 TeV muon colliders, respectively.
$^\dagger$In the BSM benchmarks, the leading Higgs-muon coupling is fixed again to its SM value, $\alpha_1=1$, while other couplings are set to zeros except for $\alpha_n$. Cuts are applied consecutively.}
\label{tab:mm45h}
\end{table}

In Tab.~\ref{tab:mm23h}, we list the SM contributions and the corresponding VBF backgrounds for $2H$ and $3H$ production cross sections at a 3 and 10 TeV muon collider. We also show the representative case of $\alpha_2=1$ and $\alpha_3=1$ in the case of $2H$ and $3H$.  Analogous results are displayed for  $4H$ and $5H$ production cross sections in Tab.~\ref{tab:mm45h}, where -- due to the computationally demanding calculation -- we have refrained from showing results for $\sigma_{\rm SM}^{\rm loop}$. However, it is clear from the $2H$ and $3H$ production cross sections in  Tab.~\ref{tab:mm23h} that SM predictions would lead to less than one event, as in the case of $3H$ production, Therefore they are not a necessary input for our analysis. Similarly, for the VBF processes we have not shown some of the results, due to a very difficult phase-space integration without some of the cuts. In general, Tabs.~\ref{tab:mm23h} and \ref{tab:mm45h} show that for all the multi-Higgs production processes up to a multiplicity of five, the SM backgrounds -- including both tree-level and loop-induced events -- are negligible.

\begin{figure}[!t]
  \includegraphics[width=0.48\textwidth]{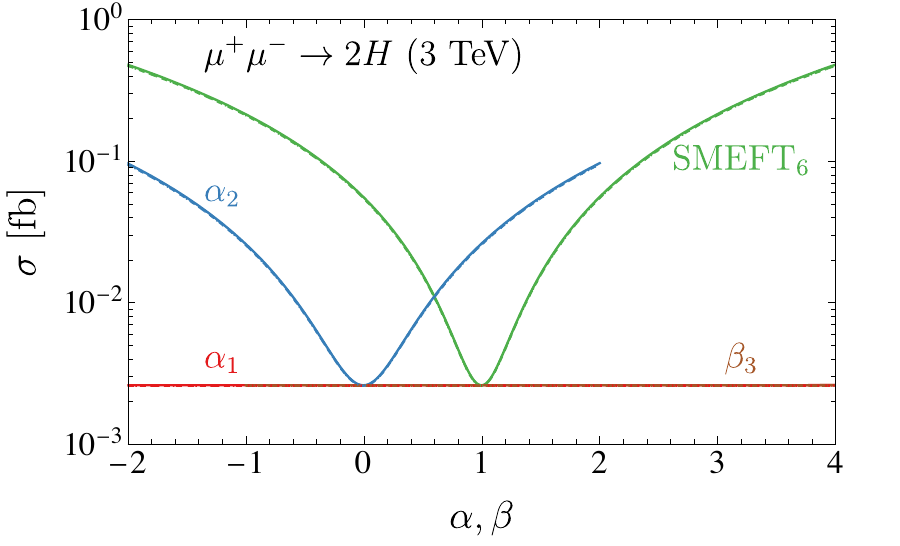}
  \includegraphics[width=0.48\textwidth]{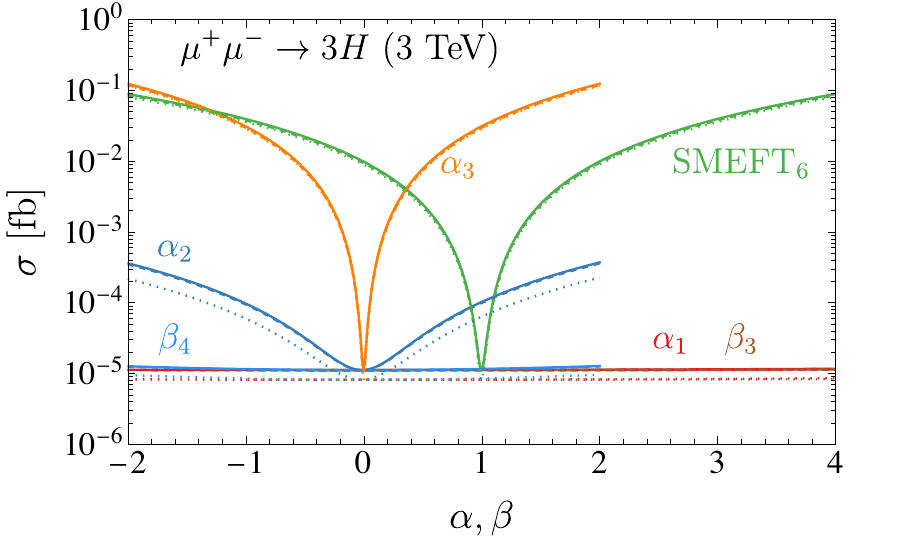}\\
  \includegraphics[width=0.48\textwidth]{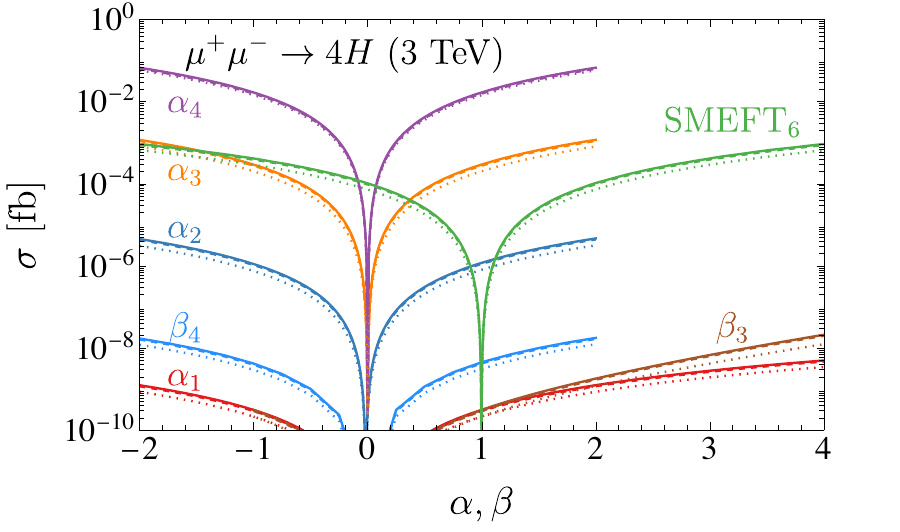}
  \includegraphics[width=0.48\textwidth]{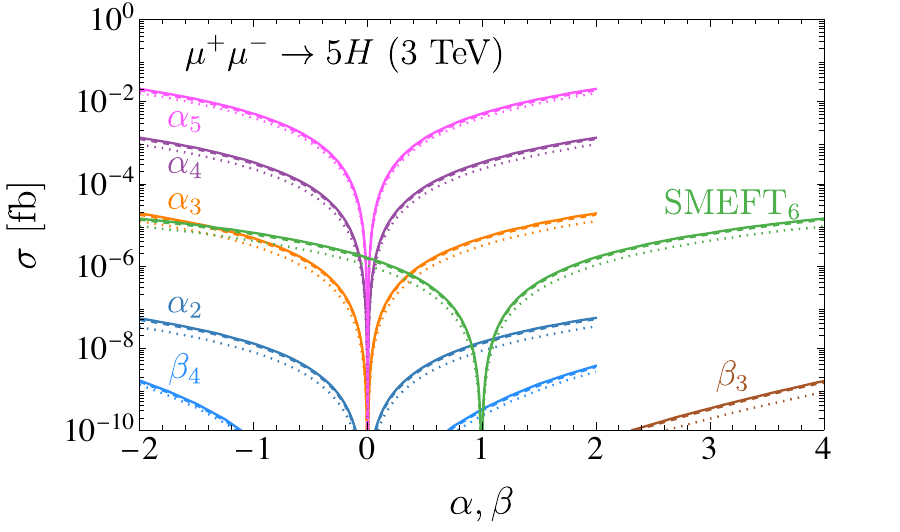}
  \caption{The cross section of $\mu^+\mu^- \to nH$ as function of the parameters $\alpha_i$ and $\beta_k$ at 3~TeV. The green curves are for $\alpha_1=1+\Delta \alpha_1$ in the $\SMEFTs$ scenario,  where the other $\alpha_i$ are given by Eq.~\eqref{eq:dim6lock}. 
  Solid lines refer to the configuration with no cuts, dashed lines to the case with $|\theta_{iB}|>10^\circ$ cuts applied and dotted lines to the case with all cuts applied.}
  \label{fig:mmnhkappa3}
\end{figure}

\begin{figure}[htb]
  \includegraphics[width=0.48\textwidth]{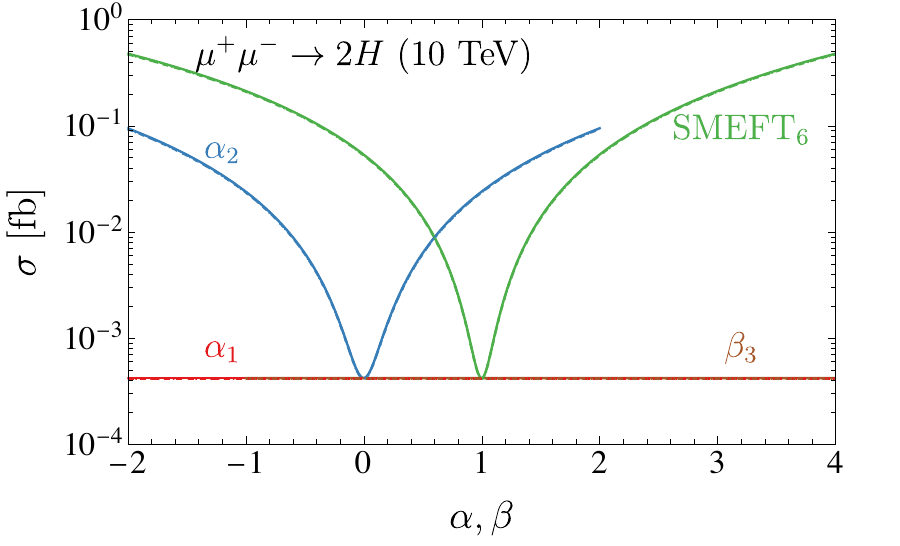}
  \includegraphics[width=0.48\textwidth]{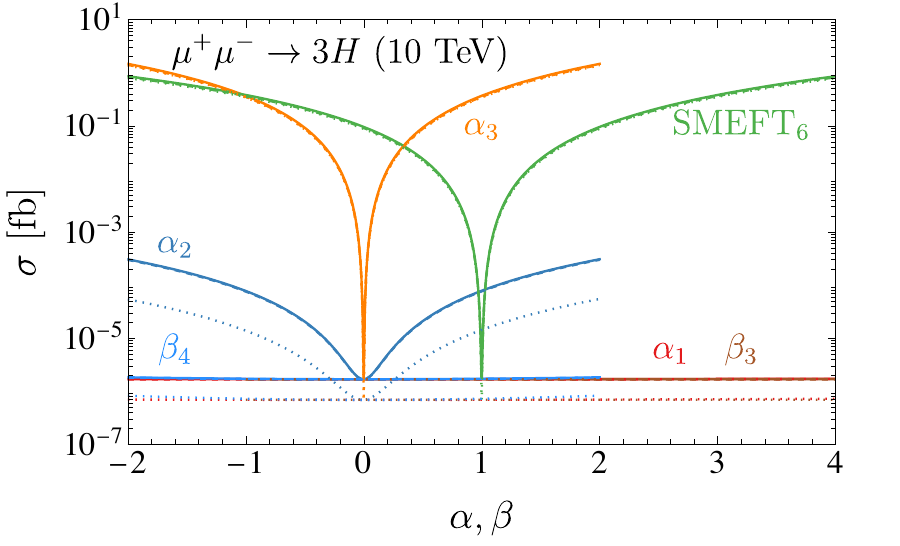}\\
  \includegraphics[width=0.48\textwidth]{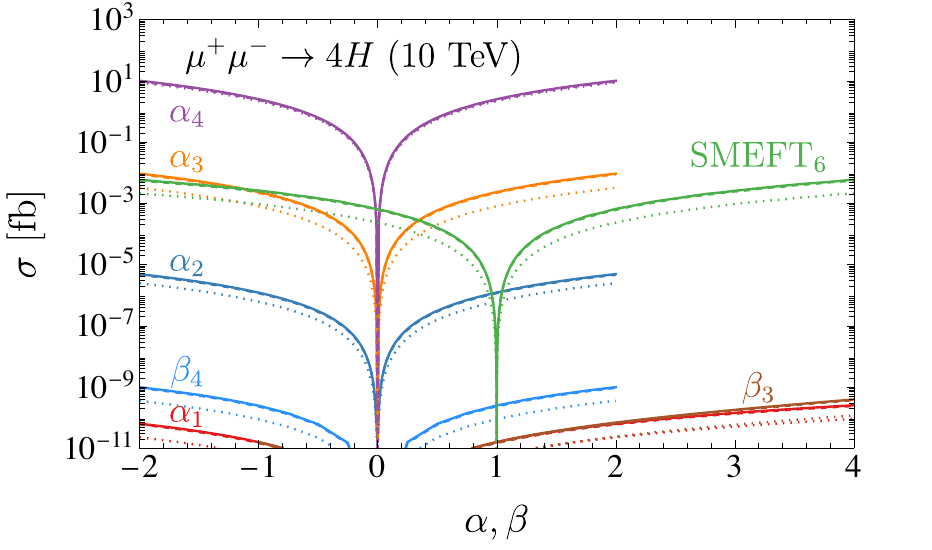}
  \includegraphics[width=0.48\textwidth]{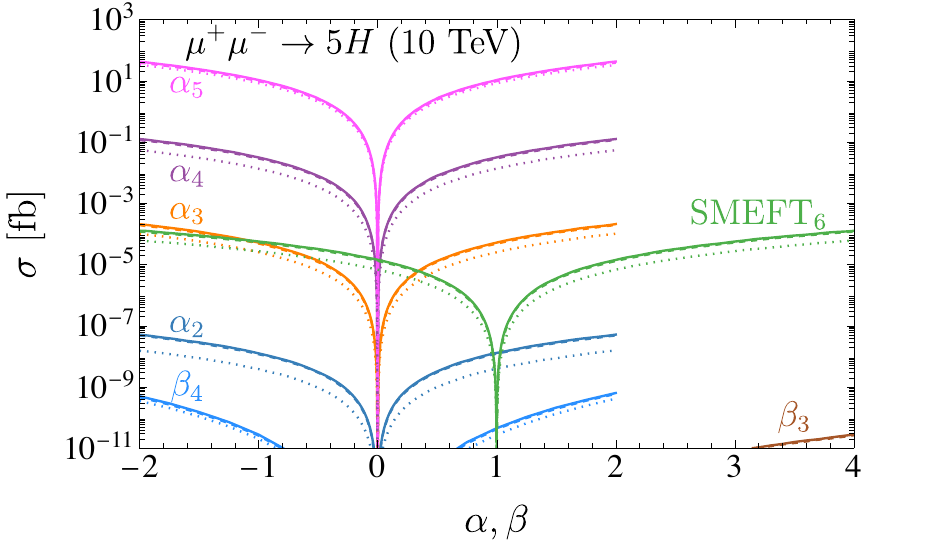}
  \caption{Same as Fig.~\ref{fig:mmnhkappa3} for 10 TeV.
  }
  \label{fig:mmnhkappa10}
\end{figure}

The most important information concerning the BSM contribution is displayed in Figs.~\ref{fig:mmnhkappa3} and \ref{fig:mmnhkappa10}, which are  for 3 and 10 TeV, respectively. In these figures we show the dependence of the cross section for $nH$ production with $2\le n\le 5$ on the $\alpha_i$ parameters.  We easily see that each $nH$ production process strongly depends on the value of $\alpha_n$, while the dependence on $\alpha_{n-k}$, with $1\le k \le n-2$, is much smaller, especially for larger values of $k$. On the other hand, the higher the multiplicity the smaller is the ratio between the dependence on any $\alpha_m$ and the corresponding $\alpha_{m-1}$ parameter, for all $m\le n$ including the leading and the dominant subleading cases $\alpha_{n}$ and  $\alpha_{n-1}$. Also, this ratio decreases at smaller energies, as the comparison between 10 and 3 TeV collision energy shows. Thus,  limits from $5H$ production on $\alpha_5$ may be degraded by very large values of $\alpha_4$, but it is sufficient to consider the case of $4H$ production to constrain $\alpha_4$, and so on up to $\alpha_3$. As a side comment, this pattern further supports the statement that the approximation in Eq.~\eqref{eq:approxmultiH} is not optimal in this context, while it was completely sufficient for deriving unitarity constraints as done in Sec.~\ref{sec:Unitary}. In Figs.~\ref{fig:mmnhkappa3} and \ref{fig:mmnhkappa10} we also show the dependence on the $\beta_i$ coefficients, which parameterize the Higgs self-coupling modifications ({\it cf.}~Eq.~\eqref{eq:LextendedK}). It is manifest that the cross section for prompt multi-Higgs production is mostly unaffected by anomalous Higgs self-couplings.

In the plots, as a green line denoted as ``SMEFT'' we also show the dependence on $\alpha_1$ in the $\SMEFTs$ scenario,  which is defined by Eq.~\eqref{eq:dim6lock}; 
 here $\alpha_1$ is the only free parameter. We see that the case of $2H$ and $3H$ is completely different from the $4H$ and $5H$. Indeed, for the former, the dependence on $\alpha_1$ is the same as the leading $\alpha_i$, which is $\alpha_n=\alpha_2(\alpha_3)$ for $2H(3H)$, while for the latter the dependence on $\alpha_1$ is the same as for $\alpha_3$, and hence subdominant. This can be understood in two different and complementary ways. Starting from HEFT, in the high-energy limit the cross sections are approximately equal to the expressions given in Eq.~\eqref{eq:approxmultiH}. For $2H$ and $3H$, if Eq.~\eqref{eq:dim6lock} is implemented, we obtain the leading dependence on $\alpha_1$ in the $\SMEFTs$ scenario. For $4H$ and $5H$ with Eq.~\eqref{eq:dim6lock},
we obtain exactly zero. In other words, the leading terms at high-energy for HEFT and $\SMEFTs$ are not the same, as the $\SMEFTs$ prediction for $4H$, and especially $5H$, is suppressed.   On the other hand, starting from SMEFT, it is clear that the maximally growing amplitude would be induced by the operator of dimension-6 for $2H$ and $3H$ and dimension-8 for  $4H$ and $5H$. Thus, in the $\SMEFTs$ scenario, the maximally growing configuration, which in general does exist in the less constrained HEFT framework, is absent, and therefore the dependence on $\alpha_i$ is reduced. This would be different, {\it e.g.}, in the $\SMEFTe$ scenario, defined by Eq.~\eqref{eq:dim8lock}.

From the previous discussion, especially at high-energy, we expect a higher sensitivity within the HEFT framework compared to the $\SMEFTs$ scenario. From our simulations based on exact matrix elements, we derive the following bounds on the $\alpha_i$ parameters in the HEFT framework:
  \begin{align}\label{eq:alpha4nH}
   & |\alpha_2|\lesssim0.4\,,& ~&|\alpha_{3}|\lesssim0.3\,,&~&|\alpha_{4}|\lesssim0.5\,,& ~&|\alpha_5|\lesssim0.9\,,&\qquad&{\rm at~3~TeV}\, ,& \\
  &|\alpha_2|\lesssim0.2\,,& ~&|\alpha_3|\lesssim0.03\,,& ~&|\alpha_4|\lesssim0.01\,,&  ~&|\alpha_5|\lesssim0.01\,,& \qquad&{\rm at~10~TeV}\, .&
  \end{align}

 For the case in which no events are expected for the SM background ($n=3,4,5$ at 3 TeV) we have used the assumption $S= 3$ in order to obtain the $2\sigma$ (95\% CL) exclusion limit. In Tab.~\ref{tab:mmnhsig} we further document the obtained results, showing the values of $S/B$ and more digits for the bounds. The signal  corresponds to $S=\sigma_{\rm BSM}-\sigma_{\rm SM}\approx \sigma_{\rm BSM}-\sigma_{\rm SM}^{\rm loop}$, while the background is given by $B=\sigma_{\rm SM}+\sigma_{\rm VBF}\approx \sigma_{\rm SM}^{\rm loop}+\sigma_{\rm VBF}\;{}$\footnote{In these formulae we have understood that both the BSM and total SM contributions include the quantity $\sigma_{\rm SM}^{\rm loop}$, which is the dominant component of the latter.}. The numbers in Tab.~\ref{tab:mmnhsig} have been determined by performing a scan in order to obtain $\mathcal S$ as close as possible to 2.  
 
It is interesting to note that for the cases with no events from the SM background, there is a fast way for obtaining an estimate of these bounds. One can use the approximation $\sigma_{\rm BSM}(\mu^+\mu^-\to nH)\approx \bar{\sigma}_{X_{n00}}$  and require at least 3 events, which leads to 
\begin{eqnarray}\label{eq:alphaninsec4}
|\alpha_n|\lesssim\sqrt{3\frac{8\Gamma(n-1)(4\pi)^{2n-3}v^{2n}}{nm_\mu^2s^{n-2}\calL}}\,,
\end{eqnarray}
where $\calL$ is the luminosity.  The corresponding bounds read $ |\alpha_3|\lesssim0.3$, $|\alpha_{4}|\lesssim0.4$, $|\alpha_{5}|\lesssim0.6$ at 3 TeV, showing how $\bar{\sigma}_{X_{n00}}$ is a good approximation for obtaining sensible bounds. A similar exercise can be repeated, if one assumes no background, also for 10 TeV or different energies by employing Eq.~\eqref{eq:alphaninsec4} and by inspecting the plot in Fig.~\ref{fig:mmnh}, where we have both displayed the cross sections leading to 3 events and plotted $\bar{\sigma}_{X_{n00}}$ for representative values of $s$ and $\alpha_n$.

Considering the $\SMEFTs$ scenario, and therefore the presence of only the dimension 6 operator, Eqs.~\eqref{eq:map} tell us that  from the measurement of $2H$ production we obtain 
\begin{eqnarray}
|\Delta \alpha_1|\lesssim 0.3 ~~\Longleftrightarrow~~\left|c^{(6)}_{\ell\varphi}/\Lambda^2\right| &\lesssim& 3\times 10^{-9} \,{\rm GeV}^{-2} \qquad{\rm at~3~TeV}\,, \label{eq:boundmultiHEFT1a}\\
|\Delta  \alpha_1|\lesssim 0.1 ~~\Longleftrightarrow~~\left|c^{(6)}_{\ell\varphi}/\Lambda^2\right| &\lesssim& 1 \times 10^{-9} \,{\rm GeV}^{-2} \qquad{\rm at~10~TeV}\,, \label{eq:boundmultiHEFT1b}
\end{eqnarray}
  while from the measurement of $3H$ production 
 \begin{eqnarray}
|\Delta  \alpha_1|\lesssim 0.7 ~~\Longleftrightarrow~~  \left|c^{(6)}_{\ell\varphi}/\Lambda^2\right| &\lesssim& 7\times 10^{-9}\, {\rm GeV}^{-2}   \qquad{\rm at~3~TeV}\,, \label{eq:boundmultiHEFT2a}\\
|\Delta  \alpha_1|\lesssim 0.05 ~~\Longleftrightarrow~~ \left|c^{(6)}_{\ell\varphi}/\Lambda^2\right| &\lesssim& 5\times 10^{-10}\, {\rm GeV}^{-2}   \qquad{\rm at~10~TeV}\,. \label{eq:boundmultiHEFT2b}
 \end{eqnarray}
 Thus, at 3 TeV the best constraints on $|\Delta \alpha_1|$ derive from $2H$ production, while at 10 TeV  from $3H$ production. For UV models with new particles $X$ interacting with the muon and/or the Higgs boson with couplings of $\mathcal{O}(1)$, the constraints in Eqs.~\eqref{eq:boundmultiHEFT1a} and \eqref{eq:boundmultiHEFT2b} tell us, respectively, that multi-Higgs production at a muon collider can probe the mass of $X$ up to 20(40) TeV at for 3(10) TeV collisions. 

Within the $\SMEFTs$ scenario, we see that the muon Yukawa coupling can be probed at the percent level via the $3H$ production at 10 TeV. At this energy, also if dimension-8 contributions with $c^{(6)}_{\ell\varphi}\simeq 1$ were present, bounds would be mildly affected. On the contrary, generically at 3 TeV, and at both energies for $2H$ production only, bounds are much less strong. In the general HEFT framework, where any $\alpha_i$ is independent, no information on $|\alpha_1|$ can be derived. Still, the HEFT framework allows to constrain also $\alpha_4$ and $\alpha_5$, for which the $\SMEFTs$ scenario is not relevant. Moreover, for the case of $4H$ and $5H$ production, a dependence on $c^{(6)}_{\ell\varphi}\simeq 1$ is present, but, unlike the case of  $2H$ and $3H$, small values of $c^{(8)}_{\ell\varphi}$ may have a non-negligible impact: therefore we refrain to extract bounds from $4H$ and $5H$ in this scenario. 
A combined study including both $c^{(6)}_{\ell\varphi}$ and $c^{(8)}_{\ell\varphi}$ dependence would be necessary.

\begin{table}[t!]
  \centering
  \begin{tabular}{c|ccc|ccc}
  \hline
  $\sqrt{s}$ & \multicolumn{3}{c|}{3 TeV}  & \multicolumn{3}{c}{10 TeV} \\
  \hline
 $n$   & bound on $|\alpha_n|$   & $S/B$  & $\mathcal{S}$     & bound on $|\alpha_n|$   & $S/B$  & $\mathcal{S}$     \\ \hline
2 & 0.42 & 1.49 & 2.05  & 0.15 & 1.16  & 2.12 \\
3 & 0.33  & -- & --  &  $2.6 \cdot 10^{-2} $ & 3.23 & 2.03 \\
4 & 0.46 & -- & -- & $1.4 \cdot 10^{-2} $ & 1.31 & 2.00 \\
5 & 0.87 & -- &  -- & $9.0\cdot 10^{-3}$ & 0.757 & 2.03  \\
\hline
  \end{tabular}
  \caption{Bounds on the signal strength of the $\bar \mu\mu H^n$ vertex ($\alpha_n$) from the $\mu^+\mu^- \to n H$ processes.  For $nH$ ($n\geq 3$) production at the 3 TeV muon collider, the SM background gives $\sim 0$ event so we do not report the value in the Table.
  }
  \label{tab:mmnhsig}
\end{table}

\subsection{Multi-gauge boson production: $\mu^+\mu^- \to mV$}
\label{sec:multiV}

In comparison with the case of multi-Higgs boson production, multi-gauge boson production channels yield much larger SM cross sections, since gauge couplings are much larger than the muon Yukawa one.
Moreover, the multi-gauge boson production involves the Higgs-muon couplings $\alpha_i$ in a non-trivial way, as can be seen in Tab.~\ref{tab:processes}. Especially in the unitary gauge, which is used in our Monte Carlo simulations, the dependence on the $\alpha_i$ parameters is not manifest from the Feynman diagrams. Instead, assuming the high-energy limit ($\sqrt{s}\gg M_W$) and using the Golstone-Boson equivalence theorem (GBET), it is much easier to understand that also the interactions between longitudinal polarizations of the $W$ and $Z$ bosons and the muon must depend on $\alpha_i$ or equivalently $c_{\ell\varphi}^{(6)}$ in the $\SMEFTs$ scenario via Eqs.~\eqref{eq:map}, see also  Appendix~\ref{sec:HEAmps}.\footnote{It is important to note that the GBET works for amplitudes, but not in general for individual vertices. Therefore, it is not expected that a vertex involving Goldstone bosons leads to Feynman rules in HEFT and SMEFT that can be related via Eq.~\eqref{eq:map}.}

Contrary to the case  of multi-Higgs production, the leading contribution in the SM originates from the tree-level diagrams. Including BSM effects, in the HEFT parameterization the multi-gauge boson production cross sections can be written as
\begin{eqnarray}
\sigma_{\rm BSM}(\mu^+\mu^-\to mV)=  \sigma_{m}(\alpha_{p},\alpha_{p-1}, \dots
,\alpha_1)\, , \label{eq:multigauge}
\end{eqnarray}
where in our parameterization the leading order SM contribution $\sigma_{\rm SM}^{\rm LO}$ is  included in it ($\alpha_1=1$, $\alpha_i=0$ for $i>1$), and $p$ is an integer index such that $p<m$.

Before discussing results obtained via  Monte Carlo simulations based on exact matrix elements, it is very instructive to look at the high-energy limit of the cross section of each of these processes, both in the HEFT framework and in the $\SMEFTs$ scenario. Using Eq.~\eqref{eq:sigma-klm}, where $i$ is the number of neutral Goldstone bosons $\phi_0$ and $j$ the number of charged Goldstone boson $\phi_+\phi_-$ pairs, we obtain at high energies
\begin{eqnarray}
\sigma_{m}(\alpha_{p},\alpha_{p-1}, \dots
,\alpha_1)\approx\bar{\sigma}_{X_{0ij}}\big|_{i+2j=m}
= \frac{(2\pi)^4}{4}|C_{0ij}(0)|^2\Phi_{0ij}\bigg|_{i+2j=m}\, , \label{eq:generalmultiV}
\end{eqnarray}
which leads to the expressions in the following. 
Namely, for $m=2$,
\begin{eqnarray}
\label{eq:approxmulti2V}
\sigma_{\rm BSM}(\mu^+\mu^- \to ZZ)
\simeq  \frac{\left(\alpha _1-1\right){}^2 m_{\mu }^2}{64 \pi  v^4}&=&\frac{v^2 }{128 \pi }\frac{(c_{\ell\varphi}^{(6)})^2}{\Lambda^4},\nonumber\\
\sigma_{\rm BSM} (\mu^+\mu^- \to WW)
\simeq \frac{\left(\alpha _1-1\right){}^2 m_{\mu }^2}{32 \pi  v^4}&=&\frac{v^2 }{64 \pi }\frac{(c_{\ell\varphi}^{(6)})^2}{\Lambda^4}, 
\end{eqnarray}
for $m=3$,
\begin{eqnarray}
\label{eq:approxmulti3V}
\sigma_{\rm BSM} (\mu^+\mu^- \to ZZZ)
\simeq \frac{3 \left(\alpha _1-1\right){}^2 s m_{\mu }^2}{2048 \pi ^3 v^6}&=&\frac{3 s }{4096 \pi ^3}\frac{(c_{\ell\varphi}^{(6)})^2}{\Lambda^4}, \nonumber\\
\sigma_{\rm BSM} (\mu^+\mu^- \to ZWW)
\simeq \frac{\left(\alpha _1-1\right){}^2 s m_{\mu }^2}{1024 \pi ^3 v^6}&=&\frac{s }{2048 \pi ^3}\frac{(c_{\ell\varphi}^{(6)})^2}{\Lambda^4}, 
\end{eqnarray}
for $m=4$,
\begin{eqnarray}
\label{eq:approxmulti4V}
\sigma_{\rm BSM} (\mu^+\mu^- \to ZZZZ)\simeq \frac{\left(-3 \alpha _1+2 \alpha _2+3\right){}^2 s^2 m_{\mu }^2}{262144 \pi ^5 v^8}&\simeq&0+\mathcal O\left(   \frac{s v^2 }{\Lambda^4}\right) \left(c_{\ell\varphi}^{(6)}\right)^2\nonumber\\
\sigma_{\rm BSM} (\mu^+\mu^- \to ZZWW)
\simeq \frac{\left(-3 \alpha _1+2 \alpha _2+3\right){}^2 s^2 m_{\mu }^2}{196608 \pi ^5 v^8}&\simeq&0+\mathcal O\left(   \frac{s v^2 }{\Lambda^4}\right)\left(c_{\ell\varphi}^{(6)}\right)^2,\nonumber\\ 
\sigma_{\rm BSM} (\mu^+\mu^- \to WWWW)
\simeq \frac{\left(-3 \alpha _1+2 \alpha _2+3\right){}^2 s^2 m_{\mu }^2}{98304 \pi ^5 v^8}&\simeq&0+\mathcal O\left(   \frac{s v^2 }{\Lambda^4}\right)\left(c_{\ell\varphi}^{(6)}\right)^2, 
\end{eqnarray}
and for $m=5$ 
\begin{eqnarray}
\label{eq:approxmulti5V}
\sigma_{\rm BSM} (\mu^+\mu^- \to ZZZZZ)
\simeq \frac{5 \left(-3 \alpha _1+2 \alpha _2+3\right){}^2 s^3 m_{\mu }^2}{50331648 \pi ^7 v^{10}}&\simeq&0+\mathcal O\left(   \frac{s v^2 }{\Lambda^4}\right)\left(c_{\ell\varphi}^{(6)}\right)^2, \nonumber\\
\sigma_{\rm BSM} (\mu^+\mu^- \to ZZZWW)
\simeq \frac{\left(-3 \alpha _1+2 \alpha _2+3\right){}^2 s^3 m_{\mu }^2}{12582912 \pi ^7 v^{10}}&\simeq&0+\mathcal O\left(   \frac{s v^2 }{\Lambda^4}\right)\left(c_{\ell\varphi}^{(6)}\right)^2, \nonumber\\
\sigma_{\rm BSM} (\mu^+\mu^- \to ZWWWW)
\simeq \frac{\left(-3 \alpha _1+2 \alpha _2+3\right){}^2 s^3 m_{\mu }^2}{18874368 \pi ^7 v^{10}}&\simeq&0+\mathcal O\left(   \frac{s v^2 }{\Lambda^4}\right)\left(c_{\ell\varphi}^{(6)}\right)^2.
\end{eqnarray}

In the case of $m=4,5$, we have stressed in the previous equations that if one would simply plug  Eq.~\eqref{eq:map} into Eq.~\eqref{eq:generalmultiV} in the $\SMEFTs$ scenario, we would obtain exactly zero. However, in reality the dominant term in the high-energy limit features a different power of $s$ in HEFT and $\SMEFTs$ and cannot be correctly captured by Eq.~\eqref{eq:generalmultiV}, which in turn is based on Eq.~\eqref{eq:sigma-klm} that has been derived within the HEFT framework. In the SMEFT framework, the leading term in energy is not the one from the dimension-6 operator for all multiplicities. Similar to the multi-Higgs production discussed in the previous section, starting from $m=4$, at least dimension-8 effects should be included, while in the $\SMEFTs$ scenario they are neglected.\footnote{We will comment later on the effect induced by the dimension-8 operators.} That is not surprising, since from an SMEFT perspective violating the relation \eqref{eq:dim6lock} means having higher-dimension operators and therefore larger growth with energy. For instance, using the GBET, for $4V$ production one can see that the diagram featuring  the $\bar \mu \mu \phi\phi\phi\phi$ vertex would be the dominant one in SMEFT at high energy. However, as can be seen in the associated Feynman rules in Appendix \ref{sec:feynmanrules}, this vertex is of dimension 8 (or higher).  Therefore,  the HEFT automatically identifies the violation of Eqs.~\eqref{eq:dim6lock} as the leading contribution. In conclusion, while for $m=2,3$ the HEFT framework and $\SMEFTs$ scenarios are expected to lead to a similar phenomenology for a given value of $|\Delta \alpha_1 |$, in the case of $m=4,5$ the situation is expected to be completely different, similarly to what has been observed also for the multi-Higgs production.  For the latter,  we  expect much less stringent bounds for the $\SMEFTs$ scenario, especially for 10 TeV and $5V$ processes, where the growth in energy is suppressed by a factor $1/s^2$.  

As done in the previous section for the case of multi-Higgs production, we use for our simulation the full matrix element calculation via {\aNLO} and {\WZ} and not only the approximations in the high-energy regime. This is crucial because, as just explained, in the $\SMEFTs$ scenario cross sections would be exactly equal to zero for $4V$ and $5V$ production in the high-energy approximation. We discuss in the following the results that have been obtained via our Monte Carlo simulations.

\medskip

As also shown in Ref.~\cite{Han:2021lnp}, the dependence on $\alpha_1$ of the $WW$ and $ZZ$ pair productions is relatively weak, mainly through the $s$-channel Higgs diagrams.
Therefore, we start our analysis from three vector boson processes $\mu^+\mu^- \to WWZ, ZZZ$, which only involves the $\alpha_1$ coupling, as indicated in Tab.~\ref{tab:processes} and in Eq.~\eqref{eq:approxmulti2V}.
\begin{table}[htb]
  \centering
\begin{tabular}{c|ccc|ccc}
\hline
$\sqrt{s}$  & \multicolumn{3}{c|}{3 TeV} &   \multicolumn{3}{c}{10 TeV}\\
\hline
 & $\alpha_1=-1$ & $\alpha_1=1$ & VBF & $\alpha_1=-1$ & $\alpha_1=1$ & VBF\\
 \hline
$\sigma$~[fb] & \multicolumn{6}{c}{$WWZ$}\\
 \hline
 No Cut & 33 & 33 & 15 & 9.7 & 9.5 & 110 \\
 $M_F>0.8\sqrt{s}$ & 33 & 33 & $1.5 \cdot 10^{-1}$ & 9.7 & 9.5 & $4.4 \cdot 10^{-1}$\\
$10^\circ < \theta_B < 170 ^\circ$ & 17 & 17 & $4.5 \cdot 10^{-2}$ & 4.2 & 4.0 & $3.5 \cdot 10^{-2}$  \\
$\Delta R_{BB} > 0.4$ &  15 & 15 &$4.3 \cdot 10^{-2}$ & 3.1 & 2.9 & $2.9 \cdot 10^{-2}$ \\
\hline
  event \#  & $1.5 \cdot 10^4$  & $1.5 \cdot 10^4$ &  43 & $3.1 \cdot 10^4$ & $2.9 \cdot 10^4$ & $2.9 \cdot 10^2$   \\
  \hline
  $S/B$ & -- &\multicolumn{2}{c|}{--} & $7.8 \cdot 10^{-2}$ & \multicolumn{2}{c}{--} \\
  $\calS$ & -- & \multicolumn{2}{c|}{--} & 13 &  \multicolumn{2}{c}{--} \\
  \hline
$\sigma$~[fb] & \multicolumn{6}{c}{$ZZZ$}\\
 \hline
 No Cut & 0.38 & $0.35$ & $9.7\cdot10^{-1}$ & 0.44 & $8.4\cdot10^{-2}$ & 8.1 \\
 $M_F>0.8\sqrt{s}$ & 0.38 & $0.35$ & $3.3\cdot10^{-3}$ & 0.44 & $8.4\cdot10^{-2}$ & $1.2\cdot 10^{-2}$\\
$10^\circ < \theta_B < 170 ^\circ$ & 0.18 & 0.16 & $1.7\cdot10^{-3}$ & 0.37 & $2.5\cdot10^{-2}$ & $2.5\cdot 10^{-3}$  \\
$\Delta R _{BB} > 0.4$ & 0.17 & 0.14 & $1.6\cdot10^{-3}$ & 0.36 & $2.2\cdot10^{-2}$ & $2.5\cdot 10^{-3}$ \\
\hline
  event \#  & 170  & 140 &  1.6 & $3.6 \cdot 10^3$ & $2.2 \cdot 10^2$ & 25   \\
  \hline
  $S/B$ & 0.19 & \multicolumn{2}{c|}{--} &14 & \multicolumn{2}{c}{--}\\
  $\calS$ &2.2 & \multicolumn{2}{c|}{--} &$1.1 \cdot 10^{2}$ & \multicolumn{2}{c}{--}\\
  \hline
\end{tabular}
  \caption{The cross sections and expected event numbers of $\mu^+\mu^- \to 3V$ at 3 and 10 TeV muon colliders, respectively. The background includes both the SM contribution ($\alpha_1=1$) and VBF. The cuts are applied consecutively.}
  \label{tab:mmvvvSM}
\end{table}

\begin{figure}[htb]
  \centering
  \includegraphics[width=0.49\textwidth]{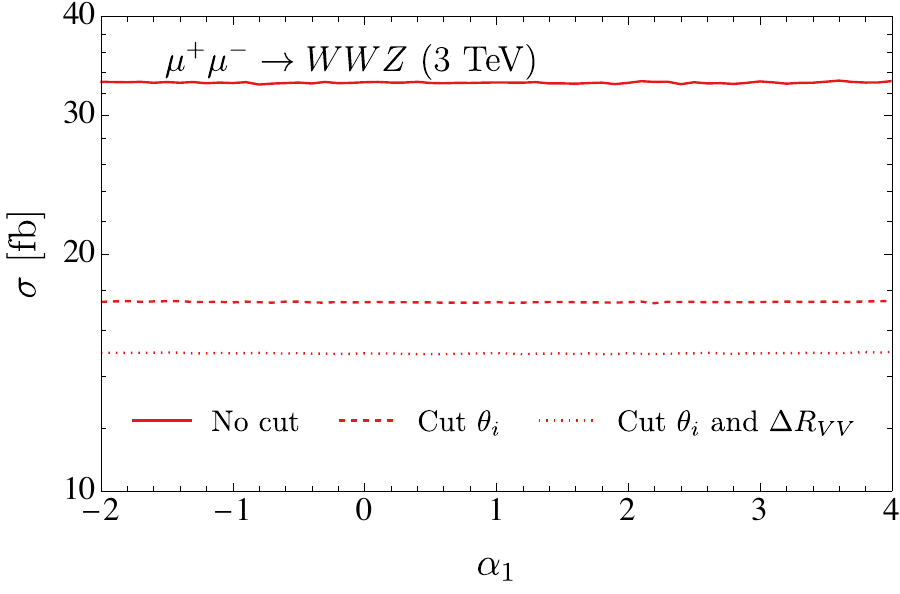}
  \includegraphics[width=0.49\textwidth]{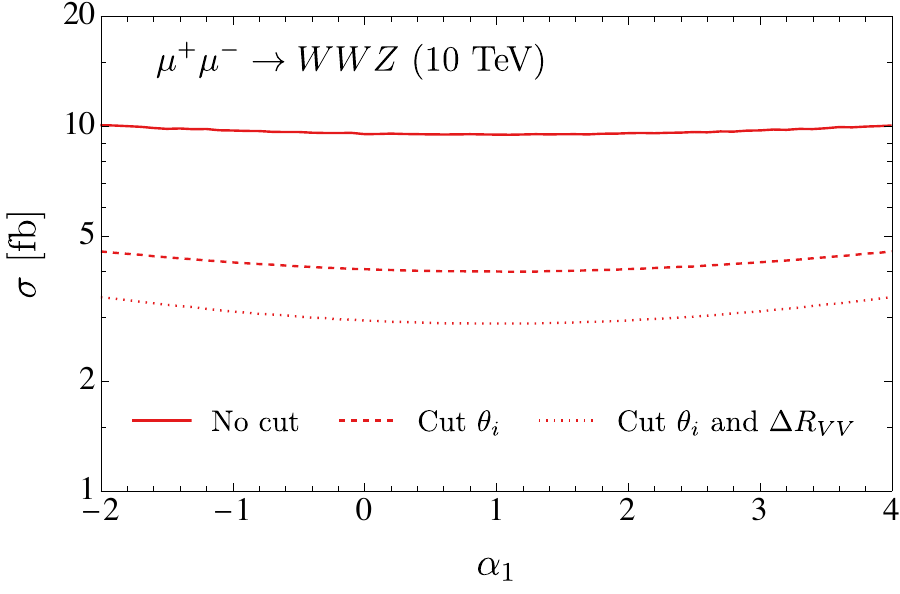}
  \caption{The cross section of $\mu^+\mu^- \to WWZ$ as a function of the form factors $\alpha_1$. }
  \label{fig:mmwwzkappa}
\end{figure}

\begin{figure}[htb]
  \centering
  \includegraphics[width=0.46\textwidth]{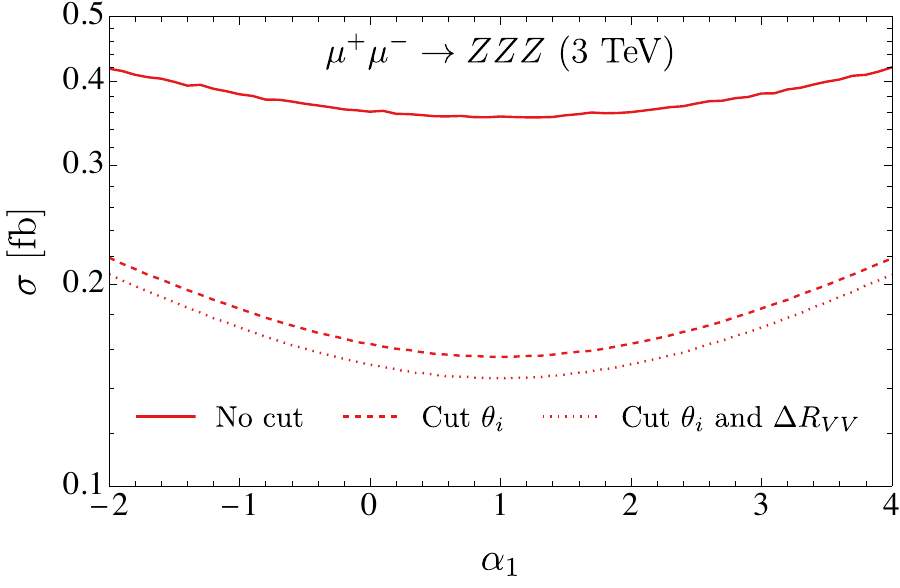}
  \includegraphics[width=0.46\textwidth]{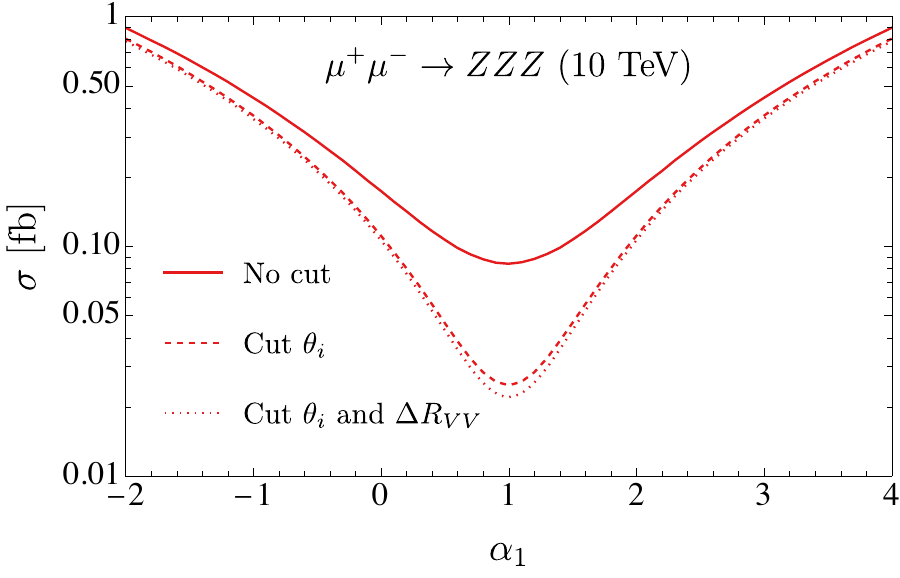}
  \caption{The cross section of $\mu^+\mu^- \to ZZZ$ as a function of the form factors $\alpha_1$.}
  \label{fig:mmzzzkappa}
\end{figure}

In Tab.~\ref{tab:mmvvvSM}, we show the results for the configuration $\alpha_1=1$ (the SM) and the representative BSM case $\alpha_1=-1$. Tab.~\ref{tab:mmvvvSM} is providing the equivalent kind of information that is reported also in, {\it e.g.}, Tab.~\ref{tab:mm23h} for multi-Higgs production.  The full cross-section dependence on $\alpha_1$  is instead shown in Fig.~\ref{fig:mmwwzkappa} and Fig.~\ref{fig:mmzzzkappa} for $WWZ$ and $ZZZ$, respectively.
From the table and the plots it is manifest that a 3 TeV collider is not very useful in this respect and cannot even distinguish the sign of $\alpha_1$. On the contrary, a 10 TeV muon collider would lead to different results. 

It is manifest that at 10 TeV the dependence on $\alpha_1$ is much larger in $ZZZ$ than in $WWZ$ production. The former can offer a ``guaranteed discovery'' on the sign of $\alpha_1$ (hence, of the muon-Higgs coupling) at a 10 TeV muon collider. As we will show in the next section, this process is in fact superior to $ZH$ production in this respect. In particular,  with ${\mathcal S} = 2.1$ and $S/B=0.14$, the quantity $|\Delta \alpha_1|$ can be constrained to 
\begin{eqnarray}
\label{eq:boundmultiZZZHEFT}
 |\Delta \alpha_1|\lesssim 0.2 \qquad {\rm~at~10~TeV}\,,
\end{eqnarray}
by $ZZZ$ production, and with  ${\mathcal S} = 2.7$ and $S/B=0.016$, to 
\begin{eqnarray}
  |\Delta \alpha_1|\lesssim 0.9 \qquad {\rm~at~10~TeV}\,,
\end{eqnarray}
by $WWZ$ production.
In the $\SMEFTs$ scenario this implies for $ZZZ$
\begin{eqnarray}
\left|c^{(6)}_{\ell\varphi}/\Lambda^2\right| \lesssim {2} \times 10^{-9} \,{\rm GeV}^{-2} \qquad{\rm at~10~TeV}\,, \label{eq:boundmultiZZZ}
\end{eqnarray}
and for $WWZ$
\begin{eqnarray}
\left|c^{(6)}_{\ell\varphi}/\Lambda^2\right| \lesssim {9} \times 10^{-9} \,{\rm GeV}^{-2} \qquad{\rm at~10~TeV}\,. \label{eq:boundmultiWWZ}
\end{eqnarray}

We notice that SM cross sections are much larger for the case of $WWZ$ production compared to $ZZZ$ production. Thus, rather than an intrinsic larger dependence on $\alpha_1$ of $(\sigma_{\rm BSM}-\sigma_{\rm SM})$,  the background from  $\sigma_{\rm SM}=\sigma_{\rm BSM}\big|_{\alpha_1=1}$ is the actual limiting factor, as can also be deduced by the comparison of Fig.~\ref{fig:mmwwzkappa} and Fig.~\ref{fig:mmzzzkappa}.

\begin{figure}[htb]
  \centering
  \includegraphics[width=0.47\textwidth]{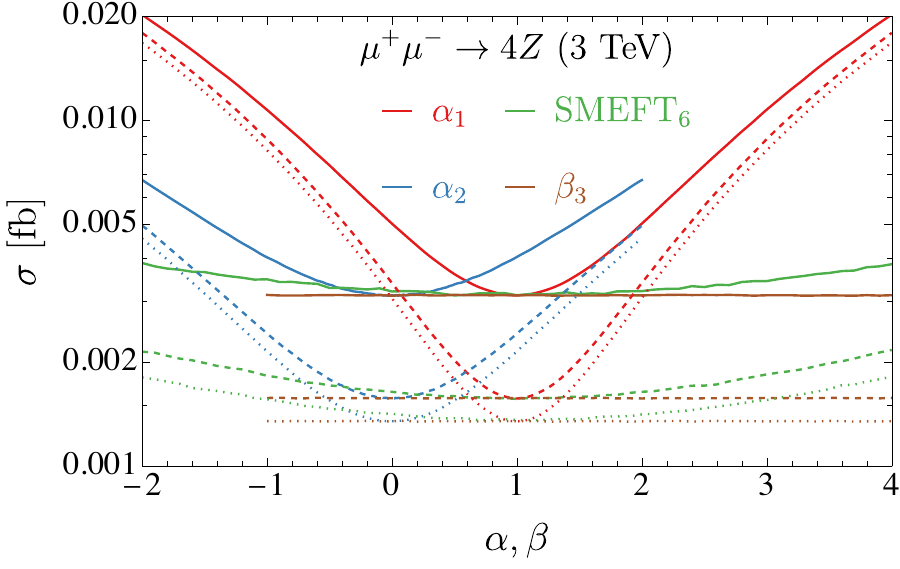}
  \includegraphics[width=0.49\textwidth]{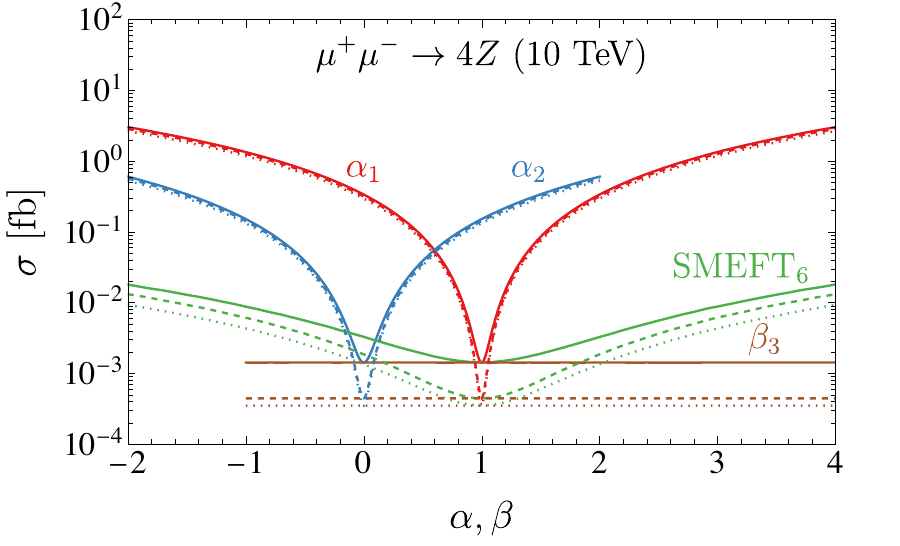}
  \caption{Same as Fig.~\ref{fig:mmnhkappa3} for $ 4Z$ production at 3 and 10 TeV, respectively.}
  \label{fig:mm4Zkappa}
\end{figure}

\begin{figure}[htb]
  \centering
  \includegraphics[width=0.49\textwidth]{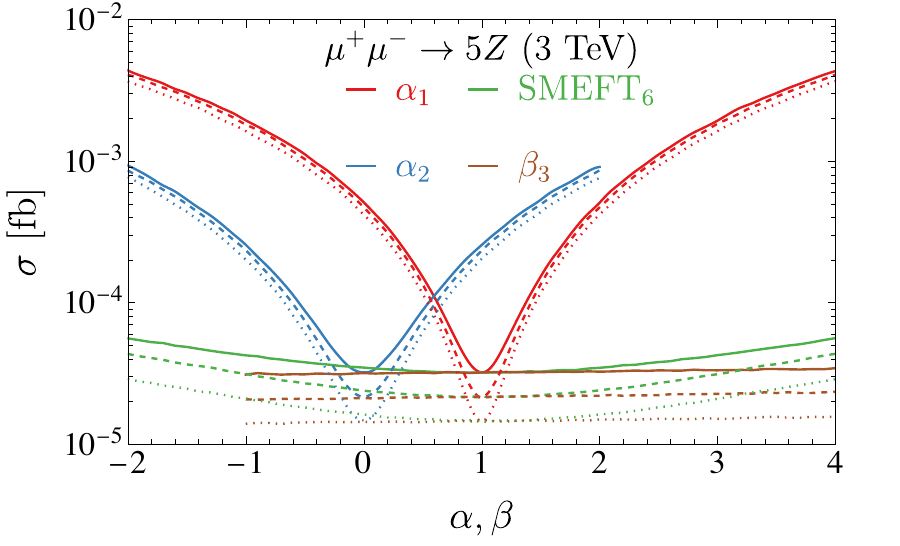}
  \includegraphics[width=0.49\textwidth]{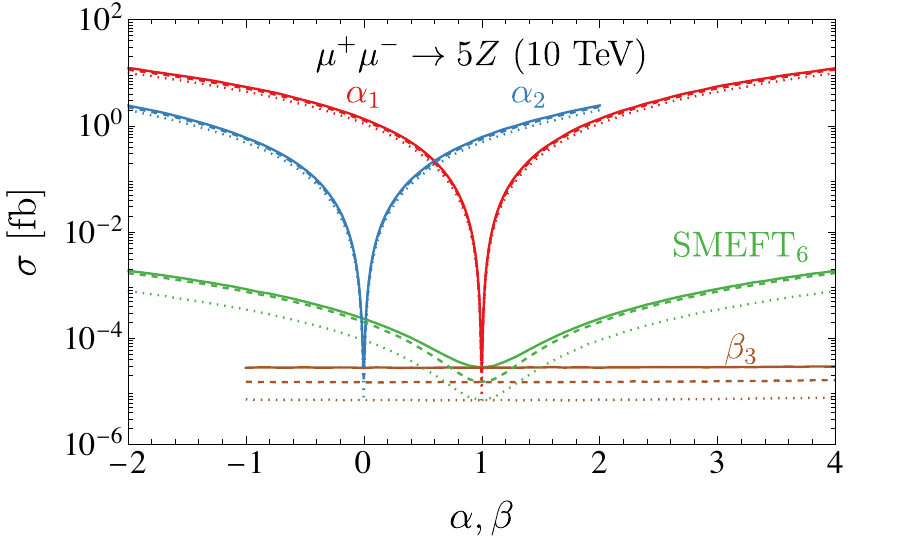}
  \caption{Same as Fig.~\ref{fig:mmnhkappa3} for $ 5Z$ production at 3 and 10 TeV, respectively.}
  \label{fig:mm5Zkappa}
\end{figure}

For the higher multiplicities considered, $4V$ and $5V$, the cross section depends not only on $\alpha_1$ but also on $\alpha_2$ (see the first row of Tab.~\ref{tab:processes}) and $\beta_3$. We show only the case of $4Z$ and $5Z$ production here in the respective Figs.~\ref{fig:mm4Zkappa} and \ref{fig:mm5Zkappa}, since they, similar to the case of $ZZZ$ production, have the largest dependence on the $\alpha_i$ parameter and the smallest SM cross sections. The rest of the $4V$, $5V$ processes are displayed in Appendix \ref{sec:furtherplots} in Figs.~\ref{fig:mm4vkappa} and \ref{fig:mm5vkappa}. 

First of all, we notice that the $4V$ and $5V$ production processes show no significant dependence on $\beta_3$, which allows us to focus on the Yukawa sector with no additional assumptions on the Higgs self-interaction. Second, we  clearly see what we have already anticipated: varying $\alpha_1$ or $\alpha_2$ independently leads to very different results w.r.t.~the one-parameter space defined by Eq.~\eqref{eq:dim6lock}, which corresponds to the $\SMEFTs$ scenario. We stress again that this is not surprising, because according to Eq.~\eqref{eq:map}, in the SMEFT a large value for $\alpha_1\,(\alpha_2)$ with $\alpha_2=0\,(\alpha_1=1)$ implies -- in order to keep at least $c_{\ell\varphi}^{(6)}$ and  $c_{\ell\varphi}^{(8)}$  of the same order -- either a small $\Lambda$ ($\Lambda\simeq v$) or, assuming a large $\Lambda$ ($\Lambda\gg v$), a very large value of $c_{\ell\varphi}^{(8)}$ w.r.t. $c_{\ell\varphi}^{(6)}$, namely  $c_{\ell\varphi}^{(8)} \simeq \frac{\Lambda^2}{v^2} c_{\ell\varphi}^{(6)}$. This is a typical case pointing to the fact that although HEFT and SMEFT can be parametrically mapped into each other,  the BSM scenarios that are correctly captured are different for each of them.
 
\begin{figure}[!t]
  \centering
  \includegraphics[width=0.49\textwidth]{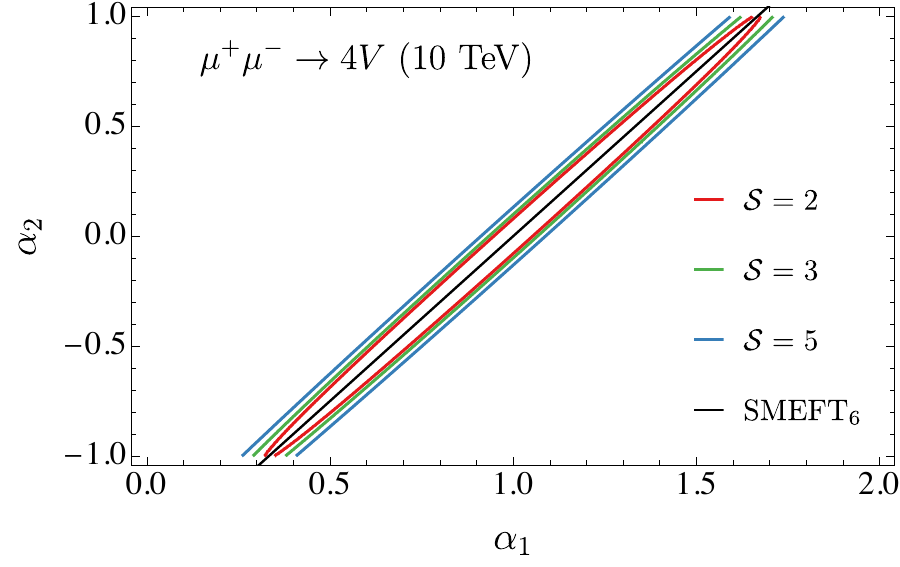}
  \includegraphics[width=0.49\textwidth]{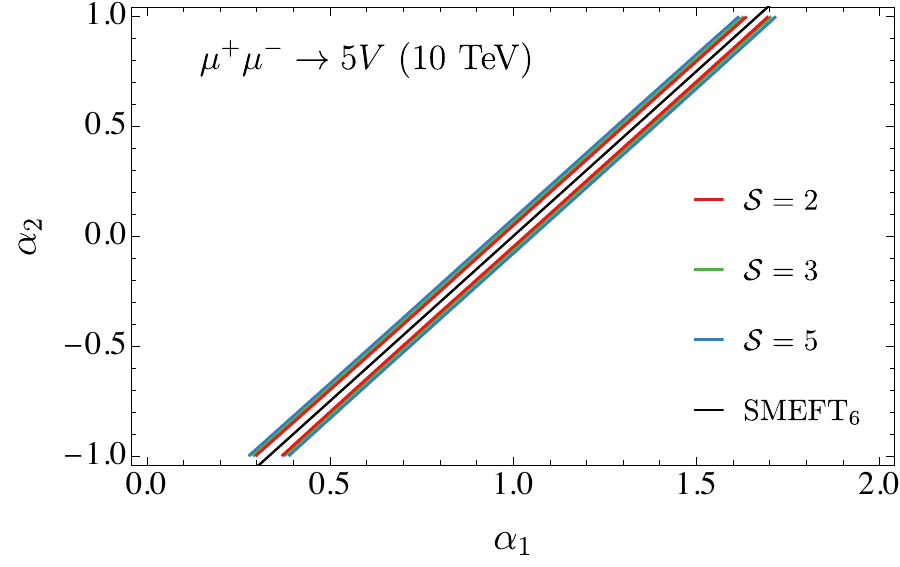}
  \caption{Contour-plots displaying the constraints in the $(\alpha_1,\,\alpha_2)$  plane from  $\mu^+\mu^- \to 4 V$ (left) and $\mu^+\mu^- \to 5 V$ (right)  production at a 10 TeV  muon collider. The red, green, and blue curves represent the ${\mathcal S}=2,\, 3,\, 5$ significances, respectively. The black solid line corresponds to the $\SMEFTs$ scenario, {\it i.e.}, Eq.~\eqref{eq:dim6lock}.}
  \label{fig:contour45V}
\end{figure}

In general, we can summarize the results obtained for $4V$ and $5V$ production as follows: In HEFT, given the large dependence on both  $\alpha_1$ and $\alpha_2$, we have to constrain the two parameters simultaneously in the $(\alpha_1,\,\alpha_2)$ plane. For processes with more $W$ bosons in the final state, the larger contribution  from the gauge interactions leads to larger SM cross sections, while for processes with more $Z$ bosons in the final state, there is a stronger dependence on the Yukawa interactions, leading to a stronger signal associated to anomalous interactions (but the overall cross sections are smaller).
 Due to either the weak signal strength and/or lack of events, these processes do not give interesting constraints on $(\alpha_1,\,\alpha_2)$ at a 3 TeV muon collider. 
 We present therefore only the $(\alpha_1,\,\alpha_2)$ contour plots for the determination of the $\alpha_1$ and $\alpha_2$ signal strengths at a 10 TeV muon collider in Fig.~\ref{fig:contour45V} for 4V (left) and 5V (right) production.   The main constraints originate from $4Z$ and $5Z$ production.

 In the plots in Fig.~\ref{fig:contour45V} we display the relation \eqref{eq:dim6lock} connecting $\alpha_1$ and $\alpha_2$ in the $\SMEFTs$ scenario as a black line. Not surprisingly, the $(\alpha_1,\alpha_2)$ bounds are very elongated ellipses aligned around the black line. The message is clear: if we fix $\alpha_1$ ($\alpha_2$) we get very strong bounds on $\alpha_2$ ($\alpha_1$) in HEFT, of the order $|\alpha_{2}|\lesssim 0.1$ ($|\Delta\alpha_{1}|\lesssim 0.1$). On the contrary, if we assume   Eq.~\eqref{eq:dim6lock}, {\it i.e.}, the $\SMEFTs$ scenario, especially for 4V, the bounds are much looser: of the order $|\Delta\alpha_{1}|\, ,|\alpha_{2}|\lesssim 1$.
 
 We cannot convert the bounds in the HEFT framework into the $\SMEFTs$ scenario by simply using Eq.~\eqref{eq:map}, since the cross section at high energy is dominated by terms that vanish if only the dimension-6 operator is present. On the other hand, if we allow for the presence of contributions proportional to $c^{(8)}_{\ell\varphi}$, we can actually try to set bounds on this coefficient. Doing this, we obtain $ \left|c^{(8)}_{\ell\varphi}/\Lambda^4\right| \leq 1.6\times10^{-14} \, {\rm GeV}^{-4}$, which for  $|c^{(8)}_{\ell\varphi}|\simeq 1$  implies $\Lambda\lesssim  3~{\rm TeV}$ and  $|c^{(8)}_{\ell\varphi}|\simeq 0.01$  implies $\Lambda\lesssim  9~{\rm TeV}$. Clearly, for collisions at 10 TeV, this condition on $\Lambda$ is pointing to HEFT as a more suitable framework for describing such dynamics, as otherwise the physics behind the muon-mass generation mechanism should be in direct reach of the collider.

\begin{figure}[!t]
  \centering
  \includegraphics[width=0.49\textwidth]{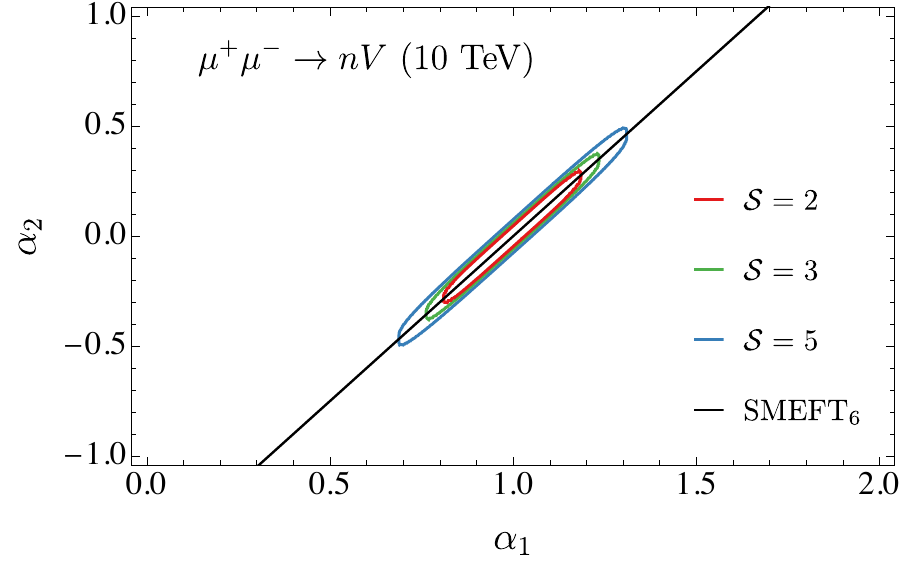}
  \caption{Same as Fig.~\ref{fig:contour45V} for all  $\mu^+\mu^- \to n V$ processes combined at a 10 TeV muon collider.}
  \label{fig:contourVV}
\end{figure}

 In Fig.~\ref{fig:contourVV}, we combine the above constraints with the one on $\alpha_1$ from $\mu^+\mu^- \to ZZZ$. While this has only a moderate impact on the bounds in the HEFT framework, especially in the parameter-space region around  $\alpha_1\sim1$ or $\alpha_2\sim0$, it has a profound impact in the case of the $\SMEFTs$ scenario. Indeed, the $ZZZ$ production does not suffer of the disparity between $\SMEFTs$ and HEFT discussed before; constraints on $\alpha_1$ and  $c^{(6)}_{\ell\varphi}$ can be one-to-one related, since there is no dependence on $\alpha_2$. Consequently, the ellipses are much less elongated, translating for the $\SMEFTs$ scenario into the bound
 \begin{eqnarray}
 |\Delta \alpha_1|\lesssim 0.2 \Longleftrightarrow \left|c^{(6)}_{\ell\varphi}/\Lambda^2\right| \lesssim {2}\times 10^{-9}\, {\rm GeV}^{-2}   \qquad{\rm at~10~TeV} \,.\label{eq:boundmultivandZZZ}
 \end{eqnarray}
 
From a HEFT perspective, the range of allowed $\Delta \alpha_1$ or  $\alpha_2$ for a given specific value of the corresponding other parameter $\alpha_2$ or $\Delta \alpha_1$ close to zero, is only mildly affected. On the other hand, values of $|\Delta \alpha_1|\gtrsim 0.3 $ and $|\alpha_2|\gtrsim 0.5 $ can be excluded, regardless of the value of  $\alpha_2$ or $\Delta \alpha_1$ , respectively.

\subsection{Higgs-associated gauge boson production: $\mu^+\mu^-\to mV+nH$}

\label{sec:multiVH} 

We now turn to the description of Higgs and gauge boson associated production, $\mu^+\mu^-\to mV+nH$ with $2 \le m+n \le 5$ and $m,n \neq 0$.
A special role in this context is covered by $\mu^+\mu^- \to ZH$ production, the  process with the largest SM cross section and the only one with $m+n=2$. As also discussed  in Ref.~\cite{Han:2021lnp}, this process is sensitive to the Higgs-muon coupling at a high-energy muon collider. The corresponding Feynman diagrams involve only the $\bar{\mu}\mu H$ interactions parameterized by $\alpha_1$;  hence, as also indicated in Tab.~\ref{tab:processes}, this process does not depend on any other $\alpha_i$, 

\begin{figure}[htb]
  \centering
  \includegraphics[width=0.49\textwidth]{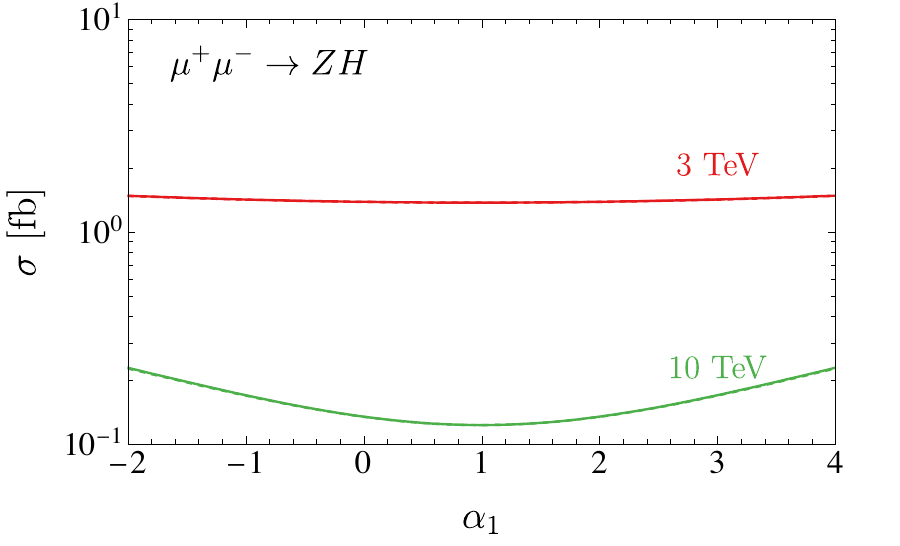}
  \caption{The $\alpha_1$ dependence of $\mu^+\mu^- \to ZH$ cross sections at a 3 (10) TeV muon collider.}
  \label{fig:mmzhkappa}
\end{figure}

The $\alpha_1$ dependence of the total cross section at a muon collider of 3 and 10 TeV is shown in Fig.~\ref{fig:mmzhkappa}.
As already observed in Ref.~\cite{Han:2021lnp}, the deviation of the $ZH$ cross section from the SM value has a symmetry $\sigma(1+\Delta\alpha_1)=\sigma(1-\Delta\alpha_1)$.
The predictions for $\alpha_1=\pm 1$ values (\emph{i.e.,} $\Delta\alpha_1=0$ and $\Delta\alpha_1=-2$) are tabulated in Tab.~\ref{tab:mmzh}, where $\alpha_1=1$ corresponds to the SM scenario. Table \ref{tab:mmzh} has a structure very similar to the one of, {\it e.g.}, Tab.~\ref{tab:mmvvvSM} for $3V$ production.  
 As explicitly demonstrated in Refs.~\cite{Han:2021lnp}, we can see that the VBF background can be well separated from the annihilation signal via the $M_F > 0.8 \sqrt {s}$ cut, which suppresses the VBF contribution by over three orders of magnitude.\footnote{
Obviously, the separation cut $\Delta R_{BB}>0.4$ is not relevant to the $2\to2$ annihilation processes.}

Via these customized cuts, we obtain the signal-to-background ratios $S/B=0.036~(0.37)$ for a 3~(10) TeV muon collider for the case $\alpha_1=-1$.
For a 3 TeV machine with the luminosity $\calL=1~\iab$, we obtain the statistical sensitivity $\calS=1.33$ for $\alpha_1=-1$.
In other words, through the $ZH$ production, a 3 TeV muon collider is not expected to measure the deviation of the Higgs-muon coupling within $|\Delta\alpha_1|=2$ in the 95\% CL and therefore discriminate $\alpha_1=1$ from $\alpha_1=-1$.
In comparison, at 10 TeV assuming the luminosity $\calL=10~\iab$, we can constrain the deviation of the Higgs-muon coupling from its SM value within
\begin{equation}
 |\Delta \alpha_1|\lesssim 0.8    \qquad{\rm at~10~TeV}  \,,
 \end{equation}
which in the $\SMEFTs$ scenario translates into
\begin{eqnarray}
\left|c^{(6)}_{\ell\varphi}/\Lambda^2\right| \lesssim 8\times 10^{-9} \,{\rm GeV}^{-2} \qquad{\rm at~10~TeV}  \,.\label{eq:boundZH}
\end{eqnarray}
As anticipated, we notice that the bounds from $ZZZ$ production at 10 TeV are superior to $ZH$, {\it cf.}~Eqs.~\eqref{eq:boundmultiZZZHEFT}  and \eqref{eq:boundmultiZZZ}. 
\begin{table}[t!]
  \centering
  \begin{tabular}{c|ccc|ccc}
  \hline
  $\sqrt{s}$ & \multicolumn{3}{c|}{3 TeV}  & \multicolumn{3}{c}{10 TeV} \\
  \hline
  $\sigma$ [fb]  & $\alpha_1=-1$        & $\alpha_1=1$  & VBF       & $\alpha_1=-1$         & $\alpha_1=1$   & VBF      \\ \hline
  No Cut                             & $1.42$ & $1.37$ & 10.0 & $1.70 \cdot 10^{-1} $ & $1.23\cdot 10^{-1}$  & 35.7\\
 $M_F>0.8\sqrt{s}$ & 1.42  & 1.37 & $8.70\cdot10^{-3}$ &  $1.69 \cdot 10^{-1} $ & $1.23 \cdot 10^{-1}$ & $1.28\cdot10^{-2}$ \\
 $10^\circ < \theta_B < 170 ^\circ$ & $1.42$ & $1.37$ & $1.88\cdot10^{-3}$ & $1.69 \cdot 10^{-1} $ & $1.23 \cdot 10^{-1}$ & $5.86 \cdot 10^{-4}$\\ \hline
  event \#  & 1420  & 1370 &  1.88 & 1690 & 1230 & 5.86   \\
  \hline
  $S/B$ & 0.036 & \multicolumn{2}{c|}{--} & 0.37 & \multicolumn{2}{c}{--}\\
  $\calS$ &1.33 & \multicolumn{2}{c|}{--} & 12.4 & \multicolumn{2}{c}{--}\\
  \hline
  \end{tabular}
  \caption{Same as Tab.~\ref{tab:mmvvvSM} for  $\mu^+\mu^- \to ZH$ at 3 and 10 TeV muon colliders.}
  \label{tab:mmzh}
\end{table}
\\

We consider now the Higgs and gauge boson associated production at higher multiplicities, $\mu^+\mu^-\to mV+nH$ with $3 \le m+n \le 5$, and $m, n \neq 0$.
This class of processes combines the advantages of multi-Higgs production and multi-gauge boson production. On the one hand, as also discussed in Ref.~\cite{Han:2021lnp}, the  Higgs and gauge boson associated production processes have a better sensitivity to the BSM physics than multi-gauge boson production processes. On the other hand, they have larger cross sections than  the purely multi-Higgs boson production.
As in the case of the  multi-gauge boson production, Eq.~\eqref{eq:multigauge},  the cross section can be parameterized as  
\begin{eqnarray}
\sigma_{\rm BSM}(\mu^+\mu^-\to mV+nH)=  \sigma_{m,n}(\alpha_{p},\alpha_{p-1}, \dots
,\alpha_1)\, , \label{eq:multiHgauge}
\end{eqnarray}
where in our parameterization the leading order SM contribution $\sigma_{\rm SM}^{\rm LO}$ is  included, and $p$ is an integer index such that $p<m+n$. Also in this case, it is instructive to look at the high-energy limit of the cross sections of these processes, which by Eq.~\eqref{eq:sigma-klm} reads
\begin{eqnarray}
\sigma_{m,n}(\alpha_{p},\alpha_{p-1}, \dots
,\alpha_1)\approx\bar{\sigma}_{X_{nij}}\big|_{i+2j=m}
=\frac{(2\pi)^4}{4}|C_{nij}(0)|^2\Phi_{nij}\bigg|_{i+2j=m}\, .
\label{eq:generalmultiHV}
\end{eqnarray}
Including also the case $m+n=2$, from which we started, Eq.~\eqref{eq:generalmultiHV} leads to
\begin{eqnarray}
\label{eq:approxmultiHV2}
\sigma_{\rm BSM}(\mu^+\mu^- \to HZ)
\simeq \frac{\left(\alpha _1-1\right){}^2 m_{\mu }^2}{32 \pi  v^4} = \frac{v^2 }{64 \pi }\frac{(c_{\ell\varphi}^{(6)})^2}{\Lambda^4} \, ,
\end{eqnarray}
and for $m+n=3$ to
\begin{eqnarray}
\label{eq:approxmultiHV3}
\sigma_{\rm BSM}(\mu^+\mu^- \to HHZ)
\simeq \frac{\left(-\alpha _1+\alpha _2+1\right){}^2 s \,m_{\mu }^2}{512 \pi ^3 v^6} =\frac{s }{4096 \pi ^3}\frac{(c_{\ell\varphi}^{(6)})^2}{\Lambda^4}\, ,\nonumber\\
\sigma_{\rm BSM}(\mu^+\mu^- \to HZZ)
\simeq \frac{\left(-\alpha _1+\alpha _2+1\right){}^2 s \,m_{\mu }^2}{512 \pi ^3 v^6}= \frac{s }{4096 \pi ^3}\frac{(c_{\ell\varphi}^{(6)})^2}{\Lambda^4} \, ,\nonumber\\
\sigma_{\rm BSM}(\mu^+\mu^- \to HWW)
\simeq \frac{\left(-\alpha _1+\alpha _2+1\right){}^2 s \,m_{\mu }^2}{256 \pi ^3 v^6}  = \frac{s }{2048 \pi ^3}\frac{(c_{\ell\varphi}^{(6)})^2}{\Lambda^4}\, ,
\end{eqnarray}
for $m+n=4$ to
\begin{eqnarray}
\label{eq:approxmultiHV4}
\sigma_{\rm BSM}(\mu^+\mu^- \to HHHZ)
\simeq \frac{\left(\alpha _1-\alpha _2+\alpha _3-1\right){}^2 s^2 m_{\mu }^2}{16384 \pi ^5 v^8}  &\simeq&0+\mathcal O\left(   \frac{s v^2 }{\Lambda^4}\right) \left(c_{\ell\varphi}^{(6)}\right)^2\, ,\nonumber\\
\sigma_{\rm BSM}(\mu^+\mu^- \to HHZZ)
\simeq \frac{3 \left(\alpha _1-\alpha _2+\alpha _3-1\right){}^2 s^2 m_{\mu }^2}{32768 \pi ^5 v^8} &\simeq&0+\mathcal O\left(   \frac{s v^2 }{\Lambda^4}\right) \left(c_{\ell\varphi}^{(6)}\right)^2\, ,\nonumber\\
\sigma_{\rm BSM}(\mu^+\mu^- \to HHWW)
\simeq \frac{3 \left(\alpha _1-\alpha _2+\alpha _3-1\right){}^2 s^2 m_{\mu }^2}{16384 \pi ^5 v^8}  &\simeq&0+\mathcal O\left(   \frac{s v^2 }{\Lambda^4}\right) \left(c_{\ell\varphi}^{(6)}\right)^2\, ,\nonumber\\
\sigma_{\rm BSM}(\mu^+\mu^- \to HZZZ)
\simeq \frac{\left(-3 \alpha _1+2 \alpha _2+3\right){}^2 s^2 m_{\mu }^2}{65536 \pi ^5 v^8}  &\simeq&0+\mathcal O\left(   \frac{s v^2 }{\Lambda^4}\right) \left(c_{\ell\varphi}^{(6)}\right)^2\, ,\nonumber\\
\sigma_{\rm BSM}(\mu^+\mu^- \to HZWW)
\simeq \frac{\left(-3 \alpha _1+2 \alpha _2+3\right){}^2 s^2 m_{\mu }^2}{98304 \pi ^5 v^8} &\simeq&0+\mathcal O\left(   \frac{s v^2 }{\Lambda^4}\right) \left(c_{\ell\varphi}^{(6)}\right)^2\, ,~~~
\end{eqnarray}
and for $m+n=5$ to
{ \allowdisplaybreaks
\begin{eqnarray}
\label{eq:approxmultiHV5}
\sigma_{\rm BSM}(\mu^+\mu^- \to HHHHZ)
\simeq \frac{\left(-\alpha _1+\alpha _2-\alpha _3+\alpha _4+1\right)^2 s^3 m_{\mu }^2}{786432
   \pi ^7 v^{10}} &\simeq&0+\mathcal O\left(   \frac{s v^2 }{\Lambda^4}\right) \left(c_{\ell\varphi}^{(6)}\right)^2\, ,\nonumber\\
\sigma_{\rm BSM}(\mu^+\mu^- \to HHHZZ)
\simeq \frac{\left(-\alpha _1+\alpha _2-\alpha _3+\alpha _4+1\right){}^2 s^3 m_{\mu }^2}{393216
   \pi ^7 v^{10}}  &\simeq&0+\mathcal O\left(   \frac{s v^2 }{\Lambda^4}\right) \left(c_{\ell\varphi}^{(6)}\right)^2\, ,\nonumber\\
\sigma_{\rm BSM}(\mu^+\mu^- \to HHHWW)
\simeq \frac{\left(-\alpha _1+\alpha _2-\alpha _3+\alpha _4+1\right){}^2 s^3 m_{\mu }^2}{196608
   \pi ^7 v^{10}} &\simeq&0+\mathcal O\left(   \frac{s v^2 }{\Lambda^4}\right) \left(c_{\ell\varphi}^{(6)}\right)^2\, ,\nonumber\\
\sigma_{\rm BSM}(\mu^+\mu^- \to HHZZZ)
\simeq \frac{\left(-6 \alpha _1+5 \alpha _2-3 \alpha _3+6\right){}^2 s^3 m_{\mu }^2}{6291456 \pi ^7 v^{10}} &\simeq&0+\mathcal O\left(   \frac{s v^2 }{\Lambda^4}\right) \left(c_{\ell\varphi}^{(6)}\right)^2\, ,\nonumber\\
\sigma_{\rm BSM}(\mu^+\mu^- \to HHZWW)
\simeq \frac{\left(-6 \alpha _1+5 \alpha _2-3 \alpha _3+6\right){}^2 s^3 m_{\mu }^2}{9437184 \pi ^7 v^{10}} &\simeq&0+\mathcal O\left(   \frac{s v^2 }{\Lambda^4}\right) \left(c_{\ell\varphi}^{(6)}\right)^2\, ,\nonumber\\
\sigma_{\rm BSM}(\mu^+\mu^- \to HZZZZ)
\simeq \frac{\left(-6 \alpha _1+5 \alpha _2-3 \alpha _3+6\right){}^2 s^3 m_{\mu }^2}{12582912 \pi ^7 v^{10}} &\simeq&0+\mathcal O\left(   \frac{s v^2 }{\Lambda^4}\right) \left(c_{\ell\varphi}^{(6)}\right)^2\, ,\nonumber\\
\sigma_{\rm BSM}(\mu^+\mu^- \to HZZWW)
\simeq  \frac{\left(-6 \alpha _1+5 \alpha _2-3 \alpha _3+6\right){}^2 s^3 m_{\mu }^2}{9437184 \pi ^7 v^{10}}&\simeq&0+\mathcal O\left(   \frac{s v^2 }{\Lambda^4}\right) \left(c_{\ell\varphi}^{(6)}\right)^2\, ,\nonumber\\
\sigma_{\rm BSM}(\mu^+\mu^- \to HWWWW)
\simeq \frac{\left(-6 \alpha _1+5 \alpha _2-3 \alpha _3+6\right){}^2 s^3 m_{\mu }^2}{4718592 \pi ^7 v^{10}} &\simeq&0+\mathcal O\left(   \frac{s v^2 }{\Lambda^4}\right) \left(c_{\ell\varphi}^{(6)}\right)^2\, . ~~~~~
\end{eqnarray}
}
Again, similarly to the case of $4V$ and $5V$ multi-gauge boson production (Eqs.~\eqref{eq:approxmulti4V} and \eqref{eq:approxmulti5V}), we observe that for $m+n=4,5$, $m,n\neq 0$ the leading contribution in HEFT is equal to zero in the $\SMEFTs$ scenario. Therefore, in this scenario, we expect for these processes a much smaller growth with energy and in turn a smaller sensitivity on anomalous Higgs-muon interactions w.r.t.~the HEFT case. Moreover, the  case $m+n=3$ exhibits a new feature that is not present in any of the multi-Higgs and multi-gauge boson production processes discussed so far. For those processes where the final-state multiplicity is equal to three (but pure Higgs or gauge boson case, $n=3 $, $m=0 $ or $n=0 $, $m=3 $) the $\SMEFTs$ scenario has the same dependence on $s$ in the high-energy limit as in the general HEFT framework. However, unlike them, in the HEFT framework the approximation in Eq.~\eqref{eq:approxmultiHV3} is equal to zero for
$\alpha_2=\Delta \alpha_1$. In other words, there is a particular direction in the $(\alpha_1, \alpha_2)$ plane, which is {\it not} the one defined by Eq.~\eqref{eq:dim6lock} in the $\SMEFTs$ scenario,\footnote{The condition $\alpha_2=\Delta \alpha_1$ in the $\SMEFTs$ scenario implies $c_{\ell\varphi}^{(6)}=0$, {\it i.e.}, the SM.} where the energy growth of the $WWH$, $ZZH$, $ZHH$ production processes is smaller than in a generic  $(\alpha_1, \alpha_2)$ configuration.  We therefore expect a smaller sensitivity on both $\alpha_1$ and $\alpha_2$ if  $\alpha_2=\Delta \alpha_1$. 

In the following we discuss the results we have obtained via the Monte Carlo simulations based on  exact matrix elements, therefore going beyond  the high-energy approximation from Eq.~\eqref{eq:generalmultiHV}. 
\begin{table}[!t]
\footnotesize
    \centering
\adjustbox{width=1.1\textwidth,center}{
    \begin{tabular}{c|cccc|cccc}
    \hline
$\sqrt{s}$  &  \multicolumn{4}{c|}{ 3 TeV } & \multicolumn{4}{c}{10 TeV} \\
\hline
$(\alpha_1,\alpha_2)$ & $(0,0)$ & (1,1) & (1,0) & VBF & $(0,0)$ & (1,1) & (1,0) & VBF   \\
\hline
$\sigma$ [fb] & \multicolumn{8}{c}{$WWH$} \\
\hline
No cut & 1.1 & 1.1 & 1.1 & 1.3 & 0.45 & 0.45 & 0.21 & 9.9\\
$M_F>0.8\sqrt{s}$   & 1.1 & 1.1 &  1.1 & $1.6\cdot 10^{-2}$ &  0.45 & 0.45 &   0.21  & $4.3\cdot10^{-2}$\\
$10^\circ < \theta_B < 170 ^\circ$ & 0.66 & 0.66 &  0.64 & $1.2\cdot 10^{-3}$ &  0.33 & 0.33 &   $9.6\cdot10^{-2}$ & $6.8\cdot10^{-4}$ \\
 $\Delta R _{BB} > 0.4$ & 0.60 & 0.60 & 0.58 & $1.2\cdot 10^{-3}$ &  0.30 & 0.30 &   $7.6\cdot10^{-2}$ & $5.9\cdot10^{-4}$\\
 event \# & 600 & 600 & 580 & 1.2 & 3000 & 3000 & 760 & 5.9\\
 $S/B$ & 0.034 & 0.034 &\multicolumn{2}{c|}{--} & 2.9 & 2.9 & \multicolumn{2}{c}{--}\\
 $\calS$ & 0.82 & 0.82 &\multicolumn{2}{c|}{--} & 61 & 61 & \multicolumn{2}{c}{--}  \\
\hline
$\sigma$ [fb] & \multicolumn{8}{c}{$ZZH$} \\
\hline
No cut & $9.1\cdot10^{-2}$ & $9.2\cdot10^{-2}$ & $8.2\cdot10^{-2}$ & 0.12 & 0.13 & 0.14 & $1.5\cdot10^{-2}$ & 1.0 \\
$M_F>0.8\sqrt{s}$  & $9.1\cdot10^{-2}$ & $9.2\cdot10^{-2}$ &  $8.2\cdot10^{-2}$ & $5.1\cdot10^{-4}$ & 0.13  & 0.14 & $1.5\cdot10^{-2}$ & $2.0\cdot10^{-3}$\\
$10^\circ < \theta_B < 170 ^\circ$ & $5.7\cdot10^{-2}$ & $5.8\cdot10^{-2}$ &  $4.8\cdot10^{-2}$ & $1.7\cdot10^{-4}$ & 0.12 & 0.12 & $6.4\cdot10^{-3}$ & $6.1\cdot10^{-4}$\\
$\Delta R _{BB} > 0.4$ & $5.3\cdot10^{-2}$ & $5.4\cdot10^{-2}$ & $4.4\cdot10^{-2}$ & $1.6\cdot10^{-4}$ & 0.12 & 0.12 & $5.4\cdot10^{-3}$ & $5.8\cdot10^{-4}$\\
event \# & 53 & 54 & 44 & 0.16 & 1200 &  1200 & 54 & 5.8 \\
 $S/B$ & 0.20 & 0.22 & \multicolumn{2}{c|}{--} & 19 & 19 & \multicolumn{2}{c}{--}\\
 $\calS$ & 1.3 & 1.4 & \multicolumn{2}{c|}{--} & 70 & 70 & \multicolumn{2}{c}{--}\\
\hline
$\sigma$ [fb] & \multicolumn{8}{c}{$ZHH$} \\
\hline
No cut & $4.2\cdot10^{-2}$ & $4.3\cdot10^{-2}$ & $3.3\cdot10^{-2}$ & $1.0 \cdot 10^{-2}$ &  0.13 & 0.13 & $6.1\cdot10^{-3}$ & $9.9\cdot10^{-2}$\\
$M_F>0.8\sqrt{s}$ & $4.2\cdot10^{-2}$ & $4.3\cdot10^{-2}$ &  $3.3\cdot10^{-2}$ & $5.6 \cdot 10^{-5}$ & 0.13 & 0.13 &   $6.1\cdot10^{-3}$ & $3.8\cdot10^{-4}$\\
$10^\circ < \theta_B < 170 ^\circ$  & $4.1\cdot10^{-2}$ & $4.2\cdot10^{-2}$ &  $3.2\cdot10^{-2}$ & $1.6 \cdot 10^{-5}$ &  0.12 & 0.12 & $6.0\cdot10^{-3}$ & $2.3\cdot10^{-4}$\\
$\Delta R _{BB} > 0.4$ & $3.5\cdot10^{-2}$ & $3.6\cdot10^{-2}$ & $2.6\cdot10^{-2}$ & $1.5 \cdot 10^{-5}$ & 0.12 & 0.12 & $4.0\cdot10^{-3}$ & $2.2\cdot10^{-4}$\\
event \# & 35 & 36 & 26 & -- & 1200 & 1200 & 40 & 2.2\\
 $S/B$ & 0.35 & 0.38 & \multicolumn{2}{c|}{--} & 27 & 27 & \multicolumn{2}{c}{--}\\
 $\calS$ & 1.6 & 1.8 & \multicolumn{2}{c|}{--} & 76 & 76 & \multicolumn{2}{c}{--}\\
\hline
    \end{tabular}
        }
    \caption{The cross sections and expected event numbers for $WWH, ZZH$ and $ZHH$ production processes at a 3 and 10 TeV muon collider, respectively. The background includes both the SM annihilation, $(\alpha_1,\alpha_2)=(1,0)$, and VBF.  The parameters $\beta_{3,4}$ are set to their SM values. The cuts are applied consecutively.}
    \label{tab:mmvvhSM}
\end{table}
First of all, we document in Tab.~\ref{tab:mmvvhSM} the impact of the phase-space cuts on the signal and the background (SM contribution and VBF) for the case $m+n=3$ ($WWH$, $ZZH$ and $ZHH$ production). Besides the SM scenario $(\alpha_1,\alpha_2)=(1,0)$, which constitutes part of the background,  two additional BSM benchmark points $(\alpha_1,\alpha_2)=(0,0)~{\rm and}~(1,1)$ are displayed. Starting from both these benchmark points, we varied either $\alpha_1$ or $\alpha_2$ by one unit up or down, confirming that, consistently with Eq.~\eqref{eq:approxmultiHV3},  the two  benchmark points return the same value for the cross section. We also see that the cuts efficiently suppress the VBF contribution, while they have similar effects for the SM, $(\alpha_1,\alpha_2)=(1,0)$, and the BSM benchmark points considered. However, while at 3 TeV signal and background are of the same order, at 10 TeV $ S \gg B$.

\begin{figure}[!t]
  \centering
  \includegraphics[width=0.48\textwidth]{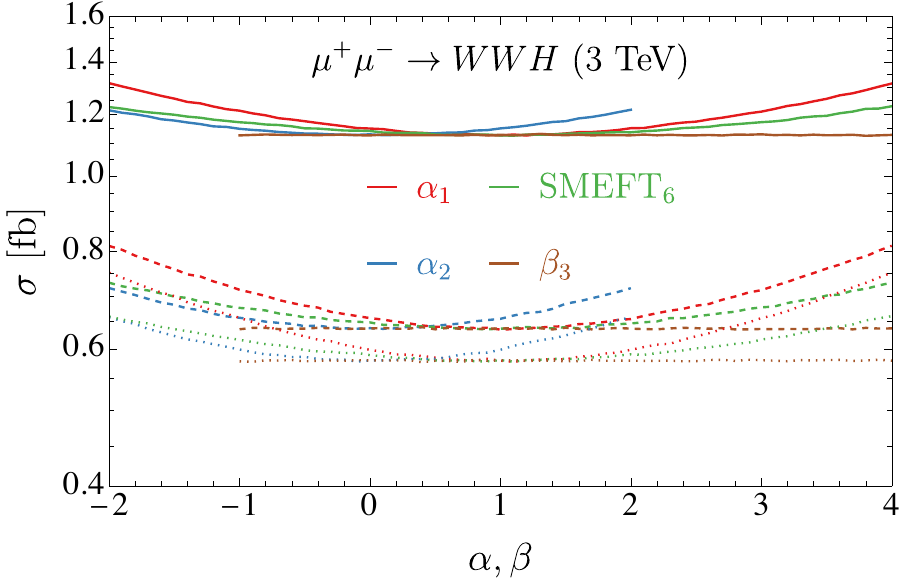}
    \includegraphics[width=0.48\textwidth]{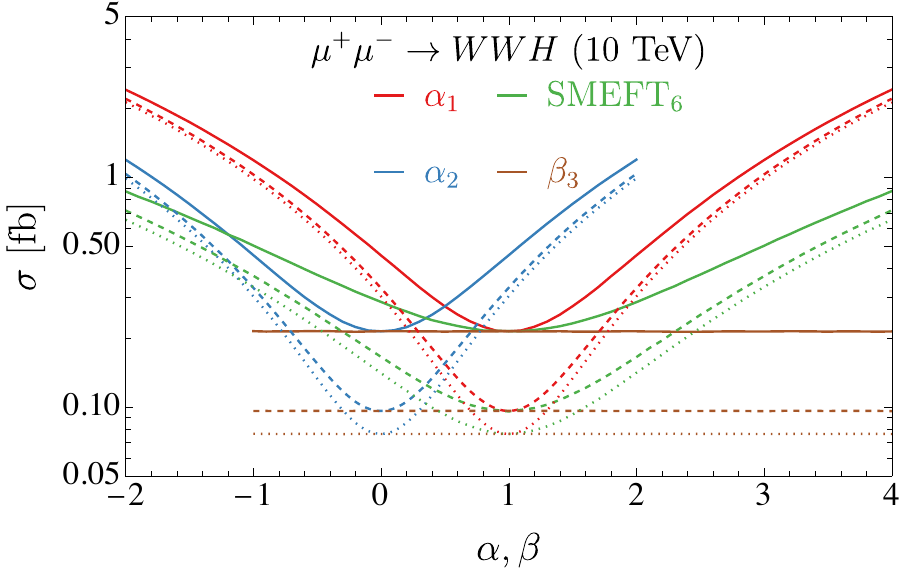}\\
  \includegraphics[width=0.48\textwidth]{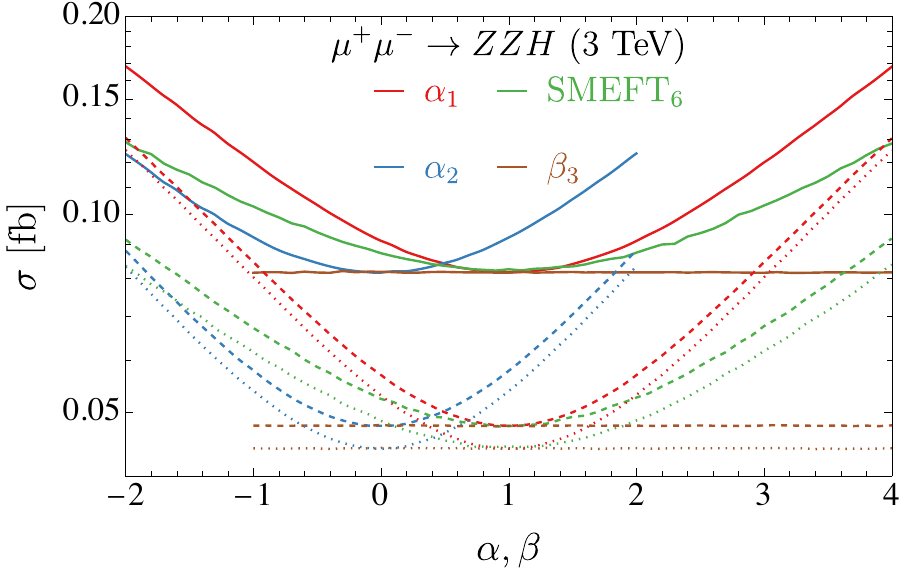}
    \includegraphics[width=0.48\textwidth]{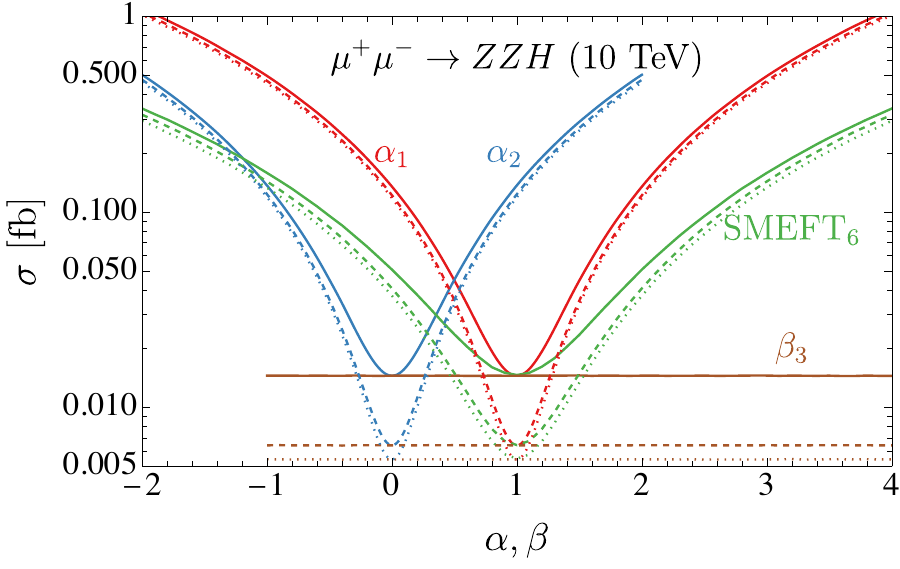}\\
  \includegraphics[width=0.47\textwidth]{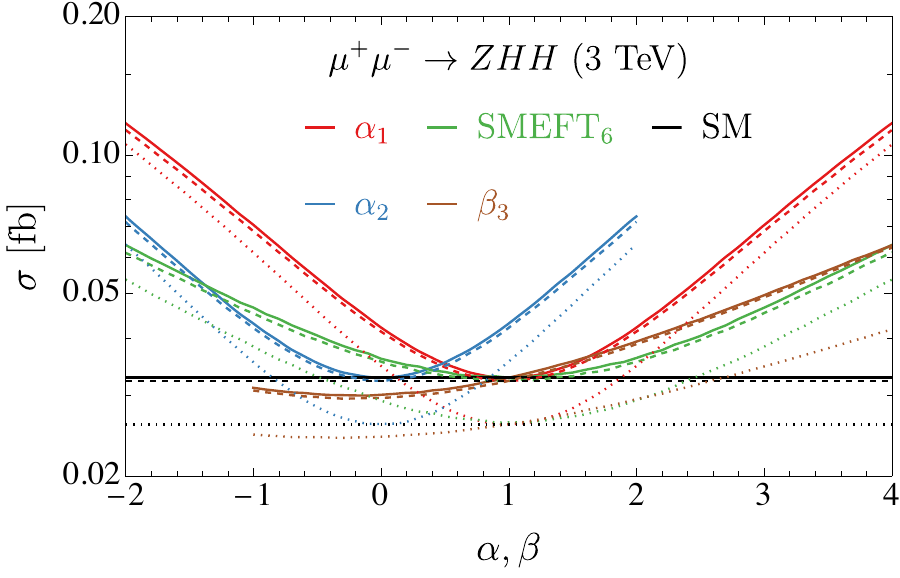}
 \hspace{1mm}    \includegraphics[width=0.5\textwidth]{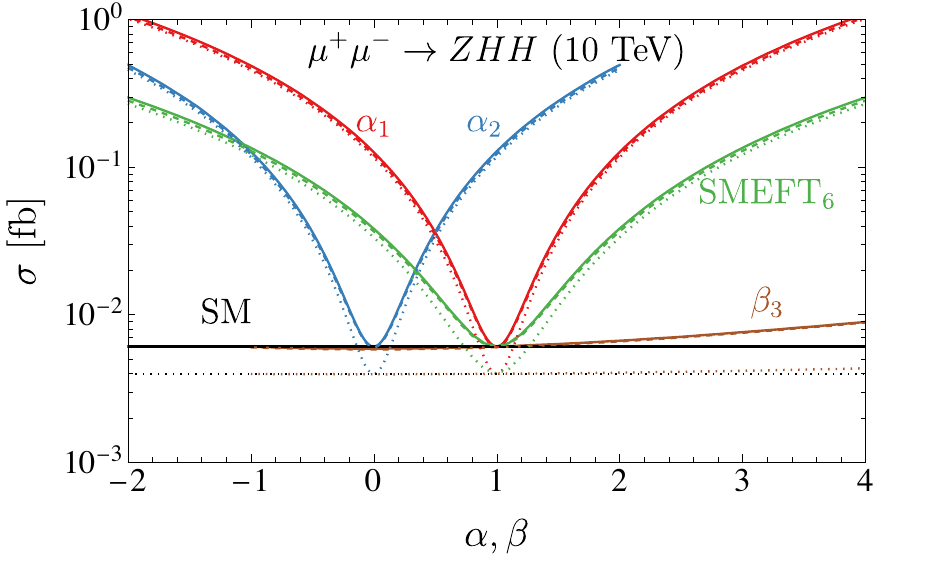}
  \caption{Same as Fig.~\ref{fig:mmnhkappa3} for $WWH$, $ZZH$, $ZHH$ production at 3 and 10 TeV, respectively.}
  \label{fig:mmvvhkappa}
\end{figure}

\begin{figure}[!t]
  \centering
  \includegraphics[width=0.49\textwidth]{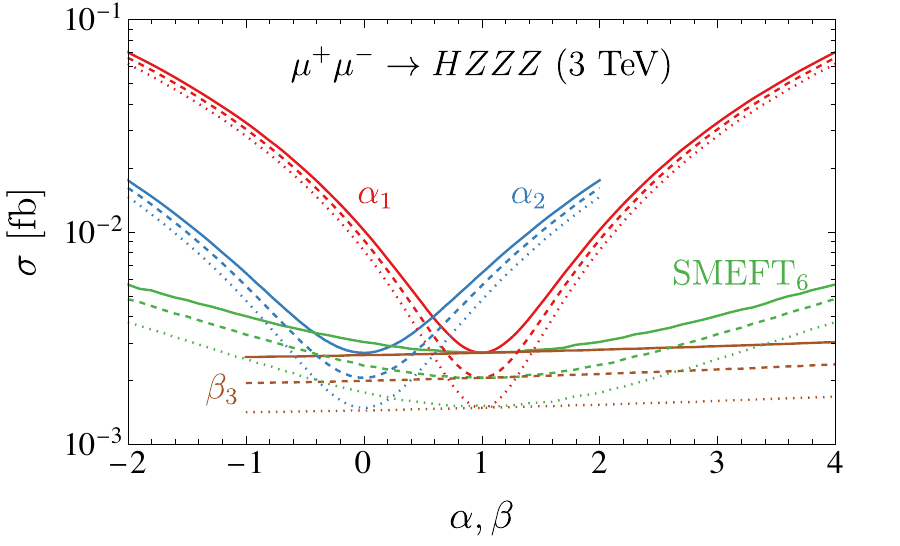}
  \includegraphics[width=0.49\textwidth]{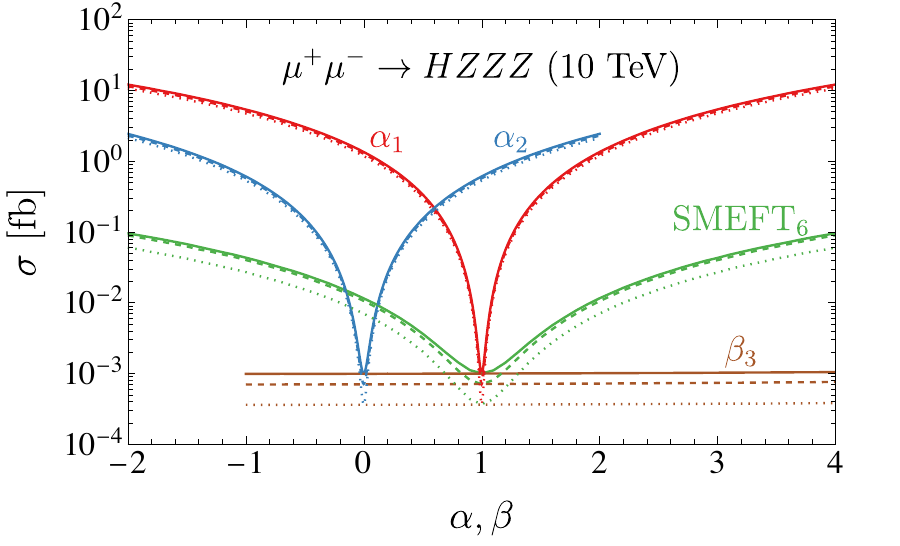}
  \caption{Same as Fig.~\ref{fig:mmnhkappa3} for $HZZZ$ production at 3 and 10 TeV, respectively. }
  \label{fig:mmzzzhkappa}
\end{figure}

We show in Fig.~\ref{fig:mmvvhkappa} the dependence of the cross sections for the $WWH$, $ZZH$ and $ZHH$ production processes on independent variations of  $\alpha_1$, $\alpha_2$ and $\beta_3$. We also show the case corresponding to the $\SMEFTs$ scenario by varying $\alpha_1$, where  $\alpha_1$ and  $\alpha_2$ are related via Eq.~\eqref{eq:dim6lock}. First, we notice that the $WWH$ and $ZZH$ production are almost independent on the value of $\beta_3$, while the $ZHH$ production shows a non-negligible dependence on $\beta_3$, especially at 3 TeV. $ZHH$ at 3 TeV is the only case for all our analyses, together with $HHZZ$, in which a possible anomalous Higgs self-coupling may have a relevant effect.  For this reason, we assume here $\beta_{3}$ to be consistent with the  SM value $\beta_3=1$. Second, we notice that as in the case of $4V$ and $5V$ production (see {\it e.g.},~Figs.~\ref{fig:mm4Zkappa} and \ref{fig:mm5Zkappa}), in the $\SMEFTs$ scenario the dependence on $\alpha_1$ is smaller than in the HEFT, but the difference is not so pronounced as in the aforementioned case. Indeed, the smaller dependence in the $\SMEFTs$ scenario is simply induced by a partial cancellation -- which instead is a total cancellation in the $4V$ and $5V$ case -- of the leading contribution in the high-energy limit (see Eq.~\eqref{eq:approxmultiHV3}), when the Eq.~\eqref{eq:dim6lock} is implemented.

The $WWH$, $ZZH$ and $ZHH$ production processes can be used to put constraints within the general $(\alpha_1,\,\alpha_2)$ plane, without any assumption on $\alpha_i$, $i\ge3$. Before discussing the constraints that can be obtained, we  notice that also the class of processes with $m+n=4$ and  $(m,n)=(3,1)$ has the same property, as can be seen from Eq.~\eqref{eq:approxmultiHV4}. We show in Fig.~\ref{fig:mmzzzhkappa} the effects of varying $\alpha_1$ and $\alpha_2$ both independently and for the case where they are related via Eq.~\eqref{eq:dim6lock}, for $ZZZH$ production. The same for $WWZH$ production is shown in Fig.~\ref{fig:mmvvvhkappa}. The pattern in this case is really the identical to $4Z$ or $5Z$ production, where the dependence on $\alpha_1$ or $\alpha_2$ exactly cancels under the assumption of the high-energy limit and  Eq.~\eqref{eq:dim6lock} as in the $\SMEFTs$ scenario.

\begin{figure}[!t]
  \centering
  \includegraphics[width=0.48\textwidth]{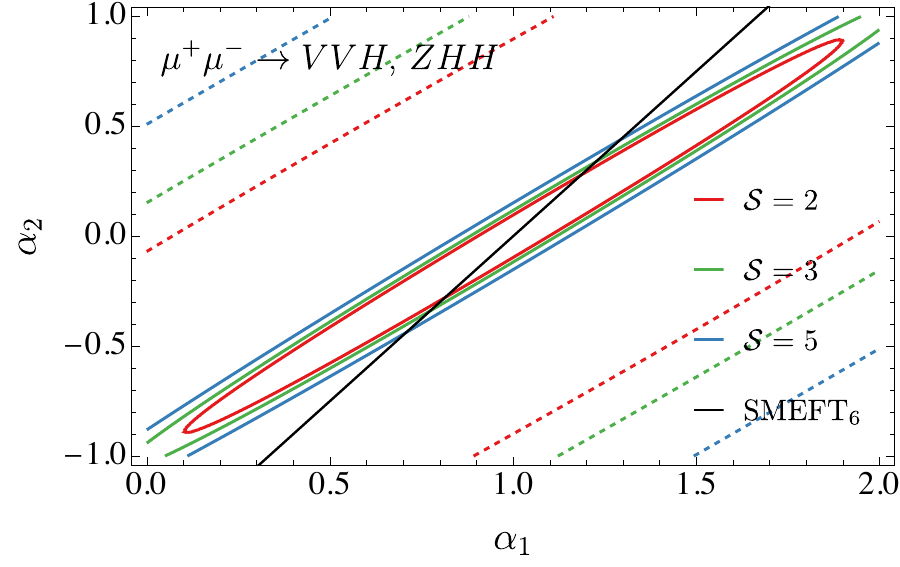}
  \includegraphics[width=0.48\textwidth]{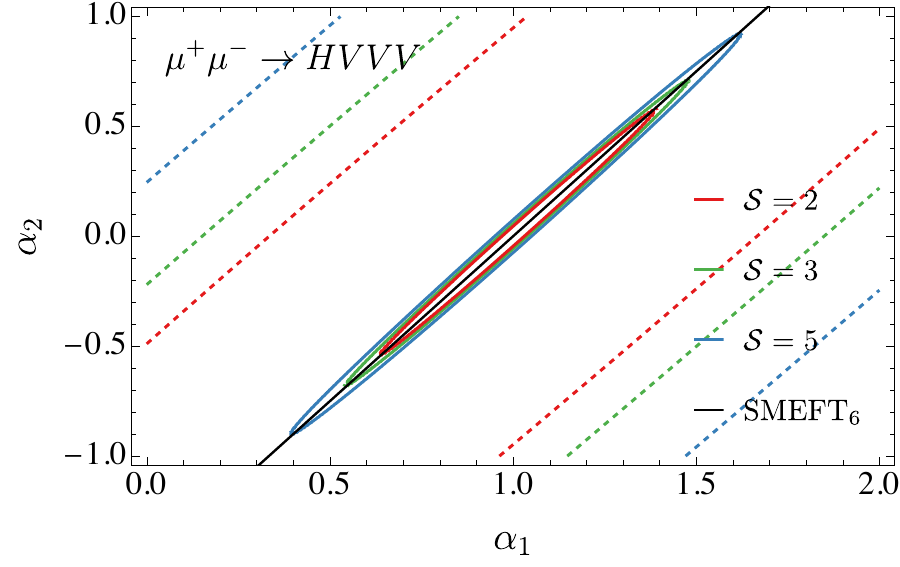}
  \caption{Same as Fig.~\ref{fig:contour45V} for the Higgs-associated gauge boson production processes for three-boson final states (left) and $\mu^+\mu^- \to 3V H$ (right) at a 3 TeV muon collider (dashed curves) and a 10 TeV muon collider (solid curves), respectively.}
  \label{fig:contour34VH}
\end{figure}

\begin{figure}[!t]
  \centering
  \includegraphics[width=0.49\textwidth]{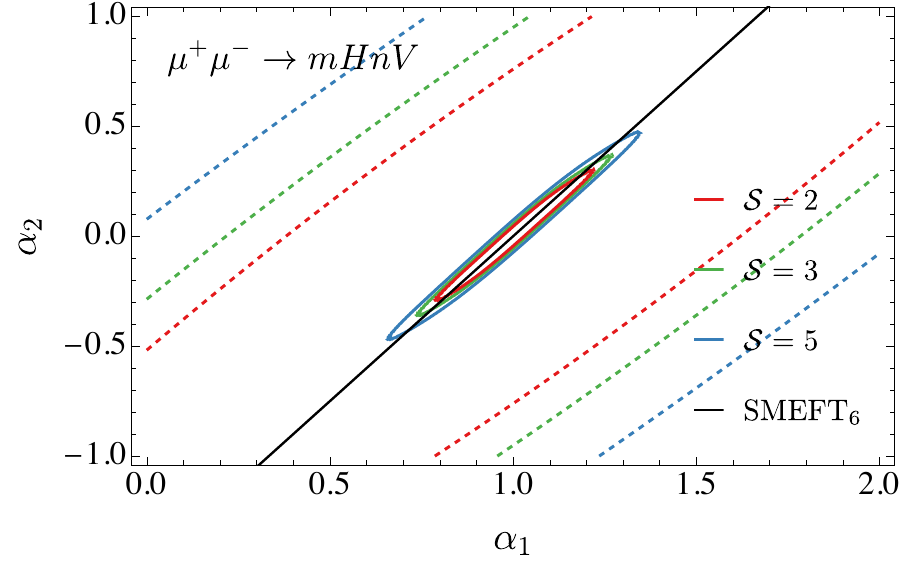}
  \caption{Same as Fig.~\ref{fig:contour45V} for the Higgs-associated gauge boson production processes that are independent of $\alpha_{i\geq 3}$ at a 3 TeV muon collider (dashed curves) and a 10 TeV muon collider (solid curves), respectively.}
  \label{fig:contourVH}
\end{figure}

We show the constraints in the $(\alpha_1,\,\alpha_2)$ plane for the case $m+n=3$ in the left plot of Fig.~\ref{fig:contour34VH} and for the case $m=3$ and $n=1$ ($3VH$) in the right one, respectively. The differences are manifest. Not only leads the $3VH$ production to stronger constraints, but the ellipses associated to different values of significance $\calS$ are here aligned around the black line associated to the $\SMEFTs$ scenario (Eq.~\eqref{eq:dim6lock}), which is not the case for $m+n=3$. As expected, in the case $m+n=3$ the ellipses are aligned around the relation $\Delta\alpha_1=\alpha_2$, which leads to the exact cancellation of the $\Delta \alpha_1$ and $ \alpha_2$ dependence of the leading term in the high-energy approximation, Eq.~\eqref{eq:approxmultiHV3}. We notice that in both plots of Fig.~\ref{fig:contour34VH}, 3 TeV collisions (dashed lines) are leading only to very loose constraints, unlike for 10 TeV collisions.

Since the $\calS$ ellipses of the left and right plots of Fig.~\ref{fig:contour34VH} are not aligned with each other, combining the information from the $WWH$, $ZZH$ and $ZHH$ processes, $m+n=3$, and all the $3VH$, improves the constraints in the $(\alpha_1,\alpha_2)$ plane. We show the corresponding constraints from the combination in  Fig.~\ref{fig:contourVH}. In particular, this plot tells us that in the $\SMEFTs$ scenario  
\begin{eqnarray}
|\Delta\alpha_1|\lesssim 0.2 \;\Longleftrightarrow\; \left|c^{(6)}_{\ell\varphi}/\Lambda^2\right| \leq 2 \times 10^{-9} \,{\rm GeV}^{-2} \qquad{\rm at~10~TeV}\,, \label{eq:bound1stcombination}
\end{eqnarray}
while at 3 TeV bounds are very loose.

\begin{figure}[!t]
  \centering
  \includegraphics[width=0.49\textwidth]{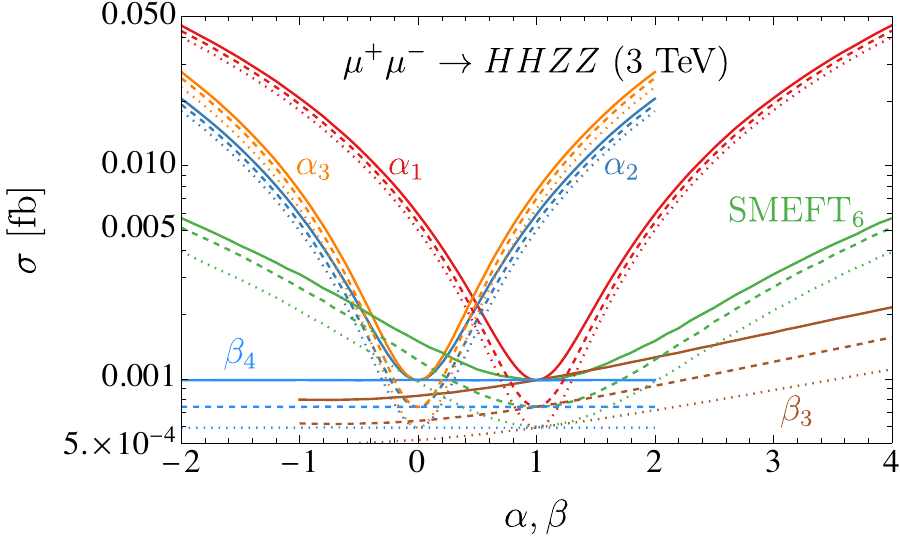}
  \includegraphics[width=0.49\textwidth]{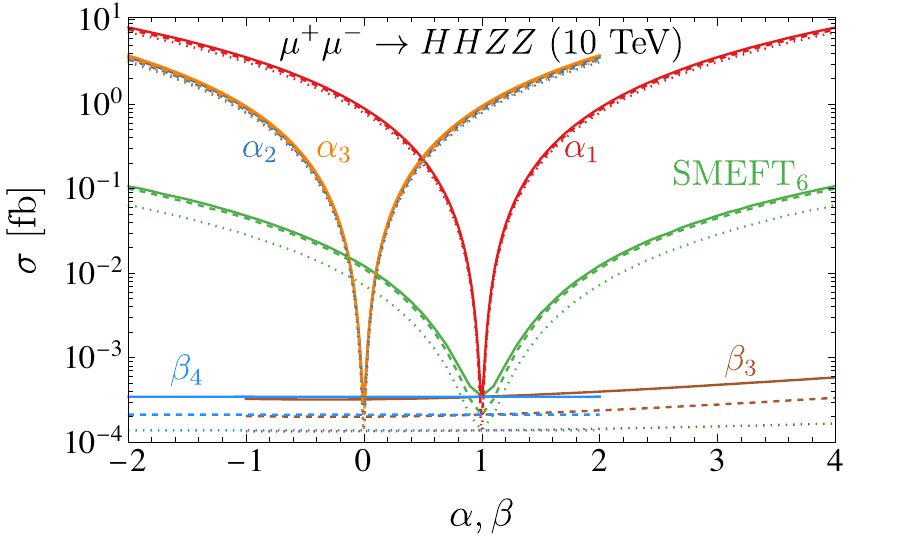}
  \caption{Same as Fig.~\ref{fig:mmnhkappa3} for $HHZZ$ production at 3 and 10 TeV, respectively.}
  \label{fig:mmzzhhkappa}
\end{figure}

\begin{figure}[!t]
  \centering
 \includegraphics[width=0.48\textwidth]{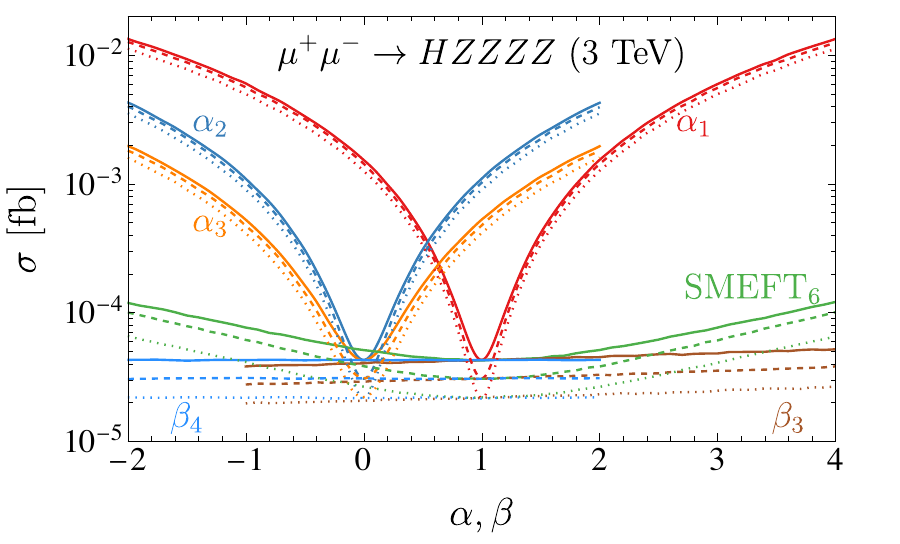}
  \includegraphics[width=0.48\textwidth]{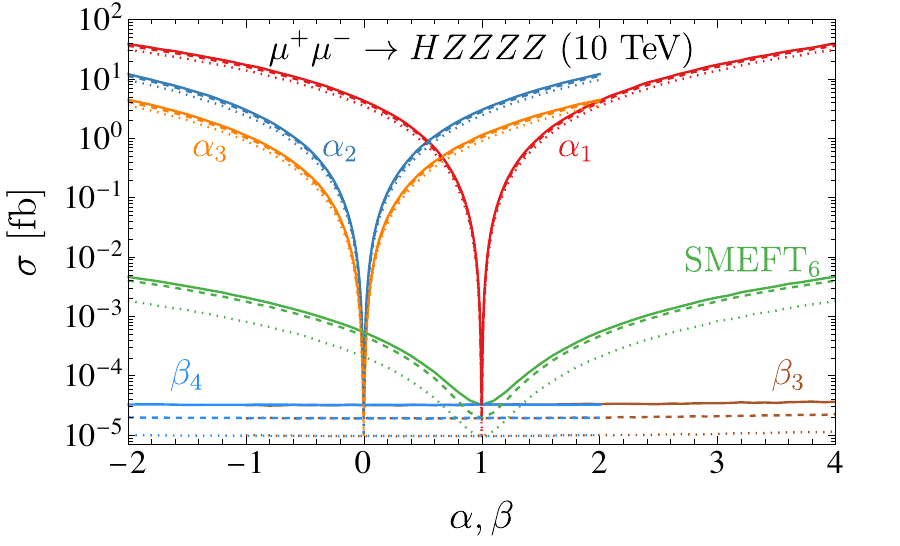}
  \includegraphics[width=0.48\textwidth]{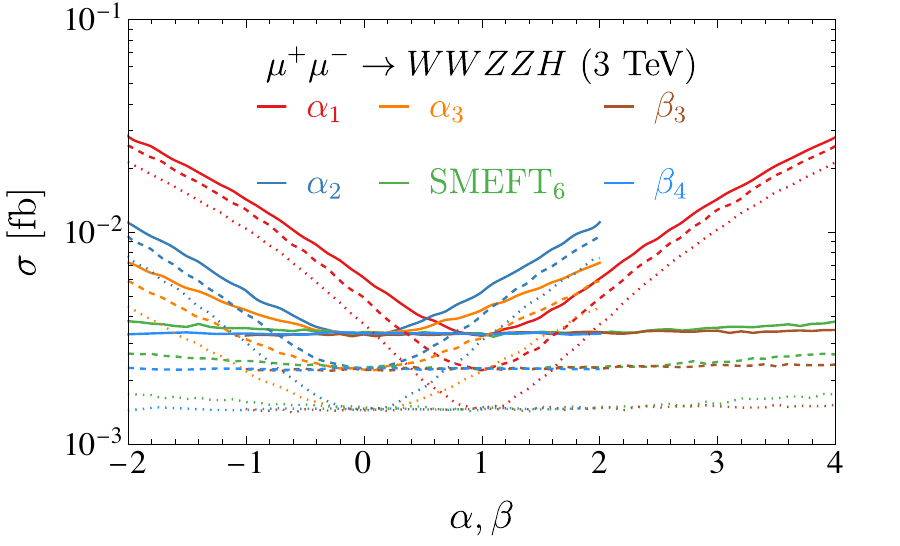}
    \includegraphics[width=0.48\textwidth]{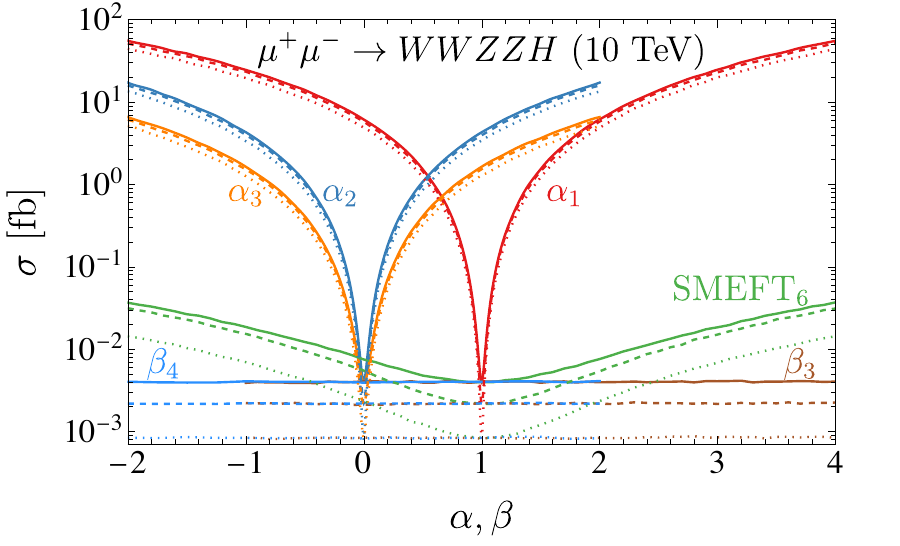}
  \includegraphics[width=0.48\textwidth]{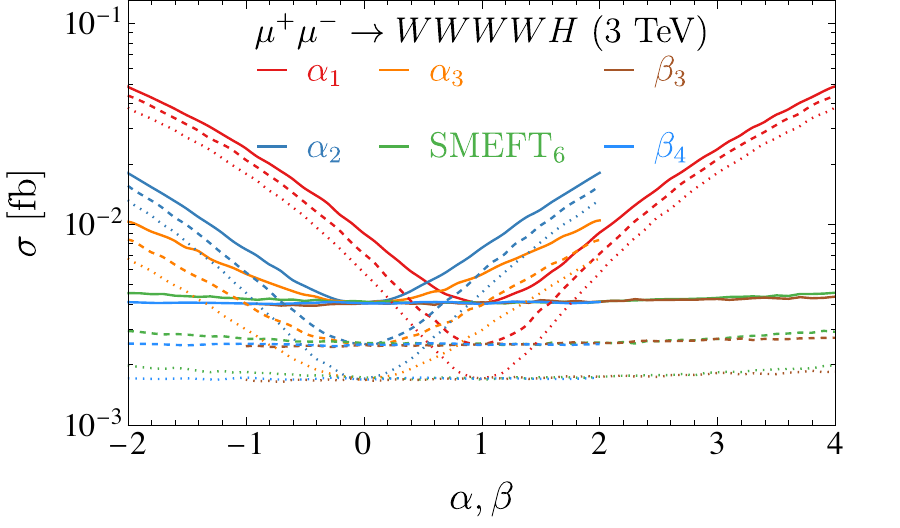}
    \includegraphics[width=0.48\textwidth]{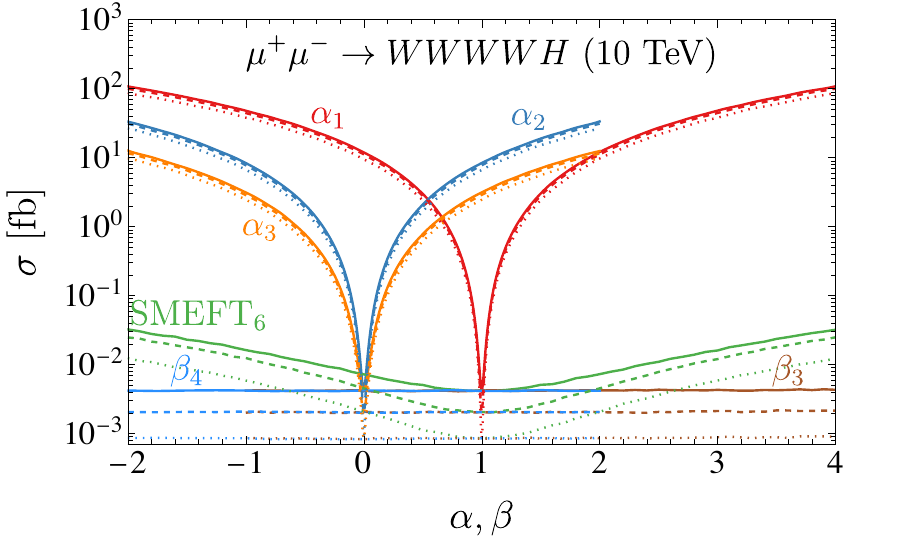}
  \caption{Same as Fig.~\ref{fig:mmnhkappa3} for $HVVVV$ production at 3 and 10 TeV, respectively.}
  \label{fig:mmvvvvhkappa}
\end{figure}

\begin{figure}[!t]
  \centering
  \includegraphics[width=0.49\textwidth]{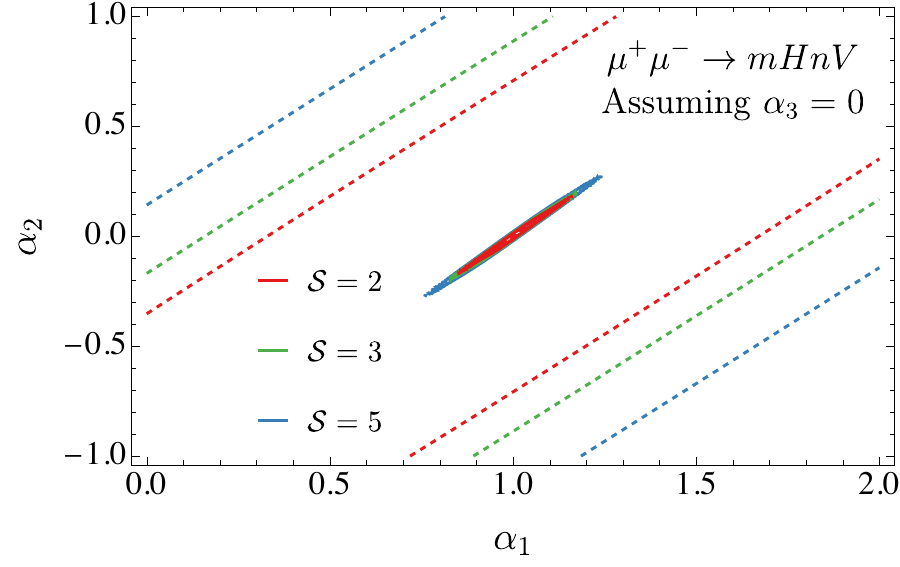}
\includegraphics[width=0.49\textwidth]{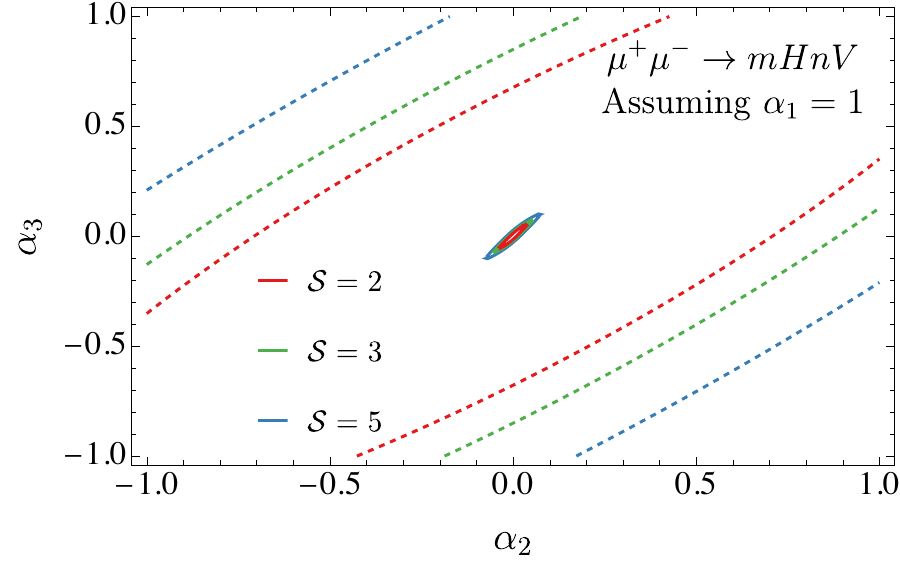}
  \caption{Same as Fig.~\ref{fig:contour45V} in the $(\alpha_1,\,\alpha_2)$ plane (left plot),  and in the $(\alpha_2,\,\alpha_3)$ plane in the right plot, respectively. All the Higgs-associated gauge boson production processes that are dependent on $\alpha_3$ at a 3 TeV muon collider (dashed curves) and a 10 TeV muon collider (solid curves) have been combined. The following assumptions are adopted: $\alpha_3=0$ (left), $\alpha_1=1$ (right). For this reason a line corresponding to  $\SMEFTs$ scenario has not been displayed. }
  \label{fig:contourVHassumption}
\end{figure}

 In the remaining $m+n=4$ and $m+n=5$ processes, $V^2H^2$, $ZH^3$, $V^4H$, $V^3H^2$, $V^2H^3$ and $ZH^4$ production, also a dependence on $\alpha_3$ appears, and for the last two classes also a dependence on $\alpha_4$ is present, as can also be seen in Eqs.~\eqref{eq:approxmultiHV4} and \eqref{eq:approxmultiHV5}.  Representative results are shown in Figs.~\ref{fig:mmzzhhkappa} and \ref{fig:mmvvvvhkappa}, while many more are collected in Figs.~\ref{fig:mmvvhhkappa}--\ref{fig:mmvvhhhkappa} in Appendix~\ref{sec:furtherplots}.
 Considering only the $V^2H^2$, $ZH^3$, $V^4H$, $V^3H^2$ production processes, {\em i.e.} excluding for the moment $V^2H^3$ and $ZH^4$,  we can set constraints either assuming $\alpha_3=0$ in the  $(\alpha_1,\,\alpha_2)$ plane (left plot of Fig.~\ref{fig:contourVHassumption}) or assuming  $\alpha_1=1$ in the  $(\alpha_2,\,\alpha_3)$ plane (right plot of Fig.~\ref{fig:contourVHassumption}). We leave the discussion of the rationale behind these two choices to Sec.~\ref{sec:combination}, where we will also employ them in the combination of different classes of processes.
 
As can be seen in Fig.~\ref{fig:contourVHassumption}, at the 10 TeV muon collider with $10~{\rm ab}^{-1}$ luminosity, all the aforementioned processes can set strong constraints, either by assuming $\alpha_3=0$ in the  $(\alpha_1,\,\alpha_2)$ plane, and especially in the  $(\alpha_2,\,\alpha_3)$ plane  by assuming $\alpha_1=1$. The strongest constraints arise from $WWWWH$ and $WWZZH$ production. On the other hand, due to the smallness of the event numbers, only the $WWHH$ and $WWWWH$ channels can provide constraints at a 3 TeV muon collider. On purpose, we did not show in both plots the $\SMEFTs$ lines,  defined by Eq.~$\eqref{eq:dim6lock}$ in the $(\alpha_1,\,\alpha_2)$ and $(\alpha_2,\,\alpha_3)$ subspaces, since in this scenario $\alpha_3=0\Longleftrightarrow (\alpha_1,\,\alpha_2)=(1,0)$ and $\alpha_1=1\Longleftrightarrow (\alpha_2,\,\alpha_3)=(0,0)$. In fact, neither are the ellipses aligned around the $\SMEFTs$ relation $\alpha_1=1+\frac{2}{3}\alpha_2$ in the left plot, nor around the  $\SMEFTs$ relation $\alpha_2=3\alpha_3$ in the right one. Rather, the direction is dictated by relations leading to the vanishing of the corresponding coefficients in Eq.~\eqref{eq:approxmultiHV4} and \eqref{eq:approxmultiHV5} for this class of processes. In the left plot, the direction is in between of $\alpha_1=1+\alpha_2$ and $\alpha_1=1+5\alpha_2/6$, which are the aforementioned relations for  Eqs.~\eqref{eq:approxmultiHV4} and \eqref{eq:approxmultiHV5}, respectively. Analogously, in the right plot the direction is between $\alpha_2=3\alpha_3/5$ and $\alpha_2=\alpha_3$.

In HEFT, we obtain the following reach
\begin{eqnarray}
{\rm for}~ \alpha_3=0\,:\qquad |\Delta\alpha_{1}|\, ,|\alpha_{2}|\lesssim 0.2  \qquad{\rm at~10~TeV}\,, \label{eq:boundHEFT12}
\end{eqnarray}
and
\begin{eqnarray}
{\rm for}~ \alpha_1=1\,:\qquad  |\alpha_2| \,, | \alpha_3 |\lesssim 0.05  \qquad{\rm at~10~TeV}\,. \label{eq:boundHEFT23}
\end{eqnarray}
These bounds cannot be translated into the $\SMEFTs$ scenario.
 We will explain in Sec.~\ref{sec:combination} why bounds for these processes are so strong by assuming $\alpha_1=1$, but at this point of the discussion it should not be a surprise anymore that fixing one of the $\alpha_i$ and letting the others float freely can in general lead to a large growth at high energies and therefore strong constraints. Also, bounds from this class of processes in the $(\alpha_1,\,\alpha_2)$ plane are stronger than in the case of purely multi-gauge-boson production, {\it cf.}~Figs.~\ref{fig:contour45V} and \ref{fig:contourVV}. On the other hand, we remind the reader that in this case the assumption $\alpha_3=0$ has been made, while in the case of multi-gauge boson production this was not necessary.  

Considering $V^2H^3$ and $ZH^4$ production, not only the $\alpha_4$ dependence emerges (see also Fig.~\ref{fig:mmvvhhhkappa} in Appendix~\ref{sec:furtherplots}) complicating an analysis, but also due to the small number of events and the weak signal strength, these processes cannot improve the constraints.

\subsection{Constraints from combinations of processes}
\label{sec:combination}

We discuss here the constraints that we can obtain by combining the information from different processes, using different underlying assumptions.

\subsubsection{Constraints on $\boldsymbol{ (\alpha_1,\,\alpha_2)}$}

We can obtain constraints in the $(\alpha_1,\,\alpha_2)$ plane, combining multi-gauge boson production and Higgs-associated gauge boson production with $\alpha_2$ constraints from $2H$ production. We remind the reader that constraints from $3Z$ production depend only on $\alpha_1$, while some of the Higgs-associated gauge boson production channels depend also on $\alpha_3$. Thus, we can follow two different approaches:
\begin{enumerate}
\item \label{assump:1} We do not make any assumption on $\alpha_3$, but we exclude all the processes that depend on it, namely: $V^2H^2$, $ZH^3$, $V^4H$, $V^3H^2$, $V^2H^3$ and $ZH^4$. 
\item \label{assump:2} We take into account all the processes and we assume that, thanks to the $3H$ measurement, $\alpha_3$ will be measured at high precision and found compatible with the SM scenario: $\alpha_3=0$.
\end{enumerate}
The second scenario is clearly reasonable for 10 TeV, and we will see that it will be instrumental also for our argumentation at 3 TeV.

\begin{figure}[htb]
  \centering
  \includegraphics[width=0.49\textwidth]{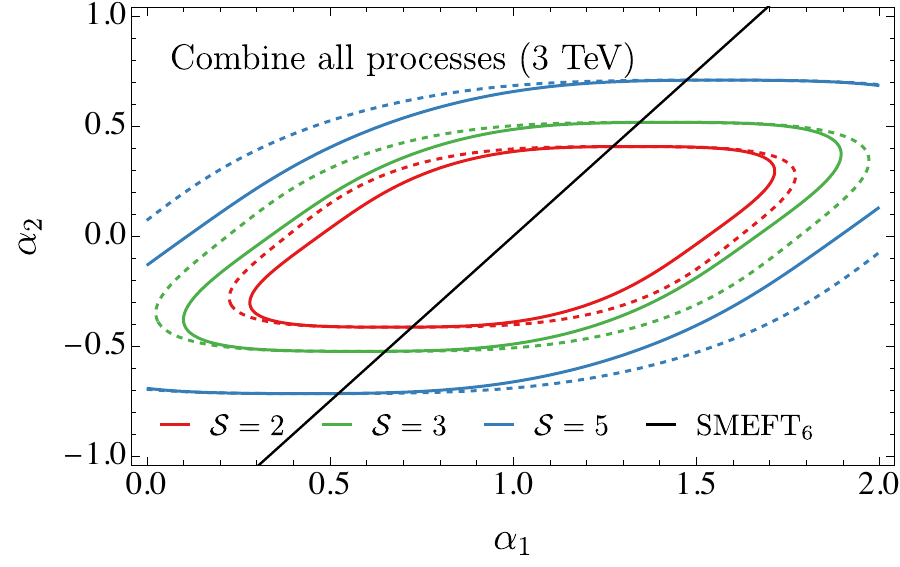}
  \includegraphics[width=0.49\textwidth]{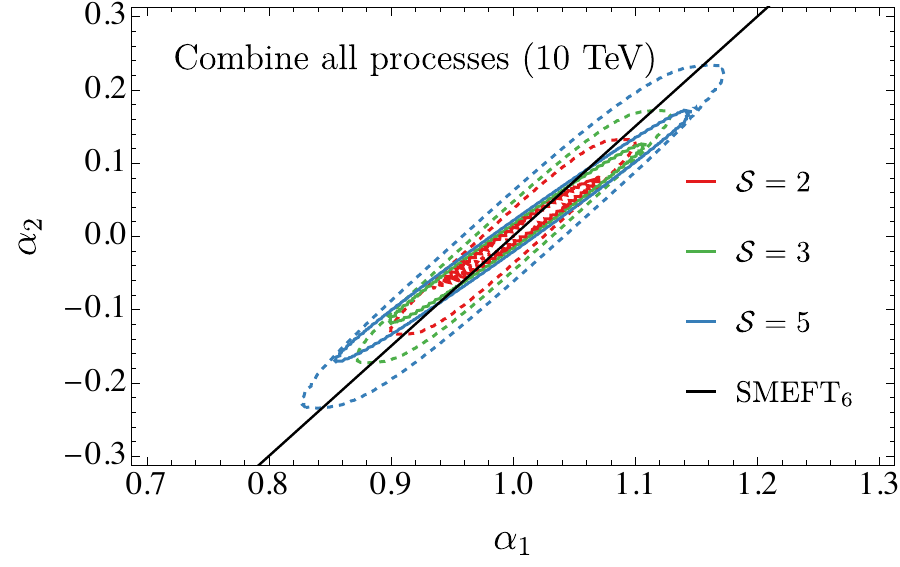}
  \caption{ Combined constraints on $(\alpha_1,\,\alpha_2)$ from di-Higgs production, multi-gauge boson production, and Higgs-associated gauge boson production processes at a 3 TeV muon collider (left) and a 10 TeV muon collider (right), respectively. The dashed curves are for the constraints with no assumptions and the solid curves includes also the processes with assumption $\alpha_3=0$. The red, green, and blue curves represent ${\mathcal S}=2,\, 3,\, 5$ significances,  respectively. The black solid line corresponds to the $\SMEFTs$ scenario, {\it i.e.}, Eq.~\eqref{eq:dim6lock}.}
  \label{fig:contoura1a2all}
\end{figure}

In Fig.~\ref{fig:contoura1a2all} we show results for Approach \ref{assump:1} as dashed lines and for Approach \ref{assump:2} as  solid lines. The plot on the left refers to the 3 TeV collider energy, while the plot on the right to the 10 TeV one.
As can be seen, we obtain the following bounds at $95\%$ for 3 TeV collisions
\begin{align}
   & |\Delta \alpha_1| \lesssim 0.75\,,&  &|\alpha_2| \lesssim 0.4&  &{\rm with~Approach~\ref{assump:1}}\,, \\
    &|\Delta \alpha_1| \lesssim 0.7\,, & & |\alpha_2| \lesssim 0.4 &  &{\rm with ~Approach~\ref{assump:2}}\,,
\end{align}
and for 10 TeV collisions
\begin{align}
   & |\Delta \alpha_1| \lesssim 0.1\,,&  &|\alpha_2| \lesssim 0.15&  &{\rm with~Approach~\ref{assump:1}}\,, \\
    &|\Delta \alpha_1| \lesssim 0.1\,, & & |\alpha_2| \lesssim 0.1 &  &{\rm with ~Approach~\ref{assump:2}}\,.
\end{align}

The bounds reported in the previous equations for $\Delta \alpha_1$ ($\alpha_2$) are valid in HEFT, regardless of the specific value assumed for the corresponding other parameter $\alpha_2$ ($\Delta \alpha_1$). Clearly, at 3 TeV and even more at 10 TeV, if $\alpha_2$ ($\Delta \alpha_1$) is fixed within the allowed range from the previous equations, the constraints on $\Delta \alpha_1$ ($\alpha_2$) can be much stronger. We notice also that while for 3 TeV the Approach \ref{assump:1} and Approach \ref{assump:2} lead to qualitatively similar constraints, at 10 TeV they are quite different for $\Delta \alpha_1$ ($\alpha_2$) if $\alpha_2$ ($\Delta \alpha_1$) is fixed. This points to the fact that all of the $V^2H^2$, $ZH^3$, $V^4H$, $V^3H^2$, $V^2H^3$ and $ZH^4$
production processes have not so much sensitivity at 3 TeV.

In the plots we have also drawn the restricted situation Eq.~\eqref{eq:dim6lock} as black line, which is formally meaningful only for the Approach \ref{assump:1}, since $\alpha_3=0$ in the $\SMEFTs$ scenario implies $\alpha_2=\Delta \alpha_1=0$. On the other hand it is interesting to note how at 3 TeV the intersection of the black line for all the ${\mathcal S}=2,\, 3,\, 5$ lines is the same for Approach \ref{assump:1} (dashed) and Approach \ref{assump:2} (solid). Assuming the $\SMEFTs$ scenario, it follows: 
\begin{eqnarray}
   |\Delta \alpha_1| \lesssim 0.25 \Longleftrightarrow |\alpha_2| \lesssim 0.4 \Longleftrightarrow \left|c^{(6)}_{\ell\varphi}/\Lambda^2\right| \lesssim 2.5\times 10^{-9}\, {\rm GeV}^{-2} \qquad{\rm at~3~TeV}\,.
\end{eqnarray}

In the case of 10 TeV, we first notice that, since the ellipses are not aligned around the $\SMEFTs$ line, in this scenario it is possible to set strong bounds. Second, the intersection with the $2\sigma$ bounds is different for the solid and dashed lines. However, as already said, only for the latter case, which is less stringent, the results are meaningful in the $\SMEFTs$ scenario and lead to
\begin{eqnarray}
    |\Delta \alpha_1| \lesssim 0.08\Longleftrightarrow  |\alpha_2| \lesssim 0.12 \Longleftrightarrow \left|c^{(6)}_{\ell\varphi}/\Lambda^2\right| \lesssim 8.0\times 10^{-10}\, {\rm GeV}^{-2} \qquad{\rm at~10~TeV}\,. \label{eq:SMEFT6comba1a2}
\end{eqnarray}

For the $\SMEFTs$ scenario, at 10 TeV bounds from $3H$ production are anyway more stringent, {\it cf.}~Eq.~\eqref{eq:boundmultiHEFT2b}, and at 3 TeV the combination of double Higgs with all just the multi-gauge boson production and Higgs-associated gauge boson production channels only slightly improves the bound from $2H$ production alone,   {\it cf.}~Eq.~\eqref{eq:boundmultiHEFT1a}.

For the general HEFT the situation is different. The $2H$ and $3H$ processes can set constraints on $\alpha_2$ and $\alpha_3$ separately, but do not provide information on $\Delta \alpha_1$. By comparing Fig.~\ref{fig:contoura1a2all} with the left plot of Fig.~\ref{fig:contourVHassumption} it is clear that at 3 TeV the combination of associated multi Higgs and gauge-boson production with $2H$ can improve the constraints in the $(\alpha_1,\,\alpha_2)$ plane and therefore on $\alpha_1$. On the contrary, at 10 TeV the $2H$ production process is not improving the constraints in the $(\alpha_1,\,\alpha_2)$ plane.

\subsubsection{Constraints on $\boldsymbol{(\alpha_2,\,\alpha_3)}$}
\label{sec:combination23}

We can obtain  $(\alpha_2,\,\alpha_3)$ constraints combining just multi-gauge boson production and Higgs-associated gauge boson production and, at 10 TeV,  constraints from $3H$ production on $\alpha_3$. At 3 TeV, since the number of events is too small, we cannot combine the information on $3H$ production with the other processes.  Besides $3H$ production, which is almost independent of $\alpha_1$, all the other processes do depend on it. Therefore we need to make an assumption, and we choose to consider  $\alpha_1=1$, {\it i.e.}, the $\bar \mu\mu H$ single-Higgs coupling has its SM value. 

Since $\alpha_1=1$ in the $\SMEFTs$ scenario implies $\alpha_2=\alpha_3=0$, with this assumption, not only cannot the bounds be translated via Eq.~\eqref{eq:dim6lock} for SMEFT dimension-6, but also we expect much stronger bounds. Indeed, similarly to what has already been discussed before for multi gauge-boson production, by fixing $\alpha_1$ and varying $\alpha_2$ and $\alpha_3$ independently we expect cross sections quickly growing  with energy. In a SMEFT framework, this is equivalent to large effects from higher-dimension operators that break not only the relation between $\alpha_1$ and $\alpha_2$ (at least dimension-8), but at the same time also the relation between $\alpha_2$ and $\alpha_3$ (at least dimension-10).

\begin{figure}[!t]
  \centering
  \includegraphics[width=0.49\textwidth]{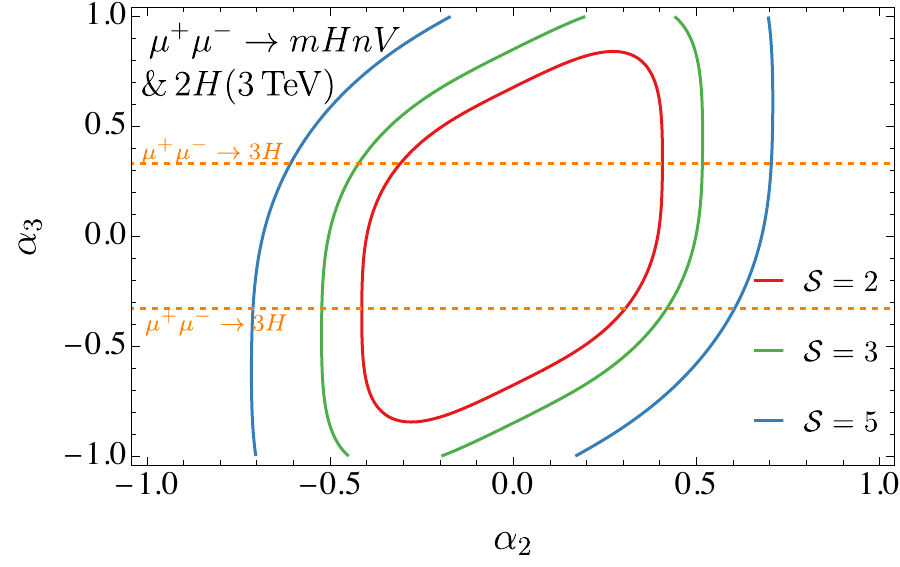}
  \includegraphics[width=0.49\textwidth]{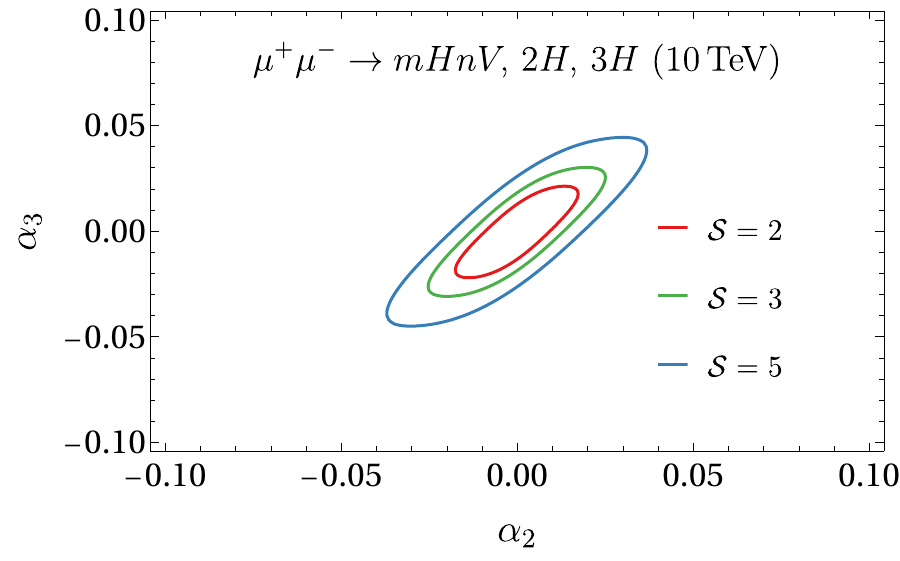}
  \caption{Same as Fig.~\ref{fig:contour45V}, but in the $(\alpha_2,\,\alpha_3)$ plane , from   Higgs-associated gauge boson production processes at a 3 TeV muon collider (left) and a 10 TeV muon collider (right), respectively.  The orange dashed lines in the left panel stand for the $95\%$ CL constraint on $\alpha_3$ from $3H$ production.}
  \label{fig:contoura2a3all}
\end{figure}

We present in Fig.~\ref{fig:contoura2a3all} the contour plots, which exhibit the following limits:
\begin{align}
    & |\alpha_2| \lesssim 0.4,& &| \alpha_3| \lesssim 0.3& &{\rm at~3~TeV}\,, \\
    &|\alpha_2| \lesssim 0.02,& & | \alpha_3| \lesssim 0.02& &{\rm at~10~TeV}\,.
\end{align}
These bounds for $\alpha_2$ ($\alpha_3$) are valid in HEFT, regardless of the specific value assumed for the other parameter $\alpha_3$ ($\alpha_2$). Clearly, at 3 TeV and even more at 10 TeV, if $\alpha_3$ ($\alpha_2$) were fixed within the allowed range from the previous equations to a specific value, the constraints on the corresponding other parameter $\alpha_2$ ($\alpha_3$) could be much stronger.

Given the shape of the contours in the left plot of Fig.~\ref{fig:contoura2a3all}  it is not surprising that the bound for $\alpha_2$ at 3 TeV is consistent with those found in the analysis in the $(\alpha_1,\,\alpha_2)$ plane in the previous section. The bound on $\alpha_3$, on the other hand, cannot compete with the one from $3H$ production, which is reported as dashed orange lines in the left plot of  Fig.~\ref{fig:contoura2a3all}.

The bounds for 10 TeV are instead very strong and derive from the condition $\alpha_1=1$ and the large growth mentioned before, which further improves the bounds originating from $3H$ production.
By comparison, from the result in Eq.~\eqref{eq:SMEFT6comba1a2} and employing the SMEFT dimension-6 relation   
we get as bound for $\alpha_3$: $|\alpha_3| \lesssim 0.4$. On the one hand, Eq.~\eqref{eq:SMEFT6comba1a2} does not have as input the $3H$ measurement. On the other hand, the comparison of dashed and solid lines in the right plot of Fig.~\ref{fig:contoura1a2all} for the $2\sigma$ bands  shows that fixing $\alpha_3$ is not having a dramatic impact on the results. The same cannot be expected for the present analysis by fixing $\alpha_1$.

\subsection{Summary of the constraints}

After the long and detailed discussion in the previous sections, we summarize here the constraints that we have presented. We explicitly separate results at 3 and 10 TeV and their interpretations within either a general HEFT framework or the $\SMEFTs$ scenario. We will tag any bound of the kind $|\Delta \alpha_1|\lesssim C$, or $|\alpha_i| \lesssim C$ with $C\ge 1$ as ``ineffective'' and we will not report them in the following. In the case of the $\SMEFTs$ scenario, we show only the bounds on $\Delta \alpha_1$. As a rule of thumb, $|\Delta \alpha_1|\lesssim C$ is equivalent to $|c^{(6)}_{\ell\varphi}/\Lambda^2|\lesssim C \times 10^{-8} ~ {\rm GeV^{-2}}$.

\subsubsection{Constraints at 3 TeV}

\paragraph{HEFT}\begin{itemize}
\item From multi-Higgs production we obtain the independent constraints:
  \begin{align}
   & |\alpha_2|\lesssim0.4\,,& ~&|\alpha_{3}|\lesssim0.3\,,&~&|\alpha_{4}|\lesssim0.5\,,& ~&|\alpha_5|\lesssim0.9\,. 
  \end{align}
  Each constraint on $\alpha_n$ originates from the corresponding $nH$ production only.
  \item Multi-gauge boson production is ineffective in setting constraints.
    \item Associated gauge and Higgs boson production ($ZH$) is ineffective in setting constraints on $\alpha_1$ independently from other $\alpha_i$. The sign of the $\bar \mu \mu H$ interaction cannot be discriminated. 
    \item The $WWH$, $ZZH$ and $ZHH$ production processes combined with the $3VH$ processes are ineffective to set constraints in the $(\alpha_1, \alpha_2)$ plane. Only by fixing $\alpha_1$  ($\alpha_2$), one can get effective constraints on $\alpha_2$  ($\alpha_1$).
    \item The previous point applies also for combining information from higher-multiplicity channels from associated gauge and Higgs boson production processes and also assuming $\alpha_3=0$ or $\Delta \alpha_1 = 0$. Effective constraints on one of the $\alpha_1$, $\alpha_2$ and $\alpha_3$ parameters can be set only when the other two are fixed.
\item Combining all the processes considered in this study we obtain, regardless of the value of the other $\alpha_i$: 
\begin{eqnarray}
 |\Delta \alpha_1| \lesssim 0.75,~~~ |\alpha_2| \lesssim 0.4\,,
 \end{eqnarray}
allowing to discriminate the sign of the $\bar \mu \mu H$ interaction, or, when assuming $\alpha_1=1$ 
 \begin{eqnarray}
  |\alpha_2| \lesssim 0.4,~~~ | \alpha_3| \lesssim 0.3\,.
 \end{eqnarray}

\end{itemize}

\paragraph{$\boldsymbol \SMEFTs$}

\begin{itemize}
\item From multi-Higgs production we obtain from $2H$ production:
  \begin{align}
   & |\Delta \alpha_1|\lesssim 0.3\,,
  \end{align}
  while we get $|\Delta \alpha_1|\lesssim 0.7$ for $3H$ production.
  Therefore, the sign of the $\bar \mu \mu H$ interaction can be discriminated.

\item Multi-gauge boson production is ineffective in setting constraints.
   \item Associated gauge and Higgs boson production ($ZH$) is ineffective in setting constraints on $\alpha_1$. Via this process, the sign of the $\bar \mu \mu H$ interaction cannot be discriminated.
\item The $WWH$, $ZZH$ and $ZHH$ production processes combined with the $3VH$ processes are ineffective to set constraints.
   \item The previous point also applies when combining information from higher-multiplicity channels from associated gauge and Higgs boson production processes
   \item Combining all the processes considered in this study that depend only on $\alpha_1$ and $\alpha_2$, but not on $\alpha_3$, we obtain
\begin{eqnarray}
 |\Delta \alpha_1| \lesssim 0.25\,.
 \end{eqnarray}

\end{itemize}

\subsubsection{Constraints at 10 TeV}
\paragraph{HEFT}

\begin{itemize}
\item From multi-Higgs production we obtain the independent constraints:
  \begin{align}
  &|\alpha_2|\lesssim0.2\,,& ~&|\alpha_3|\lesssim0.03\,,& ~&|\alpha_4|\lesssim0.01\,,&  ~&|\alpha_5|\lesssim0.01\,.
  \end{align}
    Each constraint on $\alpha_n$ originates from the corresponding $nH$ production only.

      \item In the case of multi-gauge boson production, the $ZZZ$ process alone can set the constraint on $\alpha_1$, independently from other $\alpha_i$
\begin{equation}
 |\Delta \alpha_1|\lesssim 0.2   \,.
 \end{equation}
 Therefore the sign of the $\bar \mu \mu H$ interaction can be discriminated.

\item The $4V$ and $5V$ production processes can set constraints in the $(\alpha_1, \alpha_2)$ plane, resulting in  $\Delta \alpha_1\simeq 2 \alpha_2 / 3$, and constraints on $\alpha_2$ are barely effective.

\item Combining information of $3V,4V$ and $5V$ channels, the constraints from  $4V$ and $5V$ alone are improved to 
\begin{eqnarray}
\frac{2}{3} |\alpha_2|\simeq |\Delta\alpha_1|\lesssim 0.2 \, ,
\end{eqnarray}
as in the case of    $ZZZ$ alone
    
\item Associated gauge and Higgs boson production ($ZH$) can set the following constraint on $\alpha_1$, independently from other $\alpha_i$
\begin{equation}
 |\Delta \alpha_1|\lesssim 0.8   \,.
 \end{equation}
 Therefore the sign of the $\bar \mu \mu H$ interaction can be discriminated, but the bound is less stringent than in the $ZZZ$ case.
  \item The $WWH$, $ZZH$ and $ZHH$ production processes combined with the $3VH$ production processes can set constraints in the $(\alpha_1, \alpha_2)$ plane, resulting in  $\Delta \alpha_1\simeq \alpha_2$ and
\begin{eqnarray}
|\alpha_2|\simeq |\Delta\alpha_1|\lesssim 0.5 \,.
\end{eqnarray}
\item Combining information from higher-multiplicity channels of the associated gauge and Higgs boson production processes, the previous constraint is improved to \begin{eqnarray}
|\alpha_2|\simeq |\Delta\alpha_1|\lesssim 0.2 \,.
\end{eqnarray}
\item In the previous point, if in addition $\alpha_3=0$ is assumed, the bounds are slightly improved and the correlation between $\alpha_2$ and $\Delta\alpha_1$ is much more enhanced. If instead $\Delta\alpha_1=0$ is assumed, 
\begin{eqnarray}
|\alpha_2|, |\alpha_3 |\lesssim 0.05\,,
\end{eqnarray}
with a high degree of correlation.
\item Combining all the processes considered in this study we obtain, regardless of the value of the other $\alpha_i$: 
\begin{eqnarray}
 |\Delta \alpha_1| \lesssim 0.1,~~~ |\alpha_2| \lesssim 0.1
 \end{eqnarray}
  or assuming $\alpha_1=1$ 
 \begin{eqnarray}
  |\alpha_2| \lesssim 0.02,~~~ | \alpha_3| \lesssim 0.02\,.
 \end{eqnarray}

\end{itemize}

\paragraph{$\boldsymbol \SMEFTs$}

\begin{itemize}
\item From multi-Higgs production we obtain:
  \begin{align}
   & |\Delta \alpha_1|\lesssim 0.05\,,
  \end{align}
  from $3H$ production, while $|\Delta \alpha_1|\lesssim 0.1$ for $2H$ production.
 Therefore the sign of the $\bar \mu \mu H$ interaction can be discriminated.

\item In the case of multi-gauge boson production, the $ZZZ$ process alone can set the constraint on $\alpha_1$
\begin{equation}
 |\Delta \alpha_1|\lesssim 0.2   \,.
 \end{equation}
 Therefore the sign of the $\bar \mu \mu H$ interaction can be discriminated also with this process, but the bound is less stringent than in the $HHH$  case.

\item The $4V$ and $5V$ production processes can set constraints on $\Delta \alpha_1$, and in combination with the information from $3V$ they lead to 
\begin{eqnarray}
 |\Delta\alpha_1|\lesssim 0.2 \,,
\end{eqnarray}
with no substantial improvement w.r.t.~the case of $ZZZ$ alone.

\item Associated gauge and Higgs boson production ($ZH$) can set the following constraint on $\alpha_1$
\begin{equation}
 |\Delta \alpha_1|\lesssim 0.8   \,.
 \end{equation}
 Therefore the sign of the $\bar \mu \mu H$ interaction can be discriminated with this process, but the bound is less stringent than in the $ZZZ$ and especially $HHH$ channel.
  \item The $WWH$, $ZZH$ and $ZHH$ production processes combined with the $3VH$ processes can set the constraint
 \begin{eqnarray}
|\Delta\alpha_1|\lesssim 0.5 \,.
\end{eqnarray}
\item Combining information from higher-multiplicity channels of associated gauge and Higgs boson production processes, the previous constraint is improved to \begin{eqnarray}
 |\Delta\alpha_1|\lesssim 0.2 \,.
\end{eqnarray}

 \item Combining all the processes considered in this study that depend only on $\alpha_1$ and $\alpha_2$, but not on $\alpha_3$, we obtain
\begin{eqnarray}
 |\Delta \alpha_1| \lesssim 0.08\,,
 \end{eqnarray}
which is less stringent than $3H$ alone.

\end{itemize}

\section{Conclusion and discussions}
\label{sec:conclusions}

A muon collider will provide a unique opportunity for investigating new physics effects in the muon sector. In particular, it will allow the study of the helicity-flip muon interactions associated with the fermion mass generation mechanism, which is still untested for the first two fermion generations.
In a previous paper \cite{Han:2021lnp}, we found that $\mathcal{O}(1)$ deviations in the low-energy Yukawa coupling lead to enhancements in multi-boson production at multi-TeV energies which is potentially observable at a higher energy muon collider. In this work we have extended and refined the study, focusing on vector-boson production in association with a Higgs particle as well as pure multi-Higgs production.

We have studied $\bar{\mu}\mu H^n$ anomalous interactions both in the HEFT framework, where the interactions between the muon and the Higgs boson depend on
$$\sum_{n=1}^\infty \alpha_n \frac{m_\mu}{v^n}\bar{\mu}\mu H^n\, ,$$
and in a SMEFT scenario ($\SMEFTs$) where the dominant effects arise only from a dimension-six operator $$ \frac{c_{\ell\varphi}^{(6)}}{\Lambda^{2}}
    (\varphi^\dagger\varphi)
    ({\bar \ell}_L\varphi \mu_R+\hc)\, ,$$
    with $\varphi$ being the Higgs doublet. In this scenario the quantity $c_{\ell\varphi}^{(6)}/\Lambda^{2}$ can be written in term of $\alpha_1\equiv 1+\Delta \alpha_1$, and vice-versa. The same applies  to $\alpha_2$ and $\alpha_3$, which however are not independent from $\Delta \alpha_1$, while for $i>3$ we have $\alpha_i=0$. 
    
We find that a 10 TeV muon collider can be very sensitive to probe the $\bar{\mu}\mu H^n$ anomalous interactions with $n=3,4,5$: at the level of few percents in $(\sqrt 2 m_\mu)/v $ units and at the order of 20\% for $\bar{\mu}\mu H^2$. Assuming the $\SMEFTs$ scenario, the process $\mu^+\mu^-\to 3H$ leads to an indirect measurement of the strength of the $\bar{\mu}\mu H$ interaction, the only BSM free parameter in this scenario, at the 5\% level. In the general HEFT framework, where much more freedom for the different $\bar{\mu}\mu H^n$ is possible, the precision is only slightly worse, at the level of 10\%. 
In comparison, a 3 TeV muon collider is much less powerful in testing the Higgs-muon coupling
than a 10 TeV collider. Bounds on  $\bar{\mu}\mu H^n$ with $n=2, 3,4, 5$ degrade to respectively 40\%, 30\%, 50\% and 90\%.  In the aforementioned SMEFT scenario, the strength of the $\bar{\mu}\mu H$ interaction can be probed only at the 25\% level and in HEFT at $75\%$ level.
Summarizing, this leads to:
  \begin{align}
&{ \rm HEFT ~at~3~TeV:}&
&|\Delta \alpha_1|\lesssim 0.75 \,,&~ &|\alpha_2|\lesssim0.4\,,& ~&|\alpha_3|\lesssim0.3\,,& ~&|\alpha_4|\lesssim0.5\,,&  ~&|\alpha_5|\lesssim0.9\,,\nonumber\\
  &{\rm HEFT ~at~10~TeV:}&
&|\Delta \alpha_1|\lesssim 0.1 \,,&~ &|\alpha_2|\lesssim0.2\,,& ~&|\alpha_3|\lesssim0.03\,,& ~&|\alpha_4|\lesssim0.01\,,&  ~&|\alpha_5|\lesssim0.01\,.\nonumber
  \end{align}
  and 
  \begin{align}
  &{\rm \SMEFTs ~at~3~TeV:}&
  &|\Delta \alpha_1|\lesssim 0.25 \Longleftrightarrow \left|c^{(6)}_{\ell\varphi}/\Lambda^2\right| \lesssim 2.5\times 10^{-9}\, {\rm GeV}^{-2} \,,\nonumber\\
  &{\rm \SMEFTs ~at~10~TeV:}&
  &|\Delta \alpha_1|\lesssim 0.05 \Longleftrightarrow \left|c^{(6)}_{\ell\varphi}/\Lambda^2\right| \lesssim 5\times 10^{-10}\, {\rm GeV}^{-2} \,.\nonumber
      \end{align}
Many more sensitivity bounds are reported in this work, based on different sub-classes of processes and possible assumptions on other $\alpha_i$ parameters.
We have discussed in detail the different bounds arising in the general HEFT framework, as well as in the $\SMEFTs$ scenario which can be seen as selecting a particular direction in the multidimensional parameter space of the former. The $\SMEFTs$ scenario leads to smaller growth at high energy for high multiplicities in the final state and therefore smaller sensitivity. On the other hand, depending only on one single parameter, constraints can be more stringent. The most relevant example is the case of 10 TeV where $3H$ production leads to strong constraints on $|\Delta \alpha_1|$, while in HEFT those constraints are set on $|\alpha_3|$, which is independent from $|\Delta \alpha_1|$.

The difference between the 3 and 10 TeV results are mainly due to the large growth with the energy that is induced by anomalous $\bar{\mu}\mu H^n$ interactions, especially if effects that in SMEFT would be of dimension higher than 6 are taken into account, as effectively done in the general HEFT framework. We have also studied possible limitations due to unitarity violation and concluded that at 3 and 10 TeV and for the processes studied in this work, such problems are not present. However, we notice that passing from 3 to 10 TeV can lead to a very different growth of the cross section when the final-state multiplicity is increased:  at energies as, {\it e.g.}, 30 TeV unitarity problems would become manifest.  

In order to study perturbative unitarity we have calculated high-energy limits in the HEFT framework, based on the GBET. We have provided a general formalism and explicit formulas for any process considered in this work, which have been exploited in the discussion of the results for understanding the qualitative  behavior of the Monte Carlo simulations that have been performed in order to extract bounds. At the same time we have also verified the negligible contributions from anomalous Higgs self-couplings.

In this study we have considered only total cross sections and no  advanced simulations of experimental set-ups. On the one hand, a more realistic study involving experimental aspects as vector boson tagging may degrade the expected sensitivity. On the other hand, taking into account differential information or more advanced techniques for improving the extraction of the information from the signal, may also improve our bounds.   The latter would be definitely desirable for a 3 TeV collider in order to achieve better bounds than those here. Conversely,  at the 10 TeV, our study demonstrates the great potential of a muon collider in setting bounds on anomalous muon and Higgs interactions, further supporting the project of such machine operating at this energy. In conclusion, while LHC will provide us with a first measurement of the muon-Higgs interaction, only a 10 TeV muon collider can discriminate the SM muon Yukawa sector from different mechanisms, {\em e.g.} being sensitive to the sign of the muon-Higgs coupling or deviations of a few per cent in a model-independent approach.

\subsection*{Acknowledgments}

WK, NK, and TS were supported by the Deutsche Forschungsgemeinschaft (DFG, German Research Foundation) under grant 396021762 - TRR 257. 
JRR acknowledges the support of the Deutsche
Forschungsgemeinschaft (DFG, German Research Association) under
Germany's Excellence Strategy-EXC 2121 ``Quantum Universe"-390833306 as well as by National Science Centre
(Poland) under the OPUS research project no. 2021/43/B/ST2/01778.
This work has also been funded by the Deutsche Forschungsgemeinschaft (DFG, German Research Foundation) -- 491245950.
FM acknowledges support by FRS-FNRS (Belgian National Scientific Research Fund) IISN projects 4.4503.16 and by the MUR through the PRIN 2022RXEZCJ.
TH and KX were supported by the U.S. Department of Energy under grant Nos. DE-SC0007914 and in part by Pitt PACC. Also DP and JRR are grateful for support by Pitt PACC.
The work of TH and KX was performed partly at the Aspen Center for Physics, which is supported by the U.S. National Science Foundation under Grant No. PHY-1607611 and No. PHY-2210452.
KX was also supported by the U.S. National Science Foundation under Grant No. PHY-2112829, PHY-2013791, and PHY-2310497.
EC work was supported by the European Research Council (ERC) under the European Union’s Horizon 2020 research and innovation programme (Grant agreement No. 949451).
Computational resources have been provided by the supercomputing facilities of the Universit\'e catholique de Louvain (CISM/UCL) and the Consortium des \'Equipements de Calcul Intensif en 
F\'ed\'eration Wallonie Bruxelles (C\'ECI) funded by the Fond de la Recherche Scientifique de Belgique (F.R.S.-FNRS) under convention 2.5020.11 and by the Walloon Region.
This research was done using services provided by the OSG Consortium \cite{osg07,osg09,osg3,osg4}, which is supported by the National Science Foundation awards 2030508 and 1836650.
YM, DP and JRR also acknowledge support from the COMETA COST Action CA22130.
This work is co-funded by the European Union (EU). Views and opinions expressed are however those of the author(s) only and do not necessarily reflect those of the EU or European Research Executive Agency (REA). Neither the EU nor the REA can be held responsible for them. The paper is endorsed by the International Muon Collider Collaboration (IMCC).
\clearpage

\appendix

\section{Examples for UV-complete models}
\label{sec:uvcomplete_models}

In this appendix, we will briefly discuss two simplified, {\it i.e.},  UV-(semi)complete models that exhibit characteristic features in the Yukawa sector, providing as motivation the EFT setup for our study from the point of view of model building.  The models both have the decoupling property and thus lead to a low-energy SMEFT Lagrangian.  The leading corrections occur at operator dimension six and eight, respectively.  Due to the smallness of the SM Yukawa coupling, the new effects in the Yukawa sector become a phenomenologically dominant feature.  Other SMEFT operators which emerge in such models compete with unsuppressed SM interactions and are thus accessible only by precision measurements.  The two UV models are therefore supporting, respectively, the $\SMEFTs$ and $\SMEFTe$ scenarios considered in the main text.

\subsection{Muon-Singlet Mixing}
\label{sec:model-d6}
This model extends the SM spectrum by a scalar singlet $S$ and by a Dirac fermion $E_{L,R}$ which is also an electroweak singlet, acting as a heavy partner of the right-handed muon.  Structures of this kind have been discussed at length in the literature; compare, {\it e.g.}, top-quark see-saw~\cite{Chivukula:1998wd} and custodial-symmetric Little-Higgs models~\cite{Chang:2003un}, where heavy-singlet partners appear.  The simplified model below focuses on the muon sector, and could be considered as a ``charged-lepton seesaw scenario''.

We construct the generic Lagrangian as 
\begin{align}
  \mathcal{L}
  &=
    \mathcal{L}_{\text{SM}} + \mathcal{L}_{ES}\, ,
  \\
  \mathcal{L}_{ES}
  &=
    \frac{1}{2}\left(\pd_{\mu}S\pd^{\mu}S-\Lambda_S^2 S^2\right)
   + \bar E_L(\ii\slashed D)E_L + \bar E_R(\ii\slashed D)E_R
    - \Lambda_E(\bar E_RE_L + \hc)
    \notag\\
  &\quad{}
    - \lambda_{\varphi S} (\varphi^\dagger \varphi)S
    - \left(Y_{E \varphi}\bar\ell_L \varphi E_R
    + Y_{ES} \bar E_L S\mu_R
    + \hc\, \right) - c_{\varphi S^2}(\varphi^\dagger \varphi)S^2 .
\end{align}
There are three mass scales $\Lambda_E,\Lambda_S,\lambda_{\varphi S}$ and three dimensionless couplings, two Yukawa couplings $Y_{E \varphi}$, $Y_{ES}$ and $c_{\varphi S^2}$.  After EWSB, $E$ mixes with $\mu$ and $S$ mixes
with $h$.  The mixing effects are suppressed by powers of the ratio $v/\Lambda$, where $v$ is the electroweak scale, and $\Lambda$ represents any of the three new mass scales. 

Integrating out the heavy fields, 
we obtain three operators in the low-energy EFT,
\begin{align}
  \label{eq:model-D6}
  \mathcal{L}_{ES}^{(6)}
  &=
      - \frac{\lambda_{\varphi S}^2}{2\Lambda_S^4}(\varphi^\dagger \varphi)\Box(\varphi^\dagger \varphi)
    + Y_{E \varphi}^2\frac{1}{\Lambda_E^2}(\bar\ell_L \varphi)\ii\slashed D(\varphi^\dagger\ell_L)
    - Y_{E \varphi} Y_{ES} \frac{\lambda_{\varphi S}}{\Lambda_E \Lambda_S^2}(\varphi^\dagger \varphi)(\bar\ell_L \varphi\mu_R + \hc)\notag \\
   &\quad - \frac{\lambda_{\varphi S}^2c_{\varphi S^2}}{\Lambda_S^4}(\varphi^\dagger \varphi)^3\,.
\end{align}
The first term in Eq.~\eqref{eq:model-D6} generates a small universal modification of all Higgs couplings.  This modification is challenging to detect in current collider experiments but can be constrained independently at a muon collider~\cite{Forslund:2023reu}.  The second term can be eliminated by the fermion's equations of motion.  Its effect can be included, but it is suppressed both by the SM Yukawa coupling and by the mixing parameter $v^2/\Lambda_E^2$, and thus negligible.   

The third term in~\eqref{eq:model-D6} provides the $D=6$ SMEFT operator which affects the Higgs-muon interactions in Eq.~\eqref{eq:SMEFT}.  This coefficient is not suppressed by the small SM Yukawa coupling.  Setting $\lambda_{\varphi S}=\Lambda_S=\Lambda_E=\Lambda$ for simplicity, we find 
\begin{equation}\label{eq:ModelDim6}
  c_{\ell\varphi}^{(6)} = Y_{E \varphi} Y_{ES}\,.
\end{equation}

In terms of the HEFT notation of Eq.~\eqref{eq:HiggsParameters}, the Lagrangian~\eqref{eq:model-D6} amounts to corrections to the muon mass and to the muon Yukawa coupling,
respectively,
\begin{align}
  m_\mu^{\text{D4+D6}} &= \frac{v}{\sqrt2}
          \left(y_\mu + \frac12 Y_{E \varphi} Y_{ES} \frac{v^2}{\Lambda^2}\right)\,,
  &
  y_{\mu,1}^{\text{D4+D6}}
  &= y_\mu + \frac32 Y_{E \varphi} Y_{ES} \frac{v^2}{\Lambda^2}\,. \label{eq:mumass_uv1}
\end{align}

In either framework, we should express the EFT parameters in terms of the input observables which satisfy Eq.~\eqref{eq:SM-input}, and therefore apply the counterterms in Eq.~\eqref{eq:SMEFT-ct}.  With this renormalization in place, the muon mass retains its physical meaning, and we obtain the effective Higgs-muon couplings
\begin{align}\label{eq:ModelHEFTpars}
  y_{\mu,1} &= y_{\mu} +Y_{E \varphi} Y_{ES}\frac{v^2}{\Lambda^2}\,,
  &
  y_{\mu,2} &= \frac{3v^2}{2}\frac{Y_{E \varphi} Y_{ES}}{\Lambda^2}\,,
  &
  y_{\mu,3} &= \frac{v^2}{2}\frac{Y_{E \varphi} Y_{ES}}{\Lambda^2}\,.
\end{align}
The $\alpha_i$ vertex coefficients can be read off the relations~\eqref{eq:map}, using either Eqs.~\eqref{eq:ModelDim6} or \eqref{eq:ModelHEFTpars}.

\subsection{Muon Recurrence}
\label{sec:model-d8}
As a second simplified model, we introduce heavy vector-like partners $E_{L,R}$ and $F_{L,R}$ for both the right-handed and left-handed muon components, respectively.  Such a structure emerges, {\it e.g.}, in universal extra-dimension models~\cite{Appelquist:2000nn} or in models of lepton partial compositeness.  Furthermore, we add a singlet scalar $S$ whose Higgs-field interactions softly break the Kaluza-Klein symmetry (in the compositeness picture this would be a scalar composite resonance). 
\begin{align}
  \begin{split}
  \LL =&\LL_{\text{SM}}+\LL_{EFS}\,,\\
    \LL_{EFS}=&\bar E_L \ii \slashed D E_L +\bar E_R \ii \slashed D E_R-\Lambda_E(\bar E_L E_R+\bar E_R E_L)\\
    &+\bar F_L \ii \slashed D F_L +\bar F_R \ii \slashed D F_R-\Lambda_F(\bar F_L F_R+\bar F_R F_L)\\
    &+\frac{1}{2}\left(\pd_{\mu}S\pd^{\mu}S-\Lambda_S^2 S^2\right) -\lambda_{\varphi S} \left(\varphi^{\dagger}\varphi\right)S\\
    &-Y_{ES}(\bar {\mu}_R E_L+\bar E_L \mu_R)S-Y_{FS}(\bar {\ell}_L F_R+\bar F_R \ell_L)S\\
    &-Y_{EF\varphi}(\bar {F}_L \varphi E_R+\bar E_R \varphi^{\dagger}F_L)\,.
  \end{split}
\end{align}
Other allowed dimension-4 operators in the scalar sector are not relevant for the argument and left out here for simplicity.
There are independent mass scales $\Lambda_E$, $\Lambda_F$, $\Lambda_S$, $\lambda_{\varphi S}$ and three independent new Yukawa coupling $Y_{ES}$, $Y_{FS}$, $Y_{EF\varphi}$.   After EWSB, both $E$ and $F$  mix with $\mu$. 

Integrating out the heavy fields $E,F$ and $S$, we obtain up to dimension $D=8$
\begin{align}
  \begin{split}
    \LL^{(<8)}_{EFS}=&\frac{\lambda_{\varphi S}^2}{2\Lambda_S^2}(\varphi^{\dagger}\varphi)^2-\frac{\lambda_{\varphi S}^2}{2\Lambda_S^4}(\varphi^{\dagger}\varphi)\Box(\varphi^{\dagger}\varphi)+\frac{\lambda_{\varphi S}^2}{2\Lambda_S^6}(\varphi^{\dagger}\varphi)\Box^2(\varphi^{\dagger}\varphi)\\
    &+\frac{\lambda_{\varphi S}^2Y_{FS}^2}{\Lambda_F^2\Lambda_S^4}(\varphi^{\dagger}\varphi)^2 \bar \ell_L \ii \slashed{D}\ell_L+\frac{\lambda_{\varphi S}^2Y_{ES}^2}{\Lambda_E^2\Lambda_S^4}(\varphi^{\dagger}\varphi)^2 \bar \mu_R \ii \slashed{D}\mu_R\\
    &-\frac{\lambda^2_{\varphi S}Y_{EF}Y_{FS}Y_{ES}}{\Lambda_F \Lambda_E \Lambda_S^4}(\varphi^{\dagger}\varphi)^2 (\bar \ell_L \varphi \mu_R + \operatorname{h.c.})\,.
  \end{split}
\end{align}
Applying the fermion's equations of motion to the second line, the effective Lagrangian becomes
\begin{align}
  \begin{split}   \LL^{'(<8)}_{EFS}=&\frac{\lambda_{\varphi S}^2}{2\Lambda_S^2}(\varphi^{\dagger}\varphi)^2-\frac{\lambda_{\varphi S}^2}{2\Lambda_S^4}(\varphi^{\dagger}\varphi)\Box(\varphi^{\dagger}\varphi)+\frac{\lambda_{\varphi S}^2}{2\Lambda_S^6}(\varphi^{\dagger}\varphi)\Box^2(\varphi^{\dagger}\varphi)\\
    &-\left(\frac{\lambda^2_{\varphi S}Y_{EF}Y_{FS}Y_{ES}}{\Lambda_F \Lambda_E \Lambda_S^4}-\frac{\lambda_{\varphi S}^2y_{\mu}Y_{ES}^2}{\Lambda_E^2\Lambda_S^4}-\frac{\lambda_{\varphi S}^2y_{\mu}Y_{FS}^2}{\Lambda_F^2\Lambda_S^4}\right)(\varphi^{\dagger}\varphi)^2 (\bar \ell_L \varphi \mu_R + \operatorname{h.c.})\,.
  \end{split}
  \label{eq:model-D8-EFT}
\end{align}
The first term in the first line corresponds to a modification of the Higgs self-coupling which can be absorbed in the definition of SM parameter $\lambda$.
The remaining terms in the first line again generate universal modifications to all Higgs couplings.  

In the second line of~\eqref{eq:model-D8-EFT}, we identify the dimension-8 SMEFT operator which affects the Higgs-muon interactions.   There is no contribution at $D=6$, as this is protected by the (softly broken) Kaluza-Klein symmetry. 
If we again set $\lambda_{\varphi S}=\Lambda_S=\Lambda_E=\Lambda_F=\Lambda$, we obtain the $D=8$ operator coefficient
\begin{equation}\label{eq:ModelDim8}
  c_{\ell\varphi}^{(8)} = Y_{EF}Y_{FS}Y_{ES}-y_{\mu}(Y_{ES}^2+Y_{FS}^2)\,,
\end{equation}
where the term proportional to the small parameter $y_\mu$ may and will be neglected.

In terms of the HEFT notation of  Eq.~\eqref{eq:HiggsParameters}, the Lagrangian~\eqref{eq:ModelDim8} leads to corrections to the muon mass and to the muon Yukawa coupling,
\begin{align}
  m_\mu^{\text{D4+D8}} &= \frac{v}{\sqrt2}
          \left(y_\mu + \frac12 Y_{EF}Y_{FS}Y_{ES} \frac{v^4}{\Lambda^4}\right),
  &
  y_{\mu,1}^{\text{D4+D8}}
  &= y_\mu + \frac52 Y_{EF}Y_{FS}Y_{ES} \frac{v^4}{\Lambda^4}\,. \label{eq:mumass_uv2}
\end{align}
We again express all parameters in terms of input observables. The muon mass retains its physical meaning, and the effective Higgs-muon couplings are
\begin{align}\label{eq:ModelDim8HEFTpars}
    y_{\mu,1} &= y_{\mu}+  {v^4}\frac{Y_{EF}Y_{FS}Y_{ES}}{\Lambda^4}\,, &
    y_{\mu,2} &=  \frac{5 v^4}{2} \frac{Y_{EF}Y_{FS}Y_{ES}}{\Lambda^4}\,, &
    y_{\mu,3} &= \frac{ 5 v^4}{ 2} \frac{Y_{EF}Y_{FS}Y_{ES}}{\Lambda^4}\,, \nonumber \\
  y_{\mu,4} &= \frac{ 5 v^4}{ 4} \frac{Y_{EF}Y_{FS}Y_{ES}}{\Lambda^4}\,, &
    y_{\mu,5} &= \frac{v^4}{ 4} \frac{Y_{EF}Y_{FS}Y_{ES}}{\Lambda^4}\,. 
\end{align}
As before, the $\alpha_i$ vertex coefficients can be read off the relations~\eqref{eq:map}, using either Eqs.~\eqref{eq:ModelDim8} or \eqref{eq:ModelDim8HEFTpars}.

\section{SMEFT-HEFT Wilson coefficient  matching}
\label{sec:relations}

\subsection{Yukawa Sector coefficient relations}
\label{sec:YukawaRelation}

 In the following, we give the relations for matching HEFT and SMEFT, with three independent parameters in the Yukawa sector for both of them: $\{c^{(6)}_{\ell\varphi},c^{(8)}_{\ell\varphi},c^{(10)}_{\ell\varphi}\}$ in SMEFT and $\{y_{\mu,1},y_{\mu,2},y_{\mu,3}\}$ or equivalently $\{\alpha_1,\alpha_2,\alpha_3\}$ in HEFT. We write the relations for HEFT parameters in terms of the SMEFT ones and vice versa. When we write $y_{\mu,i}$ (or $\alpha_i$) as function of $\{c^{(6)}_{\ell\varphi},c^{(8)}_{\ell\varphi},c^{(10)}_{\ell\varphi}\}$ we extend the relations to the cases $4\le i \le 6$. In the following equations, these cases are separated by the others via a line. Extending the relations to the cases $4\le i \le 6$ allows us to convert all the cross sections in the high-energy limits listed in Sec.~\ref{sec:pheno} in terms of the leading contributions in the SMEFT, beyond the $\SMEFTs$ or $\SMEFTe$ scenarios considered in the main text.\footnote{In doing so we also implicitly assume $\Lambda$ larger than the energy considered and as a consequence also the $\Lambda\rightarrow \infty$ limit. Thus, for $i\le 7$, dimension-12 or higher contributions are always suppressed, as much as dimension-10 (and dimension-8) are suppressed for $i\le 5$ ($i\le 3$). }

    \begin{itemize}
      \item SMEFT in terms of HEFT
      { \allowdisplaybreaks
        \begin{align}
          \frac{c^{(6)}_{\ell\varphi}}{\Lambda^2}&=-\frac{35\mmu}{4\sqrt 2v^3}+\frac{35}{8v^2}y_{\mu,1}-\frac{5}{2v^2}y_{\mu,2}+\frac{3}{4v^2}y_{\mu,3}\, , \\ 
          \frac{c^{(8)}_{\ell\varphi}}{\Lambda^4}&=\frac{21\mmu}{2\sqrt 2v^5}-\frac{21}{4v^4}y_{\mu,1}+\frac{4}{v^4}y_{\mu,2}-\frac{3}{2v^4}y_{\mu,3}\, , \\ 
          \frac{c^{(10)}_{\ell\varphi}}{\Lambda^6}&=-\frac{5\mmu}{\sqrt 2v^7}+\frac{5}{2v^6}y_{\mu,1}-\frac{2}{v^6}y_{\mu,2}+\frac{1}{v^6}y_{\mu,3}
        \, . \end{align}
\item HEFT in terms of SMEFT
        \begin{align}
          y_{\mu,1}&=\frac{\sqrt{2}\mmu}{v}+v^2 \frac{c^{(6)}_{\ell\varphi}}{\Lambda^2}+v^4 \frac{c^{(8)}_{\ell\varphi}}{\Lambda^4}+\frac{3v^6}{4} \frac{c^{(10)}_{\ell\varphi}}{\Lambda^6}\, , \\ 
          y_{\mu,2}&=\frac{3 v^2}{2} \frac{c^{(6)}_{\ell\varphi}}{\Lambda^2}+\frac{5v^4}{2} \frac{c^{(8)}_{\ell\varphi}}{\Lambda^4}+\frac{21v^6}{8} \frac{c^{(10)}_{\ell\varphi}}{\Lambda^6}\, , \\ 
          y_{\mu,3}&=\frac{v^2}{2} \frac{c^{(6)}_{\ell\varphi}}{\Lambda^2}+\frac{5v^4}{2} \frac{c^{(8)}_{\ell\varphi}}{\Lambda^4}+\frac{35v^6}{8} \frac{c^{(10)}_{\ell\varphi}}{\Lambda^6}\, , \\ 
\nonumber \\
          \hline
\nonumber \\          
          y_{\mu,4}&=\frac{5v^4}{4} \frac{c^{(8)}_{\ell\varphi}}{\Lambda^4}+\frac{35v^6}{8} \frac{c^{(10)}_{\ell\varphi}}{\Lambda^6}\, , \\ 
          y_{\mu,5}&=\frac{v^4}{4} \frac{c^{(8)}_{\ell\varphi}}{\Lambda^4}+\frac{21v^6}{8} \frac{c^{(10)}_{\ell\varphi}}{\Lambda^6}\, , \\ 
          y_{\mu,6}&=\frac{7v^6}{8} \frac{c^{(10)}_{\ell\varphi}}{\Lambda^6}\, , \\ 
          y_{\mu,7}&=\frac{v^6}{8} \frac{c^{(10)}_{\ell\varphi}}{\Lambda^6}
        \, . \end{align}
      \item HEFT in terms of \parameters
        \begin{align}
          y_{\mu,i}&=\frac{\sqrt 2 \mmu}{v}\alpha_i \; \text{for} \; i\in\{1,\dots,7\}
        \, . \end{align}
        }
      \item SMEFT in terms of \parameters
        \begin{align}\label{eq:SMEFTinAlphaBeta_i}
          \frac{c^{(6)}_{\ell\varphi}}{\Lambda^2}&=-\frac{35\mmu}{4\sqrt 2v^3}+\frac{35\mmu}{4 \sqrt 2v^3}\alpha_{1}-\frac{5\mmu}{\sqrt 2v^3}\alpha_{2}+\frac{3\mmu}{2\sqrt 2v^3}\alpha_{3}\, , \\ 
          \frac{c^{(8)}_{\ell\varphi}}{\Lambda^4}&=\frac{21\mmu}{2\sqrt 2v^5}-\frac{21\mmu}{2\sqrt 2v^5}\alpha_{1}+\frac{4\sqrt{2}\mmu}{v^5}\alpha_{2}-\frac{3\mmu}{\sqrt 2v^5}\alpha_{3}\, , \\ 
          \frac{c^{(10)}_{\ell\varphi}}{\Lambda^6}&=-\frac{5\mmu}{\sqrt 2v^7}+\frac{5\mmu}{\sqrt 2v^7}\alpha_{1}-\frac{2\sqrt 2\mmu}{v^7}\alpha_{2}+\frac{\sqrt 2\mmu}{v^7}\alpha_{3}\label{eq:SMEFTinAlphaBeta_f}
        \, . \end{align}
      \item \parameters in terms of HEFT
        \begin{align}
          \alpha_i&=\frac{v}{\sqrt 2 \mmu}y_{\mu,i} \; \text{for} \; i\in\{1,\dots,7\}
        \, . \end{align}
      \item \parameters  in terms of SMEFT
        \begin{align}
          \alpha_{1}&=1+\frac{v^3}{\sqrt 2 \mmu} \frac{c^{(6)}_{\ell\varphi}}{\Lambda^2}+\frac{v^5}{\sqrt 2 \mmu} \frac{c^{(8)}_{\ell\varphi}}{\Lambda^4}+\frac{3v^7}{4\sqrt 2 \mmu} \frac{c^{(10)}_{\ell\varphi}}{\Lambda^6}\, , \\ 
          \alpha_{2}&=\frac{3 v^3}{2\sqrt 2 \mmu} \frac{c^{(6)}_{\ell\varphi}}{\Lambda^2}+\frac{5v^5}{2\sqrt 2 \mmu} \frac{c^{(8)}_{\ell\varphi}}{\Lambda^4}+\frac{21v^7}{8\sqrt 2 \mmu} \frac{c^{(10)}_{\ell\varphi}}{\Lambda^6}\, , \\ 
          \alpha_{3}&=\frac{v^3}{2\sqrt 2 \mmu} \frac{c^{(6)}_{\ell\varphi}}{\Lambda^2}+\frac{5v^5}{2\sqrt 2 \mmu} \frac{c^{(8)}_{\ell\varphi}}{\Lambda^4}+\frac{35v^7}{8\sqrt 2 \mmu} \frac{c^{(10)}_{\ell\varphi}}{\Lambda^6}\, , \\ 
          \nonumber \\
          \hline
\nonumber \\     
          \alpha_{4}&=\frac{5v^5}{4\sqrt 2 \mmu} \frac{c^{(8)}_{\ell\varphi}}{\Lambda^4}+\frac{35v^7}{8\sqrt 2 \mmu} \frac{c^{(10)}_{\ell\varphi}}{\Lambda^6}\, , \\ 
          \alpha_{5}&=\frac{v^5}{4\sqrt 2 \mmu} \frac{c^{(8)}_{\ell\varphi}}{\Lambda^4}+\frac{21v^7}{8\sqrt 2 \mmu} \frac{c^{(10)}_{\ell\varphi}}{\Lambda^6}\, , \\ 
          \alpha_{6}&=\frac{7v^7}{8\sqrt 2 \mmu} \frac{c^{(10)}_{\ell\varphi}}{\Lambda^6}\, , \\ 
          \alpha_{7}&=\frac{v^7}{8\sqrt 2 \mmu} \frac{c^{(10)}_{\ell\varphi}}{\Lambda^6}
        \, . \end{align}
    \end{itemize}

\subsection{Higgs Sector coefficient relations}
\label{sec:HiggsRelations}
Analogously to what done in the previous section, we give the relations for  matching HEFT and SMEFT, with three  independent parameters in the Higgs sector for both of them: $\{c^{(6)}_{\varphi},c^{(8)}_{\varphi},c^{(10)}_{\varphi}\}$ in SMEFT and $\{f_{V,3},f_{V,4},f_{V,5}\}$ or equivalently  $\{\beta_3,\beta_4,\beta_{5}\}$ in HEFT. Also here, following a similar logic, we write the relations for HEFT parameters in terms of the SMEFT ones and vice versa, and when we write $f_{V,i}$ (or $\beta_i$) as function of $\{c^{(6)}_{\varphi},c^{(8)}_{\varphi},c^{(10)}_{\varphi}\}$ we extend the relations to the cases $6\le i \le 10$.

    \begin{itemize}
      \item SMEFT in terms of HEFT
        \begin{align}
          \frac{c^{(6)}_{\varphi}}{\Lambda^2}&=-\frac{63\lambda}{8v^2}+\frac{77}{8v^2}f_{V,3}-\frac{7}{v^2}f_{V,4}+\frac{5}{2v^2}f_{V,5}\, , \\ 
          \frac{c^{(8)}_{\varphi}}{\Lambda^4}&=\frac{45\lambda}{8v^4}-\frac{57}{8v^4}f_{V,3}+\frac{6}{v^4}f_{V,4}-\frac{5}{2v^4}f_{V,5}\, , \\ 
          \frac{c^{(10)}_{\varphi}}{\Lambda^6}&=-\frac{7\lambda}{4v^6}+\frac{9}{4v^6}f_{V,3}-\frac{2}{v^6}f_{V,4}+\frac{1}{v^6}f_{V,5}
        \, . \end{align}
      \item HEFT in terms of SMEFT
        \begin{align}
          f_{V,3}&=\lambda+v^2 \frac{c^{(6)}_{\varphi}}{\Lambda^2}+2v^4 \frac{c^{(8)}_{\varphi}}{\Lambda^4}+\frac{5v^6}{2}\frac{c^{(10)}_{\varphi}}{\Lambda^6}\, , \\ 
          f_{V,4}&=\frac{\lambda}{4}+\frac{3v^2}{2} \frac{c^{(6)}_{\varphi}}{\Lambda^2}+4v^4  \frac{c^{(8)}_{\varphi}}{\Lambda^4}+\frac{25v^6}{4}\frac{c^{(10)}_{\varphi}}{\Lambda^6}\, , \\ 
          f_{V,5}&=\frac{3v^2}{4} \frac{c^{(6)}_{\varphi}}{\Lambda^2}+\frac{7v^4}{2} \frac{c^{(8)}_{\varphi}}{\Lambda^4}+\frac{63v^6}{8} \frac{c^{(10)}_{\varphi}}{\Lambda^6}\, , \\ 
          \nonumber \\
          \hline
\nonumber \\     
          f_{V,6}&=\frac{v^2}{8} \frac{c^{(6)}_{\varphi}}{\Lambda^2}+\frac{7v^4}{4} \frac{c^{(8)}_{\varphi}}{\Lambda^4}+\frac{105v^6}{16} \frac{c^{(10)}_{\varphi}}{\Lambda^6}\, , \\ 
          f_{V,7}&=\frac{v^4}{2} \frac{c^{(8)}_{\varphi}}{\Lambda^4}+\frac{15v^6}{4} \frac{c^{(10)}_{\varphi}}{\Lambda^6}\, , \\ 
          f_{V,8}&=\frac{v^4}{16} \frac{c^{(8)}_{\varphi}}{\Lambda^4}+\frac{45v^6}{32} \frac{c^{(10)}_{\varphi}}{\Lambda^6}\, , \\ 
          f_{V,9}&=\frac{5v^6}{16} \frac{c^{(10)}_{\varphi}}{\Lambda^6}\, , \\ 
          f_{V,10}&=\frac{v^6}{32} \frac{c^{(10)}_{\varphi}}{\Lambda^6}
        \, . \end{align}
      \item HEFT in terms of \parameters
        \begin{align}
          f_{V,i}&=\lambda \beta_i \; \text{for}\; i\in\{3\dots 10\}
        \, . \end{align}
      \item SMEFT in terms of \parameters
        \begin{align}
          \frac{c^{(6)}_{\varphi}}{\Lambda^2}&=-\frac{63\lambda}{8v^2}+\frac{77\lambda}{8v^2}\beta_{3}-\frac{7\lambda}{v^2}\beta_{4}+\frac{5\lambda}{2v^2}\beta_{5}\, , \\ 
          \frac{c^{(8)}_{\varphi}}{\Lambda^4}&=\frac{45\lambda}{8v^4}-\frac{57\lambda}{8v^4}\beta_{3}+\frac{6\lambda}{v^4}\beta_{4}-\frac{5\lambda}{2v^4}\beta_{5}\, , \\ 
          \frac{c^{(10)}_{\varphi}}{\Lambda^6}&=-\frac{7\lambda}{4v^6}+\frac{9\lambda}{4v^6}\beta_{3}-\frac{2\lambda}{v^6}\beta_{4}+\frac{\lambda}{v^6}\beta_{5}
        \, . \end{align}
      \item \parameters in terms of HEFT
        \begin{align}
          \beta_i&=\frac{1}{\lambda}f_{V,i} \; \text{for}\; i\in\{3\dots 10\}
        \, . \end{align}
      \item \parameters in terms of SMEFT
        \begin{align}
          \beta_{3}&=1+\frac{v^2}{\lambda} \frac{c^{(6)}_{\varphi}}{\Lambda^2}+\frac{2v^4}{\lambda} \frac{c^{(8)}_{\varphi}}{\Lambda^4}+\frac{5v^6}{2\lambda}\frac{c^{(10)}_{\varphi}}{\Lambda^6}\, , \\ 
          \beta_{4}&=\frac{1}{4}+\frac{3v^2}{2\lambda} \frac{c^{(6)}_{\varphi}}{\Lambda^2}+\frac{4v^4}{\lambda}  \frac{c^{(8)}_{\varphi}}{\Lambda^4}+\frac{25v^6}{4\lambda}\frac{c^{(10)}_{\varphi}}{\Lambda^6}\, , \\ 
          \beta_{5}&=\frac{3v^2}{4\lambda} \frac{c^{(6)}_{\varphi}}{\Lambda^2}+\frac{7v^4}{2\lambda} \frac{c^{(8)}_{\varphi}}{\Lambda^4}+\frac{63v^6}{8\lambda} \frac{c^{(10)}_{\varphi}}{\Lambda^6}\, , \\ 
          \nonumber \\
          \hline
\nonumber \\     
          \beta_{6}&=\frac{v^2}{8\lambda} \frac{c^{(6)}_{\varphi}}{\Lambda^2}+\frac{7v^4}{4\lambda} \frac{c^{(8)}_{\varphi}}{\Lambda^4}+\frac{105v^6}{16\lambda} \frac{c^{(10)}_{\varphi}}{\Lambda^6}\, , \\ 
          \beta_{7}&=\frac{v^4}{2\lambda} \frac{c^{(8)}_{\varphi}}{\Lambda^4}+\frac{15v^6}{4\lambda} \frac{c^{(10)}_{\varphi}}{\Lambda^6}\, , \\ 
          \beta_{8}&=\frac{v^4}{16\lambda} \frac{c^{(8)}_{\varphi}}{\Lambda^4}+\frac{45v^6}{32\lambda} \frac{c^{(10)}_{\varphi}}{\Lambda^6}\, , \\ 
          \beta_{9}&=\frac{5v^6}{16\lambda} \frac{c^{(10)}_{\varphi}}{\Lambda^6}\, , \\ 
          \beta_{10}&=\frac{v^6}{32\lambda} \frac{c^{(10)}_{\varphi}}{\Lambda^6}
        \, . \end{align}
    \end{itemize}

\clearpage


\section{Feynman rules for EFT interactions}
\label{sec:feynmanrules}

We present the Feynman rules for the BSM vertices entering the amplitudes of the processes that we consider in our study. They have been generated automatically by {\sc \small FeynRules} \cite{Christensen:2008py, Alloul:2013bka}, while producing the {\sc \small UFO} models \cite{Degrande:2011ua, Darme:2023jdn} that have been employed for our simulations. Vertices involve interactions between the muon and the Higgs boson or the Higgs boson with itself, and are expressed both in terms of the HEFT and SMEFT parametrization, with the latter up to effects of dimension ten.

Vertices can depend on the properties of the external particle, including the momenta, which will be suppressed in the pictographical representation of the vertex but they will be understood as incoming and numbered following the same convention of the label of the particle. As an example:
\begin{equation*}
  \raisebox{-.45\height}{\begin{tikzpicture}\begin{feynman}
      \vertex (i1) at (-1,-1) {\(\Phi_1\)};
      \vertex (i2) at (-1,1) {\(\Phi_2\)};
      \vertex (a)  at (0,0)  ;
      \vertex (i3) at (1,1) {\(\Phi_3\)};
      \vertex (i4) at (1,-1) {\(\Phi_4\)};
      \diagram* {
        {[edges=plain]
           (a) -- [rmomentum'=\( p_1\)] (i1),
           (a) -- [rmomentum'=\( p_2\)] (i2),
           (a) -- [rmomentum'=\( p_3\)] (i3),
           (a) -- [rmomentum'=\( p_4\)] (i4),
        },
        };\end{feynman}; \end{tikzpicture}}=  \raisebox{-.45\height}{\begin{tikzpicture}\begin{feynman}
            \vertex (i1) at (-1,-1) {\(\Phi_1\)};
            \vertex (i2) at (-1,1) {\(\Phi_2\)};
            \vertex (a)  at (0,0)  ;
            \vertex (i3) at (1,1) {\(\Phi_3\)};
            \vertex (i4) at (1,-1) {\(\Phi_4\)};
            \diagram* {
              {[edges=plain]
                 (a) -- (i1),
                 (a) --  (i2),
                 (a) --  (i3),
                 (a) -- (i4),
              },
              };\end{feynman}; \end{tikzpicture}} \quad.
\end{equation*}

We are interested in processes of the form $\mu^+\mu^-\rightarrow X$, where $X$ is a  final state including only EW bosons ($W,Z$ and $H$) with up to five particles in the final state. Thus, in view of the use of the GBET, we consider vertices involving muons and scalars with a multiplicity up to seven and vertices with only scalars with a multiplicity up to  six. We use momentum conservation in order to simplify the Feynman rules of the Higgs sector and introduce the following shorthand notations $A,B,C,D$ for the combinations of momenta
\begin{eqnarray*}
  &A=&2
  {p}_1\cdot{p}_2-{p}_1\cdot{p}_3-
  {p}_1\cdot{p}_4-{p}_2\cdot{p}_3-{p}_2
  \cdot{p}_4+2 {p}_3\cdot{p}_4\, ,\\
  &B=&2
   {p}_1\cdot({p}_2+{p}_3+{p}_4)+3 ({p}_5+{p}_6)^2+2 {p}_2\cdot({p}_3+{p}_4)+2 {p}_3\cdot{p}_4+12
   {p}_5\cdot{p}_6\, ,\\
   &C=&8
   {p}_1\cdot{p}_2-2 ({p}_1+{p}_2)\cdot({p}_3+{p}_4+{p}_5+{p}_6)
   -2 {p}_3\cdot{p}_4+3
   ({p}_3+p_4)\cdot({p}_5+p_6)-2 {p}_5\cdot{p}_6\, ,\\
   &D=&3 {p}_1\cdot({p}_2+{p}_3)+2 ({p}_1+p_2+p_3)^2+3
   {p}_2\cdot{p}_3+3
   \left({p}_4\cdot{p}_5+{p}_4\cdot{p}_6
   +{p}_5\cdot{p}_6\right)\, .
\end{eqnarray*}

\includepdf[pages=-,width=\paperwidth,offset=75 -75,pagecommand={}]{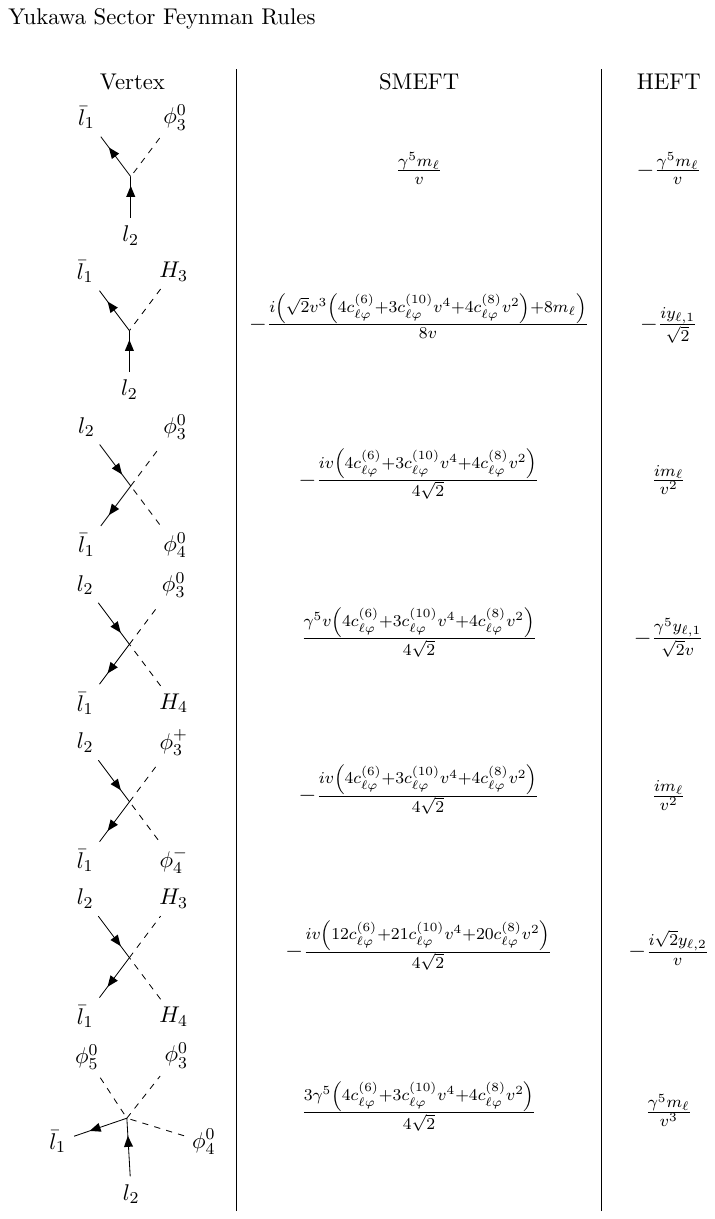}
\includepdf[pages=-,width=\paperwidth,offset=75 -75,pagecommand={}]{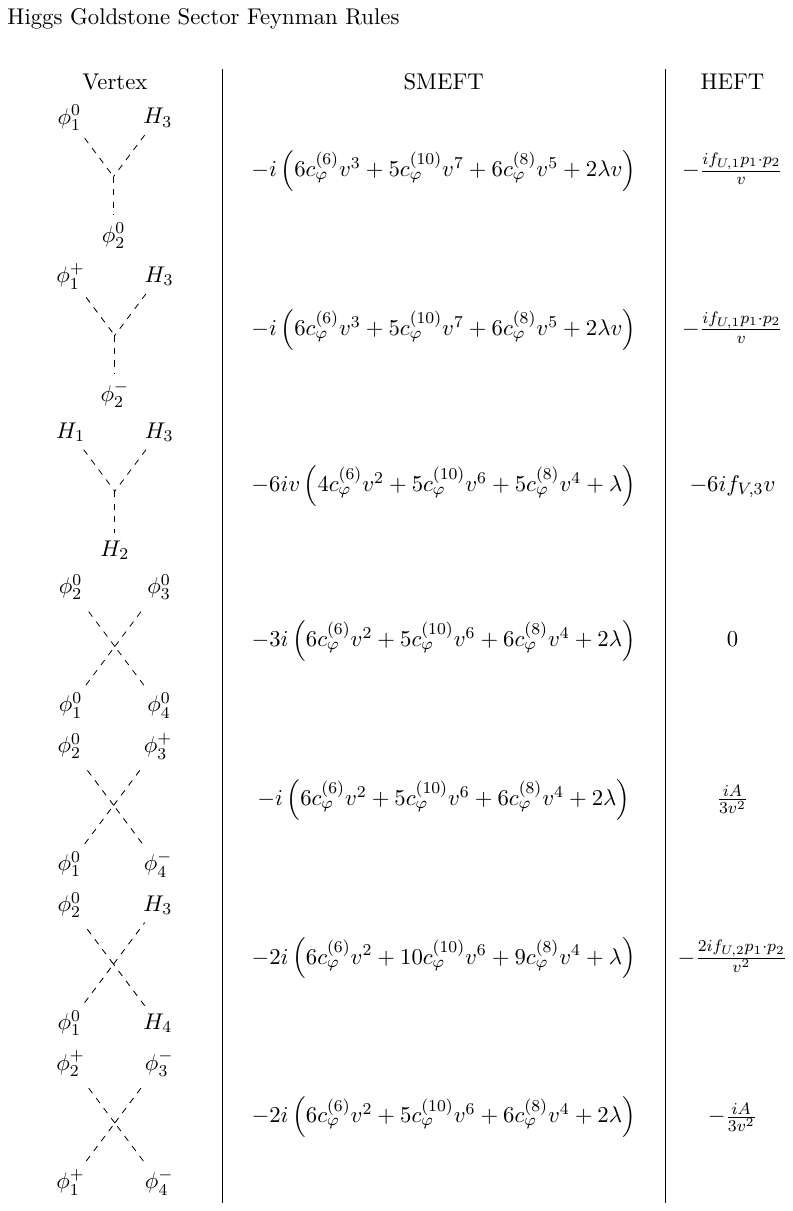}

\section{High-Energy amplitudes}

\label{sec:HEAmps}

In this section we list the explicit analytic formulae, obtained with the help of {\sc \small FeynArts} \cite{Hahn:2000kx}, for the operator coefficients $C_{nij}(0)$  entering Eq.~\eqref{eq:sigma-klm}, in the HEFT framework. We also consider some general formulae for the HEFT and SMEFT cases.

   As explained in Sec.~\ref{sec:HEHEFT}, in order to derive these coefficients, one has to start from Eq.~\eqref{eq:sdep} and thus all diagrams in Eq.~\eqref{eq:heft_diags} need to be summed. Only after this step, the dependence on all the invariants exactly cancels out, leading to Eq.~\eqref{eq:sindep}. 
   
In the Eqs.~\eqref{eq:TPFS}--\eqref{eq:SPFS}, at the end of this section, we list the coefficients $C_{nij}(0)$, with a proper normalization, for all the possibilities satisfying $N=n+i+2j$ with $2\le N\le6$. It is worth mentioning that all of the listed coefficients vanish in the SM case, since in this case  $\alpha_1=1$ and all the other $\alpha_k$ with $k>1$ are equal to zero. It is important to note that, besides the cases with $N$ external Higgs bosons, the Eqs.~\eqref{eq:TPFS}--\eqref{eq:SPFS}  do not correspond in HEFT to the diagram with only the $\bar \mu \mu \phi^N$ contact term, where with $\phi$ we generically refer to $\phi_\pm, \phi_0$ or $H$. Denoting with a hat such contribution we nevertheless provide a general formula for it, 
\begin{align}\label{eq:HEFT-Contact}
 \hat C^{\text{HEFT}}_{nlj}=\frac{v y_{\mu, n}}{\sqrt{2}}\frac{(-i)^l(-2)^j (j!)^2l! n!}{(2j+l)!v^{2j+l+n}}\binom{\lfloor \tfrac{l}{2}\rfloor+j}{j}\, .
\end{align}

 In order to avoid confusion with the imaginary part $i$ we have denoted the number of $\phi_0$, only in this formula, by $l$ and not $i$. Thus, we have $N=n+l+2j$.  In Eq.~\eqref{eq:HEFT-Contact} $\lfloor x \rfloor$ denotes the floor-function that gives the greatest integer smaller or equal to $x$. 
 
We can provide the same expression for the SMEFT Lagrangian in  Eq.~\eqref{eq:SMEFT}, which reads
\begin{align}
\begin{split}
   \hat C^{\text{SMEFT}}_{nlj}=&-\frac{m_{\mu}}{v}\Delta_{nlj}
   + \frac{1}{\sqrt 2}\sum_{k=k_N}^{\infty}\frac{c^{(2k+4)}_{\ell \varphi}}{\Lambda^{2k}}\left(\frac{v^2}{2}\right)^k
   \\
   &\qquad\qquad
   \left[\Delta_{nlj}-\frac{n! l! (j!)^2}{ 2^{-j}v^{N-1}}\binom{k}{j}\binom{2(k-j)+1-l}{n}\binom{k-j}{\lfloor \tfrac{l}{2}\rfloor}\gamma_l\right]\, ,
   \end{split}\label{eq:SMEFT-Contact}
\end{align}
with
\begin{equation}
    \Delta_{nlj}=\delta_{j0}(\delta_{n1}\delta_{l0}+i \delta_{n0}\delta_{l1})\, ,
\end{equation}
\begin{equation}
    \gamma_l =\begin{cases}
        1 \; \text{if}\;l \; \text{even}\\
        i \; \text{if}\;l \; \text{odd}
    \end{cases} \quad.
\end{equation}
 and $k_N=\operatorname{Max}\left(1,\lceil\tfrac{N-1}{2}\rceil\right)$. It is interesting to note that if all the higher-dimension operators that can contribute are taken into account, unlike what has been done in the paper for $\SMEFTs$ and $\SMEFTe$ scenarios, the contact term \eqref{eq:SMEFT-Contact} is exactly the leading contribution at high energies. In other words, calculating the squared matrix elements from Eq.~\eqref{eq:SMEFT-Contact}  leads to the exact same result as using the conversion rules between SMEFT and the HEFT for the results listed in Eqs.~\eqref{eq:TPFS}--\eqref{eq:SPFS}.
In doing so we clearly also assume no modifications in the purely Goldstone boson sector of the HEFT, {\it i.e.},
\begin{equation}
    f_{U,1}=2 \;, \; f_{U,2}=1 \;  \text{and}\;  f_{U,k}=0 \; \text{for}\; k>2\, .
\end{equation}
Thus, in SMEFT, the high energy cross sections can be approximated via 
\begin{equation}\label{eq:SMEFTCrossSection}
  \bar{\sigma}_{X_{nij}}=\frac{(2\pi)^4}{4}|\hat C^{\text{SMEFT}}_{nij}|^2\Phi_{X_{nij}}(k_1+k_2;p_1,\dots ,p_{2j+n+i})\, .
\end{equation}

In the following, as anticipated, we list the coefficients $C_{nij}(0)$, with a proper normalization, for all the possibilities satisfying $N=n+i+2j$ with $2\le N\le6$.

\medskip

\subsubsection*{Two Particle Final States}
\begin{align}
\label{eq:TPFS}
  \begin{array}{cc|c}
    \multicolumn{2}{c}{X_{klm}} & |C_{nij}(0)|^2 \tfrac{v^4}{m_{\mu}^2} \\
    \hline
    H & H & 4 \alpha _2^2 \\
   H & \phi _0 & \left(1-\alpha _1\right){}^2 \\
   \phi _0 & \phi _0 & \left(1-\alpha _1\right){}^2 \\
   \phi _+ & \phi _- & \left(1-\alpha _1\right){}^2 \\
\end{array}
\end{align}
\subsubsection*{Three Particle Final States}
\begin{align}
  \begin{array}{ccc|c}
    \multicolumn{3}{c}{X_{klm}} & |C_{nij}(0)|^2 \tfrac{v^6}{m_{\mu}^2} \\
    \hline
    H & H & H & 36 \alpha _3^2 \\
 H & H & \phi _0 & 4 \left(1-\alpha _1+\alpha
   _2\right){}^2 \\
 H & \phi _0 & \phi _0 & 4 \left(1-\alpha _1+\alpha
   _2\right){}^2 \\
 H & \phi _+ & \phi _- & 4 \left(1-\alpha _1+\alpha
   _2\right){}^2 \\
 \phi _0 & \phi _0 & \phi _0 & 9 \left(1-\alpha
   _1\right){}^2 \\
 \phi _0 & \phi _+ & \phi _- & \left(1-\alpha
   _1\right){}^2 \\
  \end{array}
\end{align}
\subsubsection*{Four Particle Final States}
\begin{align}
\label{eq:Xklm-4}
  \begin{array}{cccc|c}
    \multicolumn{4}{c}{X_{klm}} & |C_{nij}(0)|^2 \tfrac{v^8}{m_{\mu}^2} \\
    \hline
    H & H & H & H & 576 \alpha _4^2 \\
    H & H & H & \phi _0 & 36 \left(1-\alpha _1+\alpha
      _2-\alpha _3\right){}^2 \\
    H & H & \phi _0 & \phi _0 & 36 \left(1-\alpha
      _1+\alpha _2-\alpha _3\right){}^2 \\
    H & H & \phi _+ & \phi _- & 36 \left(1-\alpha
      _1+\alpha _2-\alpha _3\right){}^2 \\
    H & \phi _0 & \phi _0 & \phi _0 & 9 \left(3 -3 \alpha
      _1+2 \alpha _2\right){}^2 \\
    H & \phi _0 & \phi _+ & \phi _- & \left(3 -3 \alpha
      _1+2 \alpha _2\right){}^2 \\
    \phi _0 & \phi _0 & \phi _0 & \phi _0 & 9 \left(3 -3
      \alpha _1+2 \alpha _2\right){}^2 \\
    \phi _0 & \phi _0 & \phi _+ & \phi _- & \left(3 -3
      \alpha _1+2 \alpha _2\right){}^2 \\
    \phi _+ & \phi _+ & \phi _- & \phi _- & 4 \left(3 -3
      \alpha _1+2 \alpha _2\right){}^2 \\
\end{array}
\end{align}
\subsubsection*{Five Particle Final States}
\begin{align}
  \begin{array}{ccccc|c}
    \multicolumn{5}{c}{X_{klm}} & |C_{nij}(0)|^2 \tfrac{v^{10}}{m_{\mu}^2} \\
    \hline
    H & H & H & H & H & 14400 \alpha _5^2 \\
     H & H & H & H & \phi _0 & 576 \left(1-\alpha _1+\alpha
       _2-\alpha _3+\alpha _4\right){}^2 \\
     H & H & H & \phi _0 & \phi _0 & 576 \left(1-\alpha
       _1+\alpha _2-\alpha _3+\alpha _4\right){}^2 \\
     H & H & H & \phi _+ & \phi _- & 576 \left(1-\alpha
       _1+\alpha _2-\alpha _3+\alpha _4\right){}^2 \\
     H & H & \phi _0 & \phi _0 & \phi _0 & 36 \left(6 -6
       \alpha _1+5 \alpha _2-3 \alpha _3\right){}^2 \\
     H & H & \phi _0 & \phi _+ & \phi _- & 4 \left(6 -6
       \alpha _1+5 \alpha _2-3 \alpha _3\right){}^2 \\
     H & \phi _0 & \phi _0 & \phi _0 & \phi _0 & 36 \left(6 
     -6 \alpha _1+5 \alpha _2-3 \alpha _3\right){}^2 \\
     H & \phi _0 & \phi _0 & \phi _+ & \phi _- & 4 \left(6 
     -6 \alpha _1+5 \alpha _2-3 \alpha _3\right){}^2 \\
     H & \phi _+ & \phi _+ & \phi _- & \phi _- & 16 \left(6 
     -6 \alpha _1+5 \alpha _2-3 \alpha _3\right){}^2 \\
     \phi _0 & \phi _0 & \phi _0 & \phi _0 & \phi _0 & 225 \left(3
       -3 \alpha _1+2 \alpha _2\right){}^2 \\
     \phi _0 & \phi _0 & \phi _0 & \phi _+ & \phi _- & 9 \left(3
       -3 \alpha _1+2 \alpha _2\right){}^2 \\
     \phi _0 & \phi _+ & \phi _+ & \phi _- & \phi _- & 4 \left(3
       -3 \alpha _1+2 \alpha _2\right){}^2 \\
\end{array}
\end{align}
\subsubsection*{Six Particle Final States}
\begin{align}
\label{eq:SPFS}
  \begin{array}{cccccc|c}
    \multicolumn{6}{c}{X_{klm}} & |C_{nij}(0)|^2 \tfrac{v^{12}}{m_{\mu}^2} \\
    \hline
    H & H & H & H & H & H & 518400 \alpha _6^2 \\
    H & H & H & H & H & \phi _0 & 14400 \left(1-\alpha
      _1+\alpha _2-\alpha _3+\alpha _4-\alpha _5\right){}^2 \\
    H & H & H & H & \phi _0 & \phi _0 & 14400 \left(1
    -\alpha _1+\alpha _2-\alpha _3+\alpha _4-\alpha
      _5\right){}^2 \\
    H & H & H & H & \phi _+ & \phi _- & 14400 \left(1
    -\alpha _1+\alpha _2-\alpha _3+\alpha _4-\alpha
      _5\right){}^2 \\
    H & H & H & \phi _0 & \phi _0 & \phi _0 & 324 \left(10 
    -10 \alpha _1+9 \alpha _2-7 \alpha _3+4 \alpha
      _4\right){}^2 \\
    H & H & H & \phi _0 & \phi _+ & \phi _- & 36 \left(10 
    -10 \alpha _1+9 \alpha _2-7 \alpha _3+4 \alpha
      _4\right){}^2 \\
    H & H & \phi _0 & \phi _0 & \phi _0 & \phi _0 & 324 \left(10
      -10 \alpha _1+9 \alpha _2-7 \alpha _3+4 \alpha
      _4\right){}^2 \\
    H & H & \phi _0 & \phi _0 & \phi _+ & \phi _- & 36 \left(10
      -10 \alpha _1+9 \alpha _2-7 \alpha _3+4 \alpha
      _4\right){}^2 \\
    H & H & \phi _+ & \phi _+ & \phi _- & \phi _- & 144 \left(10
      -10 \alpha _1+9 \alpha _2-7 \alpha _3+4 \alpha
      _4\right){}^2 \\
    H & \phi _0 & \phi _0 & \phi _0 & \phi _0 & \phi _0 & 2025
      \left(5 -5 \alpha _1+4 \alpha _2-2 \alpha
      _3\right){}^2 \\
    H & \phi _0 & \phi _0 & \phi _0 & \phi _+ & \phi _- & 81
      \left(5 -5 \alpha _1+4 \alpha _2-2 \alpha
      _3\right){}^2 \\
    H & \phi _0 & \phi _+ & \phi _+ & \phi _- & \phi _- & 36
      \left(5 -5 \alpha _1+4 \alpha _2-2 \alpha
      _3\right){}^2 \\
    \phi _0 & \phi _0 & \phi _0 & \phi _0 & \phi _0 & \phi _0 &
      2025 \left(5 -5 \alpha _1+4 \alpha _2-2 \alpha
      _3\right){}^2 \\
    \phi _0 & \phi _0 & \phi _0 & \phi _0 & \phi _+ & \phi _- & 81
      \left(5 -5 \alpha _1+4 \alpha _2-2 \alpha
      _3\right){}^2 \\
    \phi _0 & \phi _0 & \phi _+ & \phi _+ & \phi _- & \phi _- & 36
      \left(5 -5 \alpha _1+4 \alpha _2-2 \alpha
      _3\right){}^2 \\
    \phi _+ & \phi _+ & \phi _+ & \phi _- & \phi _- & \phi _- &
      324 \left(5 -5 \alpha _1+4 \alpha _2-2 \alpha
      _3\right){}^2 \\
\end{array}
\end{align}

\medskip

\section{Dependencies on $\alpha_i$ and $\beta_k$ for additional processes}
\label{sec:furtherplots}
In this Appendix we display additional plots showing the dependence on $\alpha_i$ and $\beta_k$ parameters for all the processes that have been studied but not documented in the main text. 

\begin{figure}[t]
  \centering
  \includegraphics[width=0.49\textwidth]{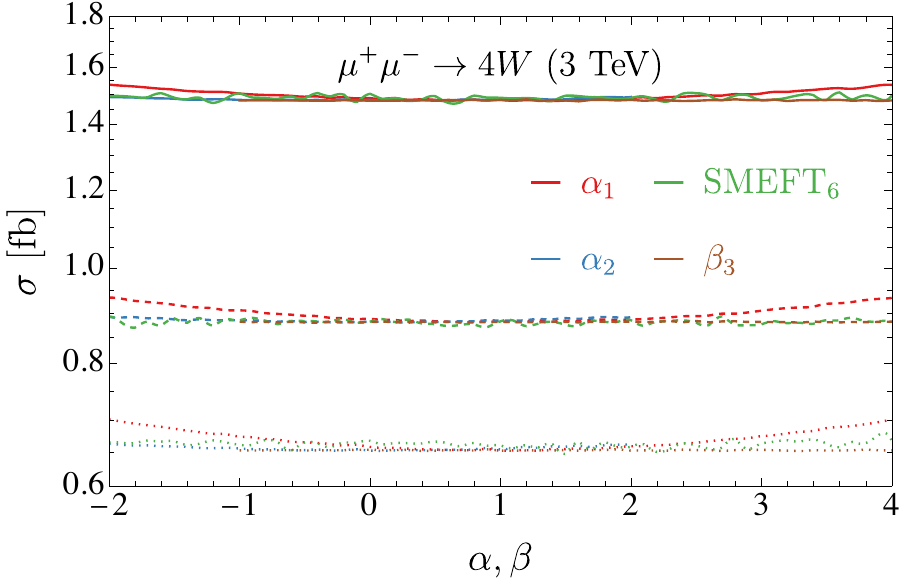}
  \includegraphics[width=0.49\textwidth]{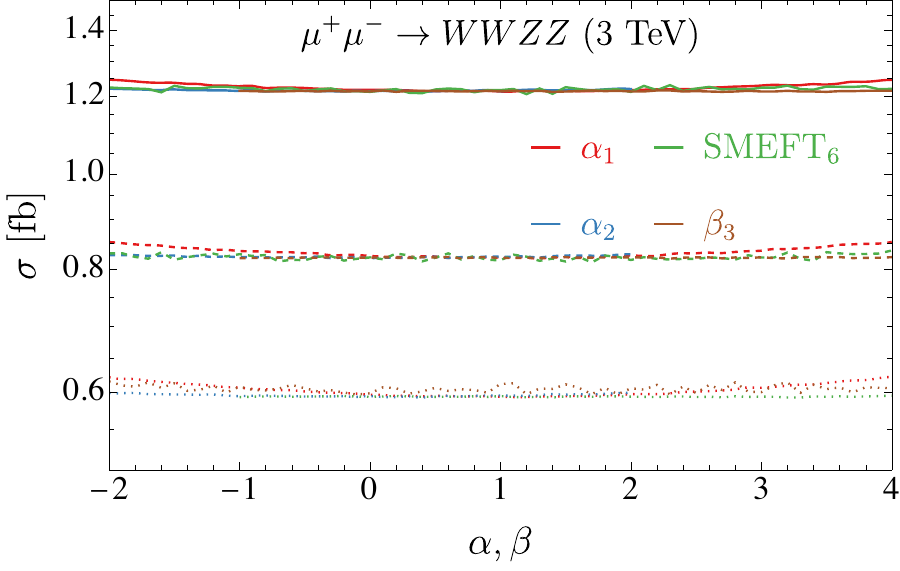}
  \includegraphics[width=0.49\textwidth]{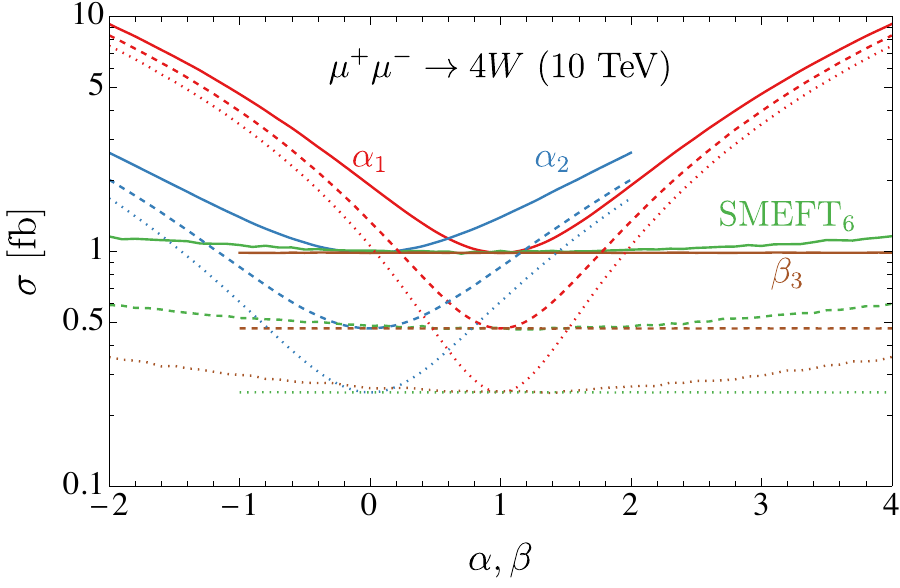}
  \includegraphics[width=0.49\textwidth]{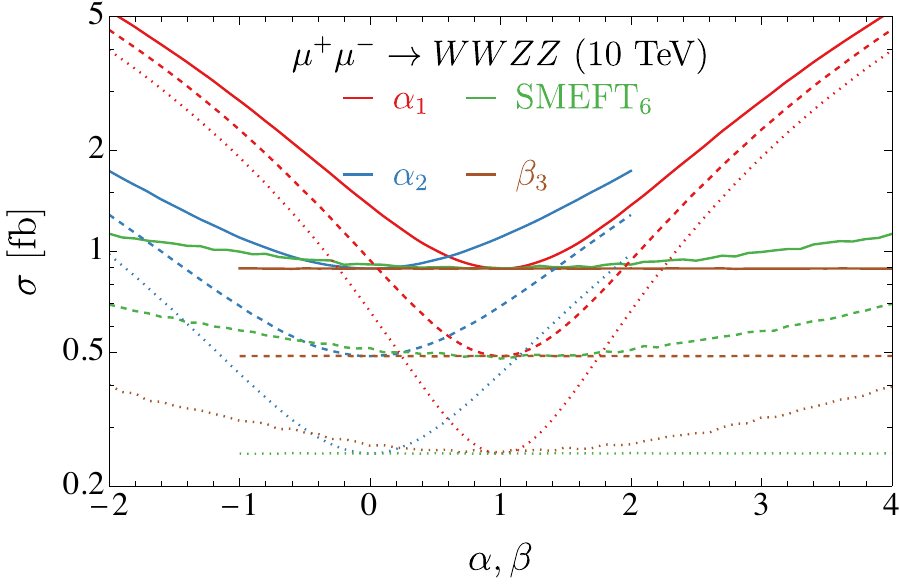}
  \caption{Same as Fig.~\ref{fig:mmnhkappa3} for the $4V$ production processes that are not showed in the main text,  at 3 and 10 TeV.}
  \label{fig:mm4vkappa}
\end{figure}

\begin{figure}[!t]
  \centering
  \includegraphics[width=0.46\textwidth]{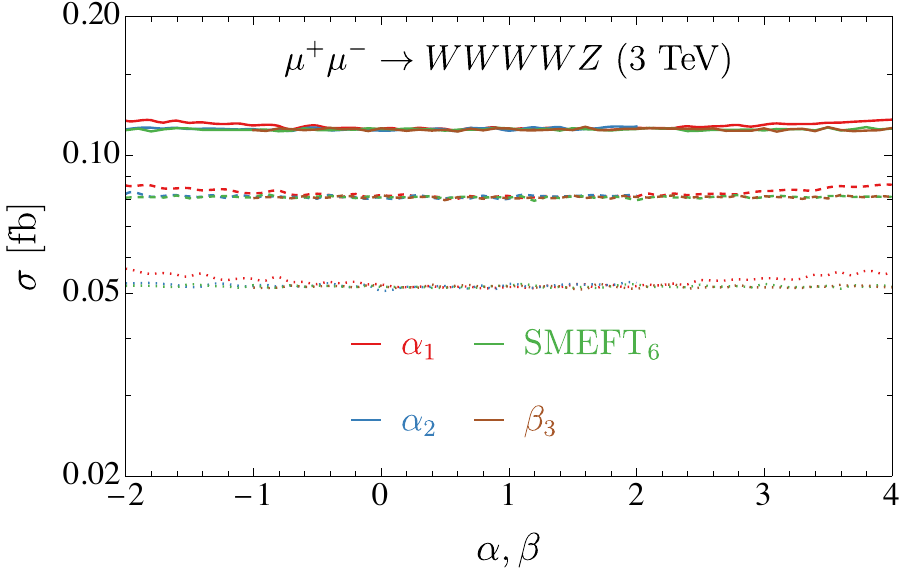}
  \hspace{5mm}\includegraphics[width=0.46\textwidth]{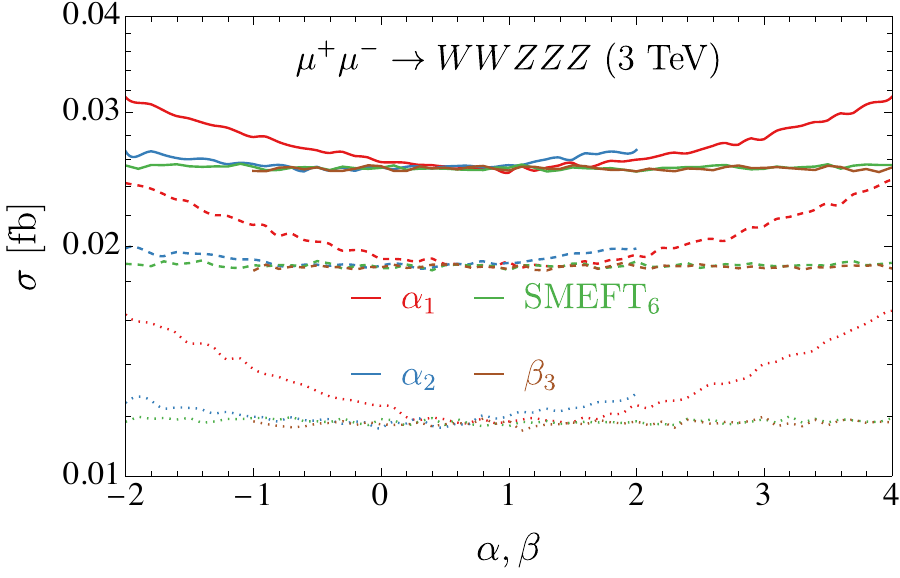}\\
  \includegraphics[width=0.49\textwidth]{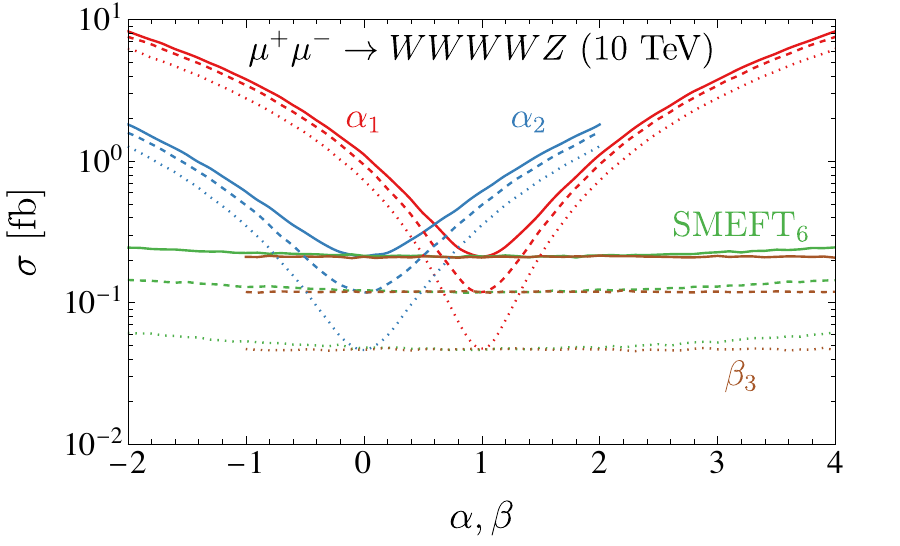}
  \includegraphics[width=0.49\textwidth]{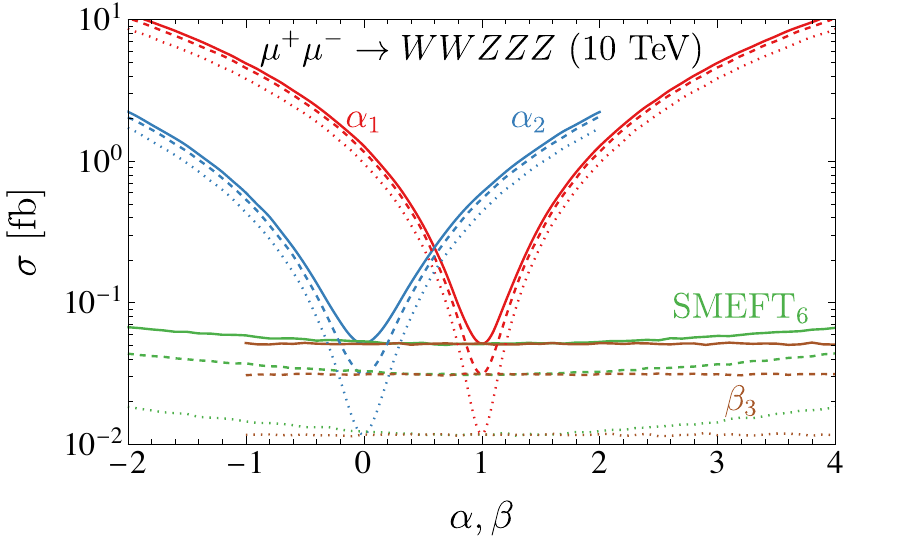}
  \caption{Same as Fig.~\ref{fig:mmnhkappa3} for the $5V$ production processes that are not showed in the main text,  at 3 and 10 TeV.}
  \label{fig:mm5vkappa}
\end{figure}

\begin{figure}[!t]
  \centering
  \includegraphics[width=0.47\textwidth]{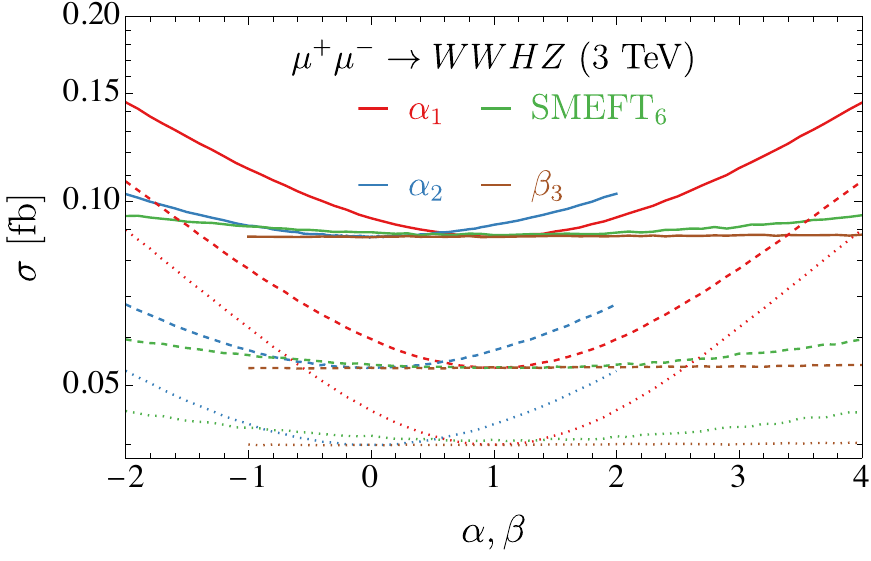}
  \includegraphics[width=0.49\textwidth]{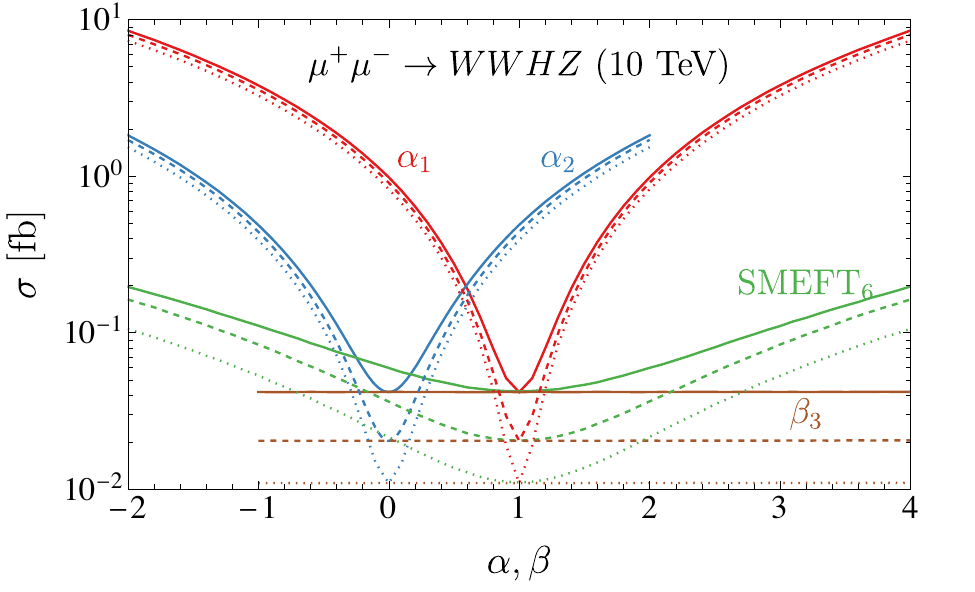}
  \caption{Same as Fig.~\ref{fig:mmnhkappa3} for the $HV^3$ production processes that are not showed in the main text,  at 3 and 10 TeV.}
  \label{fig:mmvvvhkappa}
\end{figure}

\begin{figure}[!t]
  \centering
  \includegraphics[width=0.48\textwidth]{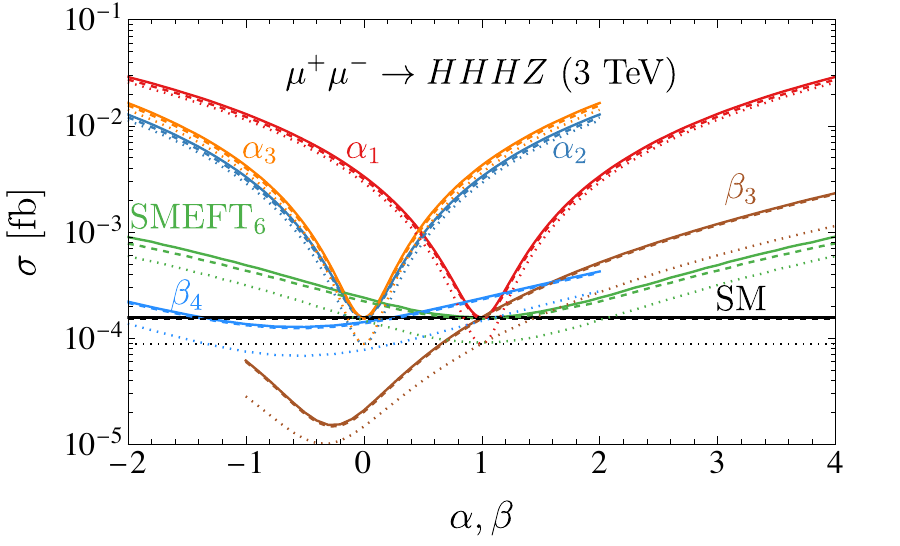}
  \includegraphics[width=0.48\textwidth]{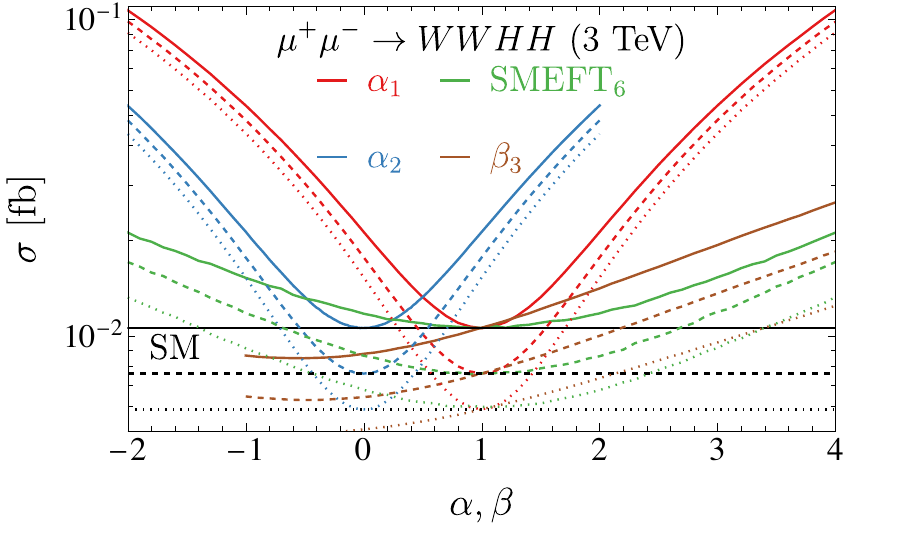}
  \includegraphics[width=0.48\textwidth]{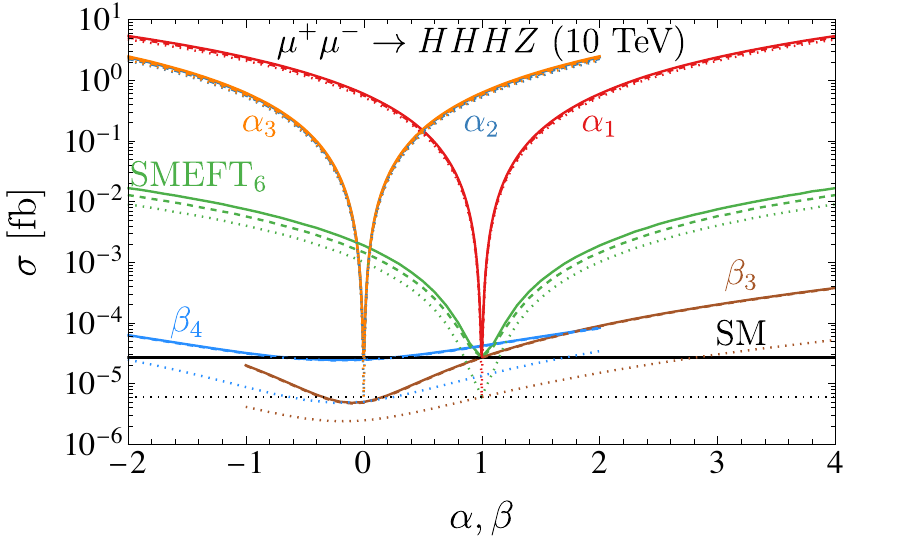}
  \includegraphics[width=0.48\textwidth]{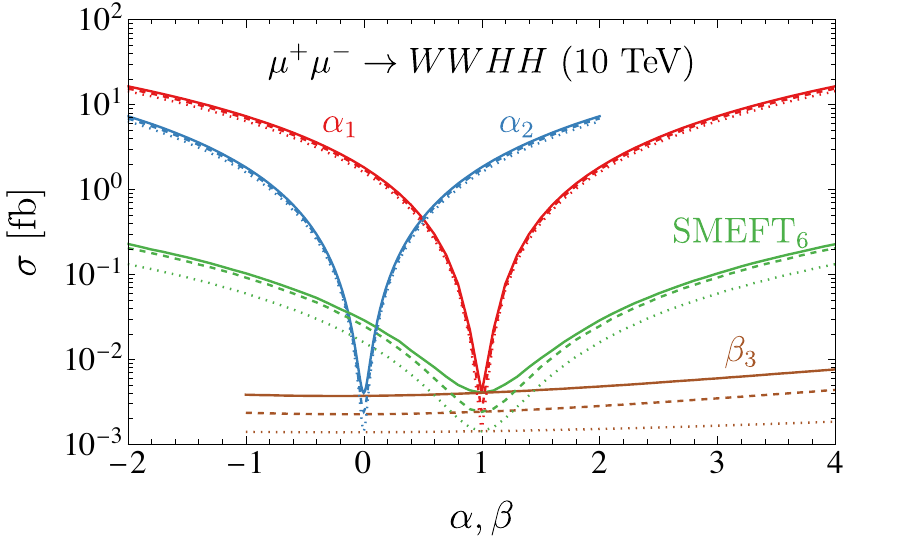}
  \caption{Same as Fig.~\ref{fig:mmnhkappa3} for the $V^2H^2$ and $ZH^3$ production processes,  at 3 and 10 TeV.}
  \label{fig:mmvvhhkappa}
\end{figure}

\begin{figure}[!t]
  \centering
  \includegraphics[width=0.49\textwidth]{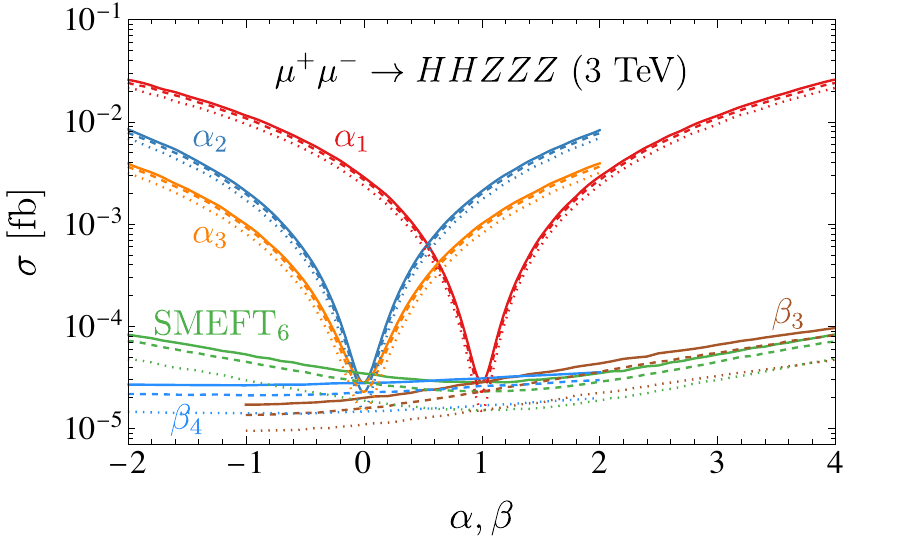}
  \includegraphics[width=0.49\textwidth]{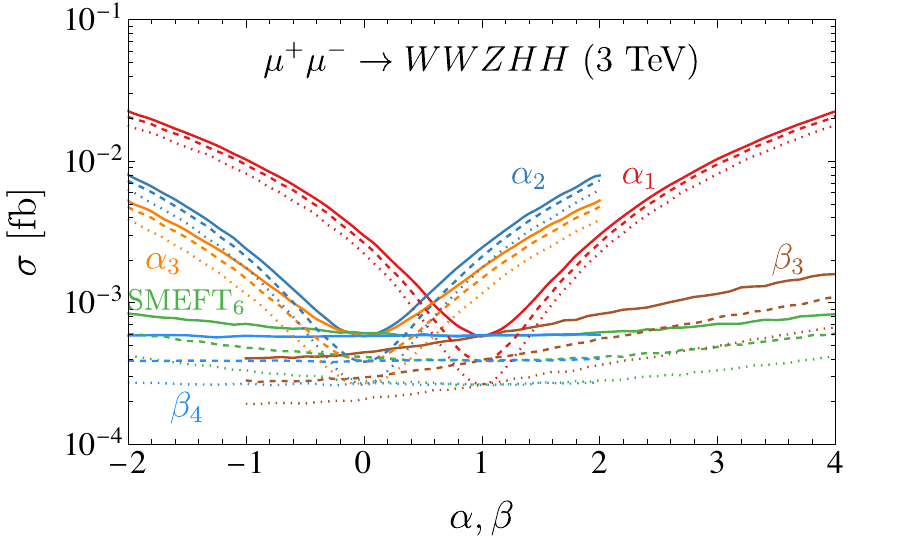}
  \includegraphics[width=0.49\textwidth]{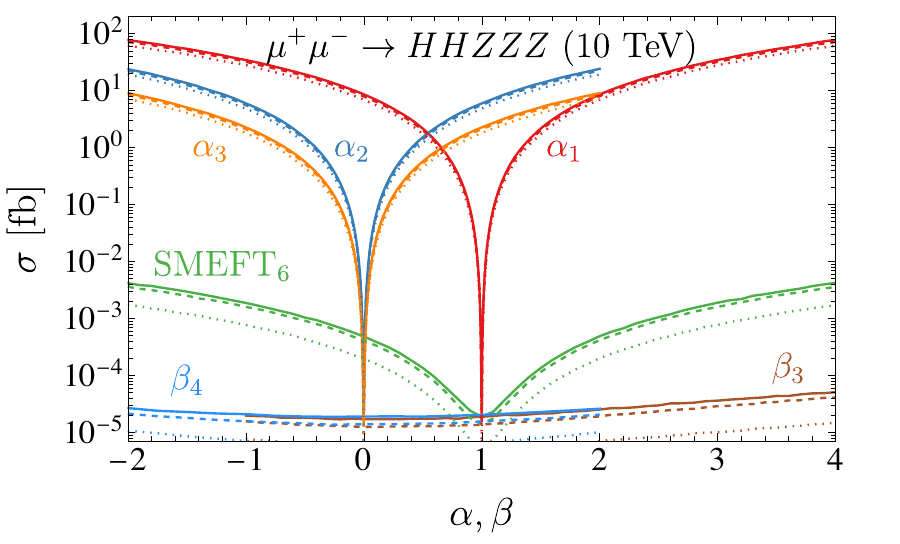}
  \includegraphics[width=0.49\textwidth]{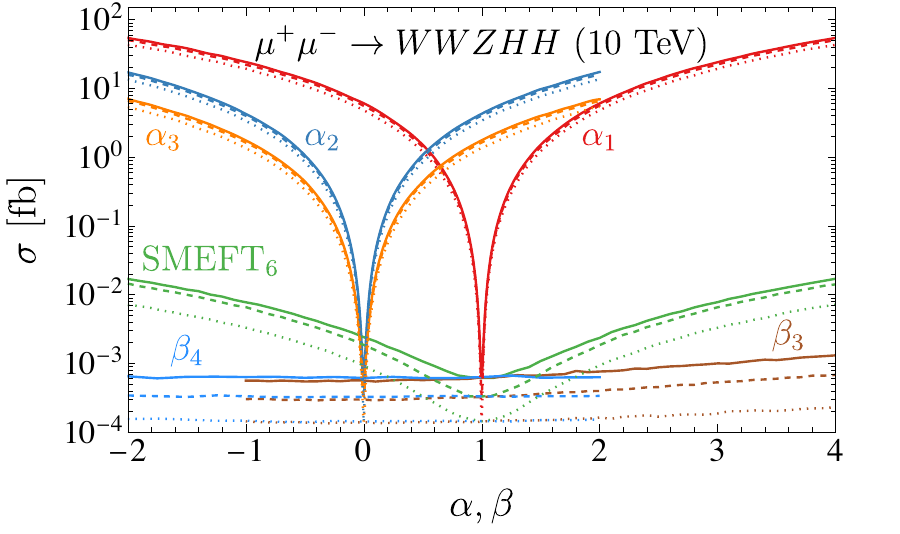}
  \caption{Same as Fig.~\ref{fig:mmnhkappa3} for the  $V^3H^2$ production processes,  at 3 and 10 TeV.}
  \label{fig:mmvvvhhkappa}
\end{figure}

\begin{figure}[!t]
  \centering
  \includegraphics[width=0.49\textwidth]{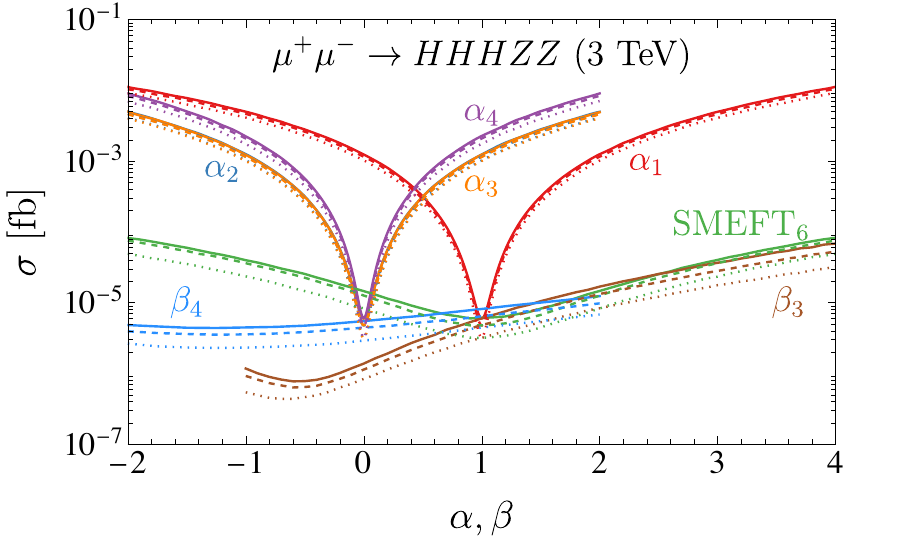}
    \includegraphics[width=0.49\textwidth]{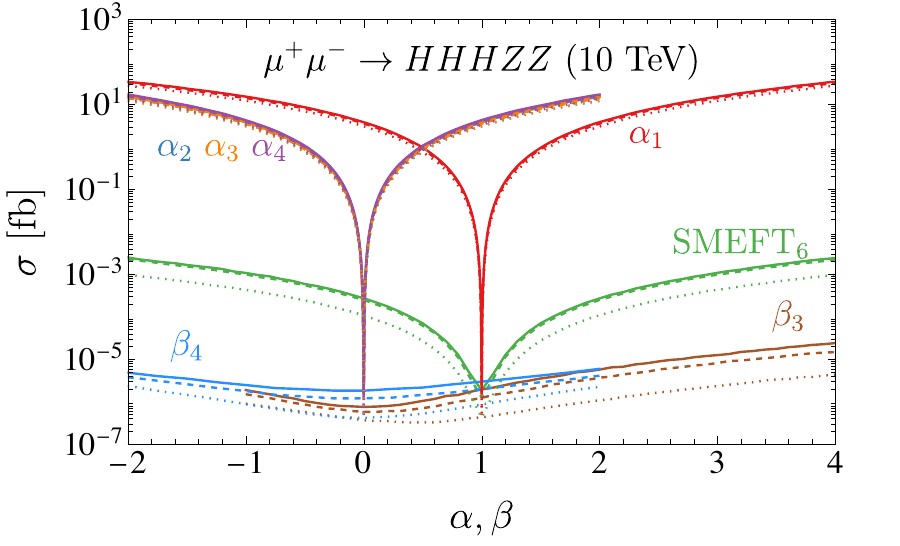}
  \includegraphics[width=0.49\textwidth]{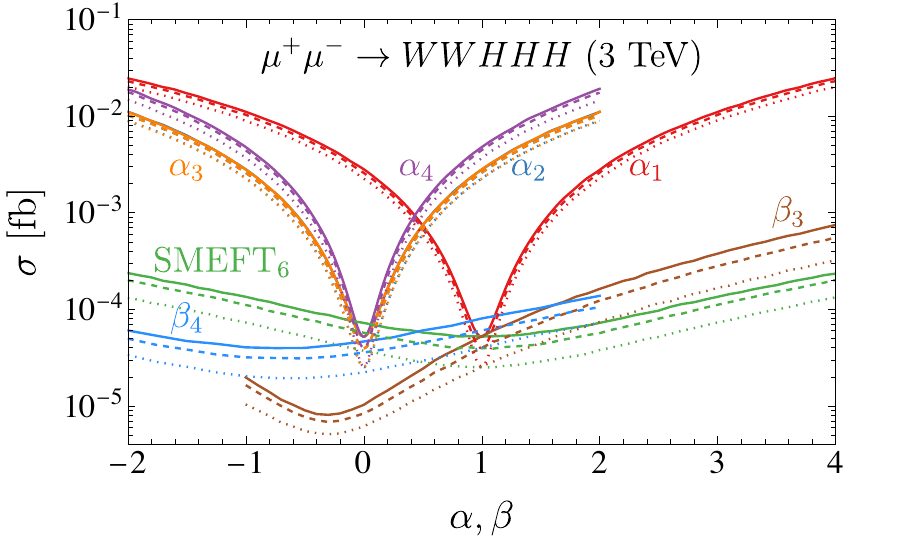}
    \includegraphics[width=0.49\textwidth]{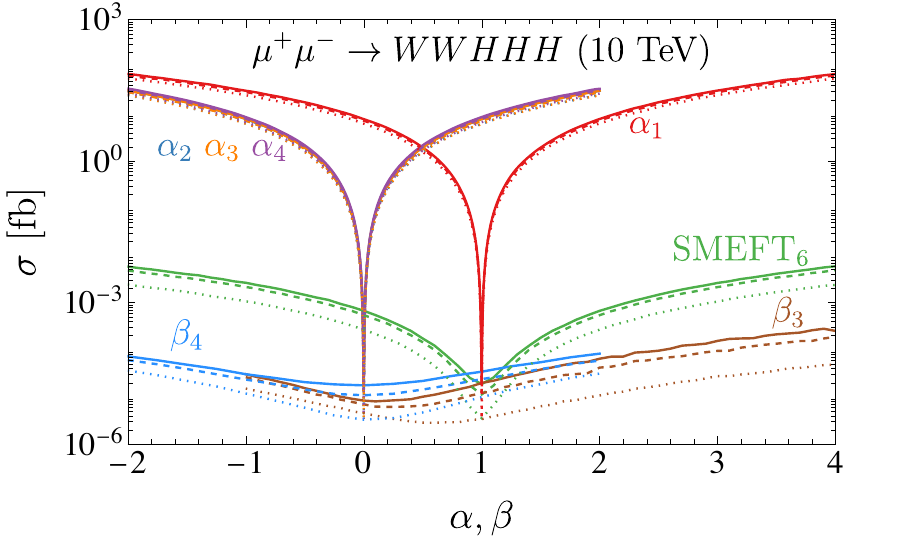}
  \includegraphics[width=0.49\textwidth]{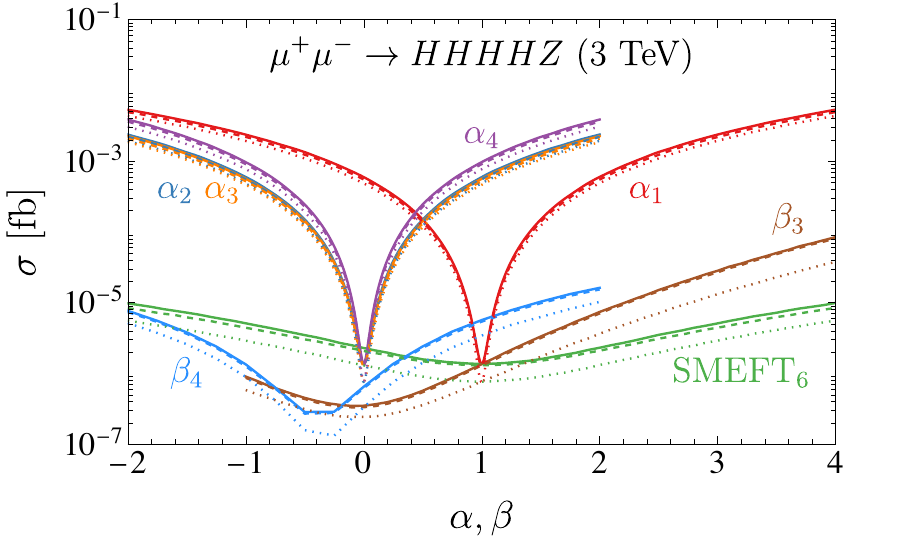}
    \includegraphics[width=0.49\textwidth]{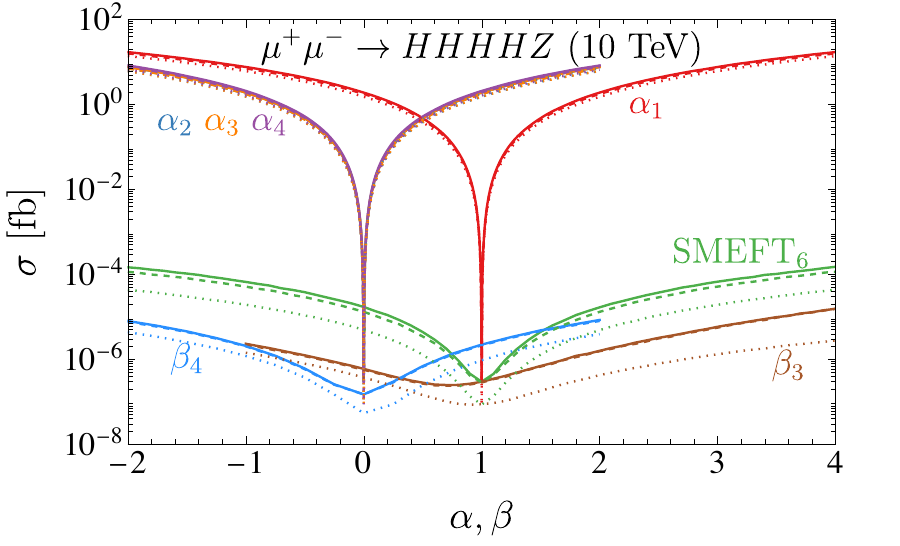}
  \caption{Same as Fig.~\ref{fig:mmnhkappa3} for the $V^2H^3$ and $ZH^4$ production processes,   at 3 and 10 TeV.}
  \label{fig:mmvvhhhkappa}
\end{figure}

\clearpage

\bibliographystyle{JHEP}
\bibliography{ref}

\end{document}